\documentclass[11pt]{article}

\usepackage[paper=letterpaper, margin=1in]{geometry}

\usepackage{ifthen}

\ifthenelse{\isundefined{\GenerateShortVersion}}
{
\def\LongVersion{}
\def\LongVersionEnd{}
\long\def\ShortVersion#1\ShortVersionEnd{}
}
{
\def\ShortVersion{}
\def\ShortVersionEnd{}
\long\def\LongVersion#1\LongVersionEnd{}
}

\newcommand{\Ignore}[1]{\ignorespaces}

\usepackage{amsmath, amssymb}

\usepackage{amsthm}

\usepackage{ifpdf}

\usepackage{authblk}

\ifpdf
\usepackage[pdftex]{graphicx}
\else
\usepackage[dvips]{graphicx}
\fi

\usepackage[%
ocgcolorlinks,%
linkcolor=blue,%
filecolor=blue,%
citecolor=blue,%
urlcolor=blue]{hyperref}

\usepackage{enumitem}

\usepackage{xcolor}

\usepackage{mdframed}

\LongVersion 
\LongVersionEnd 

\ShortVersion 
\usepackage{times}
\ShortVersionEnd 

\ShortVersion 
\renewcommand{\paragraph}[1]{\par\noindent\textbf{#1}}
\ShortVersionEnd 

\newtheorem{theorem}{Theorem}[section]
\newtheorem{lemma}[theorem]{Lemma}
\newtheorem{observation}[theorem]{Observation}
\newtheorem{corollary}[theorem]{Corollary}
\newtheorem{proposition}[theorem]{Proposition}

\newtheorem*{observation*}{Observation}

\theoremstyle{definition}
\newtheorem{criterion}{Criterion}
\newtheorem*{remark*}{Remark}
\theoremstyle{plain}

\usepackage{subcaption}

\usepackage{algpseudocode}
\usepackage{algorithm}
\algtext*{EndWhile} 
\algtext*{EndIf} 
\algtext*{EndFor} 

\floatname{algorithm}{Pseudocode}

\usepackage{tikz}
\usetikzlibrary{decorations.pathmorphing}
\def\CH{\tikz\fill[scale=0.4](0,.35) -- (.25,0) -- (1,.7) -- (.25,.15) -- 
cycle;}

\usepackage{multirow}

\newcommand{\Integers}{\mathbb{Z}}
\newcommand{\Reals}{\mathbb{R}}
\newcommand{\Ex}{\mathbb{E}}
\renewcommand{\Pr}{\mathbb{P}}
\newcommand{\Predicate}{\ensuremath{\ell}}
\newcommand{\Problem}{\ensuremath{\mathcal{P}}}
\newcommand{\Alg}{\ensuremath{\mathcal{A}}}
\newcommand{\SelfStabAlg}{\ensuremath{\mathcal{A}_{\mathit{ST}}}}
\newcommand{\Detect}{\ensuremath{\mathtt{Detect}}}
\newcommand{\PhaseProcedure}{\ensuremath{\mathtt{Phase}}}
\newcommand{\PPS}{\ensuremath{\mathtt{PPS}}}
\newcommand{\Degree}{\mathrm{d}}
\newcommand{\Distance}{\delta}
\newcommand{\DecidedSet}{\mathit{D}}
\newcommand{\UnDecidedSet}{\overline{\DecidedSet}}
\newcommand{\ContentSet}{\Gamma}
\newcommand{\UnContentSet}{\overline{\ContentSet}}
\newcommand{\HoldSymbol}{\mathit{\hbar}}
\newcommand{\Multisets}{\mathcal{M}}

\newcommand{\ContConfGraphs}{\mathcal{CCG}}
\newcommand{\SatMultisets}{\mathcal{T}}
\newcommand{\Cores}{\SatMultisets^{*}}

\newcommand{\DepthSupGraph}{\mathrm{dep}}

\newcommand{\Matched}{\mathit{Mat}}
\newcommand{\UnMatched}{\mathit{UnM}}
\newcommand{\InMIS}{\mathit{IN}}
\newcommand{\OutMIS}{\mathit{OUT}}

\newcommand{\OutReg}{\mathtt{out}}
\newcommand{\Step}{\mathtt{step}}
\newcommand{\Peers}{\mathtt{peers}}
\newcommand{\fieldPPS}{\mathtt{pps}}
\newcommand{\fieldOut}{\mathtt{g\_out}}
\newcommand{\AdvancePPS}{\mathtt{Advance\_PPS}}
\newcommand{\true}{\mathtt{true}}
\newcommand{\false}{\mathtt{false}}
\newcommand{\InMsg}{\mathtt{InMsg}}
\newcommand{\OutMsg}{\mathtt{OutMsg}}
\newcommand{\Wait}{\mathtt{wait}}

\newcommand{\LogDeg}{\operatorname{ld}}
\newcommand{\SSMM}{\ensuremath{\mathtt{SSMM}}}
\newcommand{\Finalize}{\ensuremath{\mathtt{Finalize}}}
\newcommand{\ForceConsistency}{\ensuremath{\mathtt{Force\_Consistency}}}
\newcommand{\MatchedToOther}{\mathit{E}}
\newcommand{\m}{\mathit{m}}
\newcommand{\fieldConsistency}{\mathtt{con}}
\newcommand{\fieldMM}{\mathtt{mm}}
\newcommand{\ChosenNeighbor}{\mathit{chosen}}
\newcommand{\NIL}{\ensuremath{nil}}
\newcommand{\MatchingRequest}{\mathit{Mreq}}
\newcommand{\Accept}{\mathit{Acc}}
\newcommand{\Active}{\mathit{active}}

\newcommand{\Sect}{Sec.}
\newcommand{\Thm}{Thm.}
\newcommand{\Lem}{Lem.}
\newcommand{\Obs}{Obs.}
\newcommand{\Cor}{Cor.}
\newcommand{\Prop}{Prop.}
\newcommand{\Fig}{Fig.}
\newcommand{\Eq}{Eq.}

\begin{document}

\title{Fully Adaptive Self-Stabilizing Transformer for LCL Problems}

\author{Shimon Bitton}
\affil{%
Technion - Israel Institute of Technology.
\texttt{shimonbt1@gmail.com}}
\author{Yuval Emek}
\affil{%
Technion - Israel Institute of Technology.
\texttt{yemek@technion.ac.il}}
\author{Taisuke Izumi}
\affil{%
Osaka University.
\texttt{izumi.taisuke.ist@osaka-u.ac.jp}}
\author{Shay Kutten}
\affil{%
Technion - Israel Institute of Technology.
\texttt{kutten@technion.ac.il}}

\date{}

\maketitle

\begin{abstract}
This paper introduces the first self-stabilizing transformer for local
problems that is time efficient under a constrained bandwidth model.
The transformer is applicable to a wide class of locally checkable labeling
(LCL) problems, converting a given fault free synchronous algorithm $\Alg$
that satisfies certain conditions into a self-stabilizing synchronous
algorithm $\SelfStabAlg$ for the same problem.
The key feature of the transformer is that $\SelfStabAlg$ is \emph{fully
adaptive} in the sense that its time complexity is expressed as a
(logarithmic) function of the number $k$ of nodes that suffered faults,
possibly at different times, since the last legal configuration and the
degree bound $\Delta$, irrespective of the size of the graph.
Other novel characteristics of the transformer include its small message
size overhead, its applicability to randomized algorithms, and the fact that
$\SelfStabAlg$ is anonymous and size-uniform.
By applying the transformer to known algorithms (or simple variants thereof)
for some classic LCL problems, we obtain anonymous size-uniform
self-stabilizing algorithms for these problems whose expected stabilization
time is
$O (\log (k + \Delta))$,
thus improving the state-of-the-art in various different aspects.
\end{abstract}

\section{Introduction}
\label{section:introduction}
Introduced in the seminal paper of Dijkstra~\cite{dijkstra1982self},
\emph{self-stabilization} is a fundamental and extensively studied approach to
fault tolerance requiring that the system recovers from any combination of
transient faults and stabilizes back to a legal configuration (see
\cite{dolev2000self, altisen2019introduction}
for textbooks).
The main performance measure for self-stabilizing algorithms is their
stabilization run-time that is typically measured as a function of the size of
the system:
the more processors are in the system, the longer is the time it takes for the
algorithm to stabilize.
However, while self-stabilizing systems are guaranteed to recover from
\emph{any} number of transient faults, the common scenario is that over a
limited time interval, only \emph{few} faults occur.
Although a small number of faults can seriously hinder the operation of the
whole system \cite{lamport1987distribution}, one may hope that the system
recovers from them faster than the recovery time from a larger number of
faults, regardless of the total number of processors.

With this hope in mind, we turn to the notion of \emph{fully adaptive
run-time} that expresses the recovery time of a self-stabilizing algorithm as
a function of the number of faulty nodes (supporting also dynamic
\emph{topology changes}), rather than the size of the graph $G$ on which it
runs.
Our main contribution is a generic transformer that can be applied to a wide
class of \emph{locally checkable labeling (LCL)} problems \Problem{}
\cite{naor1995can}, converting a given fault free synchronous algorithm
\Alg{} for \Problem{} into a self-stabilizing synchronous algorithm
\SelfStabAlg{}.
For an LCL problem \Problem{} to belong to this class, it should meet certain
\emph{eligibility criteria} imposed on the LCL predicate over which \Problem{}
is defined and on the fault free algorithm \Alg{}.\footnote{%
We emphasize that the synthesis of $\SelfStabAlg$ from $\Alg$, as well
as the analysis of $\SelfStabAlg$'s performance guarantees, are derived
in an automated (black-box) fashion that rely only on the generic eligibility
criteria.}

Given that $\Problem$ is eligible, our transformer ensures that $\SelfStabAlg$
is efficient in terms of its expected stabilization run-time bound.
Specifically, this bound scales logarithmically with the
number $k$ of nodes that experienced transient faults since the last legal
configuration and the graph's global degree bound $\Delta$, irrespective of
the size of $G$;\footnote{%
The transient faults are not assumed to take place simultaneously and may
occur at different times --- see \Sect{}~\ref{section:model-and-preliminaries}.
In this regard, we use the term \emph{fully} adaptive run-time as a
distinction from works analyzing the adaptive run-time of algorithms, where it
is often assumed that all faults/changes occur at the same time.}
it also depends on certain parameters of the eligibility criteria that can
often be regarded as constants (see
\Sect{}~\ref{section:introduction:concrete-problems} for concrete
examples).\footnote{%
In the language of \cite{DBLP:conf/focs/KuttenP95}, the self-stabilizing
algorithm $\SelfStabAlg$ is \emph{tightly fault local}.}
Moreover, the message size overhead of $\SelfStabAlg$ is small (determined by
parameters of the eligibility criteria), which means that our transformer is
suitable for limited bandwidth models (cf.\ CONGEST \cite{Peleg2000}).
Refer to \Thm{}\ \ref{theorem:edges:main} and
\ref{theorem:nodes:main} for the exact expressions of
$\SelfStabAlg$'s stabilization run-time and message size bounds.

Another appealing feature of our transformer is that it is applicable to
a strong self-stabilizing setup, making weak assumptions on the information
that is hard-wired into the nodes' memory (and is thus protected from
adversarial manipulations):
For our transformer to work, the only piece of information that is required to
be hard-wired into the nodes' memory is the global degree bound
$\Delta$.\footnote{%
Actually, this is needed only if the fault free algorithm $\Alg$ relies on the
knowledge of $\Delta$.}
Any other piece of information that the algorithm designer may wish to store
in the nodes' memory, including node IDs or a bound on the graph size, is
subject to adversarial manipulations.
As such, we restrict the transformer's scope to \emph{anonymous} and
\emph{size-uniform} fault free algorithms $\Alg$, which translates to an
anonymous and size-uniform self-stabilizing algorithm $\SelfStabAlg$.
Combined with its fully adaptive run-time, we conclude that $\SelfStabAlg$ can
also work in an \emph{infinite} (bounded degree) graph $G$ (see
\Sect{}~\ref{section:infinite-faults} for more details).

The analysis of our transformer relies on a new parameter of the LCL problem
$\Problem$, referred to as the \emph{propagation radius}.
Intuitively, this parameter bounds the radius to which a single fault may
affect a local checking mechanism
\cite{afek1990memory, awerbuch1991self, naor1995can, korman2010proof,
goos2016locally}
(see \Sect{}~\ref{section:LCL-eligibility-criteria} for more details).
We hope that this new parameter will be useful for the investigation of LCL
problems in other contexts as well.\footnote{%
Refer to \Sect{}~\ref{section:related-work} for a discussion of the
connections between the propagation radius and related notions such as
\emph{labeling radius} \cite{mayer1995local},
\emph{contamination radius}
\cite{ghosh1996fault, turau2018computing},
and
\emph{mending radius} \cite{balliu2021local}.}

While generic transformers, that compile a fault free algorithm into a
self-stabilizing one, have been the topic of many works
\cite{%
KatzPerry,
awerbuch1991distributed,
awerbuch1991self,
DBLP:journals/cjtcs/DolevH97,
kutten1997time,
afek2002local,
ghosh2002scalable,
azar2003distributed,
burman2005asynchronous,
beauquier20061,
lenzen2009local,
kohler2012fault,
balliu2021local}
(see \Sect{}~\ref{section:related-work} for an elaborated discussion), to the
best of our knowledge, we develop the first local transformer that comes with
appealing guarantees for both the stabilization run-time and the message size
overhead.\footnote{%
The term local transformer refers in this regard to a transformer that
produces algorithms whose stabilization run-time is not low-bounded by the
graph's diameter.}
This is also the first local transformer that is not restricted to
deterministic fault free algorithms, regardless of the message size overhead.
Furthermore, our transformer is the first one to (provably) guarantee a fully
adaptive stabilization run-time, irrespective of the graph's size and diameter
or any probabilistic assumption on the distribution of faults.
Finally, we are unaware of any existing transformer that produces anonymous
size-uniform self-stabilizing algorithms or self-stabilizing algorithms that
are applicable (out of the box) to infinite graphs.

\subsection{Concrete LCL Problems}
\label{section:introduction:concrete-problems}
To demonstrate the applicability of the new transformer, we establish the
eligibility of eight LCL problems, transforming known fault free algorithms
(or simple variants thereof) for these problems into self-stabilizing
algorithms.
The eight LCL problems consist of four \emph{distributed node problems} and the
corresponding four \emph{distributed edge problems}, derived from the
distributed node problems by projecting them on the line graph of a given
graph, as listed here:
\begin{itemize}

\item
Distributed node problem:
\emph{maximal independent set (MIS)}
\cite{Luby1986simple, AlonBI1986fast}. \\
Definition:
the output of each node
$v \in V(G)$
is
$f(v) \in \{ 1, 2 \}$;
the nodes $v$ that output
$f(v) = 1$
form an independent set;
if a node
$v \in V(G)$
outputs
$f(v) = 2$,
then there exists at least one neighbor $u$ of $v$ with
$f(u) = 1$. \\
Corresponding distributed edge problem:
\emph{maximal matching (MM)}
\cite{IsraeliI1986fast}.

\item
Distributed node problem:
\emph{node $c$-coloring}.
\cite{ColeV1986deterministic} \\
Definition:
the output of each node
$v \in V(G)$
is
$f(v) \in \{ 1, \dots, c \}$;
the nodes $v$ that output
$f(v) = i$
form an independent set for each
$1 \leq i \leq c$. \\
Corresponding distributed edge problem:
\emph{edge $c$-coloring}
\cite{KarloffS1987efficient}.

\item
Distributed node problem:
\emph{maximal node $c$-coloring}
\cite{Luby1986simple}. \\
Definition:
the output of each node
$v \in V(G)$
is
$f(v) \in \{ 1, \dots, c \}$;
the nodes $v$ that output
$f(v) = i$
form an independent set for each
$1 \leq i \leq c - 1$;
if a node
$v \in V(G)$
outputs
$f(v) = c$,
then there exists at least one neighbor $u_{i}$ of $v$ with
$f(u_{i}) = i$
for each
$1 \leq i \leq c - 1$. \\
Corresponding distributed edge problem:
\emph{maximal edge $c$-coloring}.

\item
Distributed node problem:
\emph{incremental node $c$-coloring}. \\
Definition:
the output of each node
$v \in V(G)$
is
$f(v) \in \{ 1, \dots, c \}$;
the nodes $v$ that output
$f(v) = i$
form an independent set for each
$1 \leq i \leq c - 1$;
if a node
$v \in V(G)$
outputs
$f(v) = i$,
$2 \leq i \leq c$,
then $v$ admits at least
$i - 1$
neighbors $u$ with
$f(u) \leq i - 1$. \\
Corresponding distributed edge problem:
\emph{incremental edge $c$-coloring}.

\end{itemize}
Notice that maximal node $c$-coloring (resp., maximal edge $c$-coloring) and
incremental node $c$-coloring (resp., incremental edge $c$-coloring) are
natural generalizations of MIS (resp., MM),
derived by fixing
$c = 2$.
The former problem is also a natural generalization of node
$(\Delta + 1)$-coloring
(resp., edge
$(2 \Delta - 1)$-coloring),
derived by fixing
$c = \Delta + 2$
(resp.,
$c = 2 \Delta$).

The performance guarantees of the self-stabilizing algorithms generated by our
transformer for the aforementioned eight LCL problems are listed in
Table~\ref{table:transformer:concrete-problems}.
The reader may notice that the fully adaptive run-time bounds of (all but one
of) our self-stabilizing algorithms are logarithmic in the ``natural instance
parameters'', i.e., in $k$ and $\Delta$ (in some cases, this holds as long as
$c = O (1)$).
To the best of our knowledge, these are the first self-stabilizing algorithms
for any non-trivial distributed problem that admit (provable) fully adaptive
run-time bounds whose dependency on $k$ is sub-linear (see
\Sect{}~\ref{section:related-work} for a more elaborated discussion of the
place of our algorithms within the existing literature on self-stabilizing
algorithms).

\begin{table}
{\centering
\resizebox{\textwidth}{!}{%
\begin{tabular}{|l|l|l|l|}
\hline
Problem & Fully adap.\ run-time & Message size & Comments \\
\hline\hline
MIS &
$O (\log (k + \Delta))$ &
$O (\log\log (\Delta))$ &
inc.\ node $2$-coloring
\\
\hline
node
$\lceil (1 + \epsilon) \Delta \rceil$-coloring &
$O (\log (k + \Delta))$ &
$O (\log (\Delta))$ &
\\
constant
$\epsilon > 0$ &
& &
\\
\hline
node
$(\Delta + 1)$-coloring &
$O (\log (k + \Delta))$ &
$O (\Delta \log\log (\Delta))$ &
max.\ node
$(\Delta + 2)$-coloring
\\
\hline
max.\ node $c$-coloring &
$O (\log (k + \Delta))$ &
$O (c \log\log (\Delta))$ &
reduction from MIS
\\
$2 \leq c \leq \Delta + 2$ &
& &
\\
\hline
inc.\ node $c$-coloring &
$O (c \log (k) + c^{2} \log (\Delta))$ &
$O (\log (c) + \log\log (\Delta))$ &
\\
$2 \leq c \leq \Delta + 2$ &
& &
\\
\hline\hline
MM &
$O (\log (k + \Delta))$ &
$O (\log\log (\Delta))$ &
inc.\ edge $2$-coloring
\\
\cline{2-4}
&
$O (\log (k))$ &
$O (1)$ &
not using the transformer
\\
\hline
edge
$\lceil (2 + \epsilon) \Delta \rceil$-coloring &
$O (\log (k + \Delta))$ &
$O (\log (\Delta))$ &
\\
constant
$\epsilon > 0$ &
& &
\\
\hline
edge
$(2 \Delta - 1)$-coloring &
$O (\Delta \log (k + \Delta))$ &
$O (\Delta)$ &
max.\
$(2 \Delta)$-coloring
\\
\cline{2-4}
&
$O (\log (k + \Delta))$ &
$O (\Delta^{2} \log\log (\Delta))$ &
max.\
$(2 \Delta)$-coloring
\\
\hline
max.\ edge $c$-coloring &
$O (c \log (k + \Delta))$ &
$O (c + \log\log (\Delta))$ &
\\
\cline{2-4}
$2 \leq c \leq 2 \Delta$ &
$O (\log (k + \Delta))$ &
$O (c \Delta \log\log (\Delta))$ &
line graph simulation
\\
\hline
inc.\ edge $c$-coloring &
$O (c \log (k) + c^{2} \log (\Delta))$ &
$O (c + \log\log (\Delta))$ &
\\
$2 \leq c \leq 2 \Delta$ &
& &
\\
\hline
\end{tabular}
}
\par}
\caption{\label{table:transformer:concrete-problems}
The fully adaptive run-time and message size of the self-stabilizing
algorithms developed in this paper.
All algorithms are applicable to general graphs.}
\end{table}

Moreover, the algorithms presented in the current paper are among the first
self-stabilizing algorithms for non-trivial LCL problems that are anonymous
and size-uniform.
As established in \cite{KothapalliSOS2006}, for each one of the problems
listed in Table~\ref{table:transformer:concrete-problems}, when restricted to
an $n$-node cycle graph, any anonymous (fault free) algorithm with constant
size messages requires
$\Omega (\log n)$
time in expectation.
This implies that any anonymous self-stabilizing algorithm with constant size
messages requires
$\Omega (\log k)$
time in expectation to stabilize from
$k \leq n$
transient faults.
Since the message size bounds in
Table~\ref{table:transformer:concrete-problems} reduce to
$O (1)$
when $\Delta$ is fixed (as is the case in cycle graphs), we conclude that the
fully adaptive run-time bounds of our self-stabilizing algorithms are
asymptotically optimal in terms of their (logarithmic) dependency on $k$.

\subsection{Paper's Outline}
\label{section:introduction:outline}
The remainder of this paper is organized as follows.
First, in \Sect{}~\ref{section:model-and-preliminaries}, we present the
computational model together with some related definitions that serve us
throughout the paper.
Next, an informal overview of our generic transformer is provided in
\Sect{}~\ref{section:transformer-overview}.
After the informal overview, we turn to the main part of the paper, where the
transformer is formally developed, starting with introducing the transformer's
key module in \Sect{}~\ref{section:probabilistic-phase-synchronization}.
The eligibility criteria imposed on the LCL predicate are then presented in
\Sect{}~\ref{section:LCL-eligibility-criteria}.
While \Sect{} \ref{section:probabilistic-phase-synchronization} and
\ref{section:LCL-eligibility-criteria} are common to both node and edge
problems, the rest of the transformer's components diverge between the two and
are therefore presented separately:
the transformer for distributed edge problems is developed and analyzed in
\Sect{}~\ref{section:transformer-edges} and the transformer for distributed
node problems, that turns out to include a few additional complications, is
developed and analyzed in \Sect{}~\ref{section:transformer-nodes}.

In \Sect{}~\ref{section:simulation-line-graph}, we explain how the
self-stabilizing algorithms produced by our transformer for distributed node
problems can be simulated on the line graph of a given graph, providing an
alternative --- usually less efficient in terms of the message size --- way to
develop self-stabilizing algorithms for distributed edge problems.
While it is clear from the theorems stated in \Sect{}\
\ref{section:transformer-edges} and \ref{section:transformer-nodes} that the
guarantees of the self-stabilizing algorithms produced by our transformer hold
against any combination of finitely many transient faults in (finite or
countably infinite) graphs, in \Sect{}~\ref{section:infinite-faults} we show
that under certain conditions, they also hold against infinitely many
transient faults.

The eligibility of the problems listed in
\Sect{}~\ref{section:introduction:concrete-problems} is established in
\Sect{} \ref{section:concrete-problems-edges} and
\ref{section:concrete-problems-nodes} together with the bounds presented in
Table~\ref{table:transformer:concrete-problems} for the performance guarantees
of the self-stabilizing algorithms produced by our transformer.
One of the rows in Table~\ref{table:transformer:concrete-problems}
corresponds to an improved MM algorithm that has a lot of resemblance with the
MM algorithm produced via the transformer (in particular, it also relies on
the module presented in
\Sect{}~\ref{section:probabilistic-phase-synchronization}), but nevertheless,
includes a small ``tailor-made twist'';
this algorithm and its analysis are presented in
\Sect{}~\ref{section:improved-MM}.
On the negative side, we discuss in
\Sect{}~\ref{section:non-eligibile-problems} a certain type of LCL problems
that are not eligible for our transformer.
We conclude in \Sect{}~\ref{section:related-work} with a related work
comparison and further discussion.

\section{Computational Model and Preliminaries}
\label{section:model-and-preliminaries}

\paragraph{Multisets.}
Throughout, we represent a multiset $M$ over a finite ground set $S$ as a
vector
$M \in \Integers_{\geq 0}^{S}$
so that $M(s)$ indicates the multiplicity of
$s \in S$.
The size of $M$ is defined to be the $1$-norm of its vector representation,
denoted by
$|M| = \sum_{s \in S} M(s)$.
The empty multiset over $S$, denoted by $\emptyset$, is the unique multiset of
size $0$.
For
$d \in \Integers_{\geq 0}$,
let $\Multisets_{d}(S)$ denote the collection of all multisets over $S$ of
size $d$ and let
$\Multisets(S) = \bigcup_{d = 0}^{\infty} \Multisets_{d}(S)$.

An element 
$s \in S$
is considered to be included in a multiset
$M \in \Multisets(S)$,
denoted by
$s \in M$,
if
$M(s) \geq 1$.
Given two multisets
$M, M' \in \Multisets(S)$,
the relation
$M \subseteq M'$
(or
$M' \supseteq M$)
holds if
$M(s) \leq M'(s)$
for each
$s \in S$;
the relation
$M \subset M'$
(or
$M' \supset M$)
holds if
$M \subseteq M'$
and
$M \neq M'$.
We note that the $\subseteq$ relation is a partial order that induces a
lattice on $\Multisets(S)$.

\paragraph{Graphs.}
Throughout, the term graph refers to an undirected graph, whereas the term
digraph refers to a directed graph.
We denote the node set and edge set of a graph/digraph $G$ by $V(G)$ and
$E(G)$, respectively.

Consider a graph $G$.
The \emph{line graph} of $G$, denoted by $L(G)$, is the graph
defined by setting
$V(L(G)) = E(G)$
and
$E(L(G)) = \{ \{ e, e' \} \mid e, e' \in E(G), |e \cap e'| = 1 \}$.
The set of neighbors of a node
$v \in V(G)$
is denoted by
$N_{G}(v) = \{ u \in V(G) \mid \{ u, v \} \in E(G) \}$
and the degree of $v$ is denoted by
$\Degree_{G}(v) = |N_{G}(v)|$.
The set of neighbors of an edge
$e \in E(G)$
is denoted by
$N_{G}(e) = N_{L(G)}(e)$
and the degree of $e$ is denoted by
$\Degree_{G}(e) = |N_{G}(e)|$.
Given a positive integer $\Delta$, let $\mathcal{U}_{\Delta}$ denote the
collection of all finite and countably infinite graphs whose node degrees are
up-bounded by $\Delta$.
Let
$\mathcal{U} = \bigcup_{\Delta = 1}^{\infty} \mathcal{U}_{\Delta}$
denote the collection of all finite and countably infinite graphs with finite
degrees.

The (hop-)distance in $G$ between nodes
$v, v' \in V(G)$
is denoted by
$\Distance_{G}(v, v')$.
This notation is generalized to node subsets
$U, U' \subseteq V(G)$
by defining
$\Distance_{G}(U, U') = \min_{v \in U, v' \in U'} \Distance_{G}(v, v')$,
adhering to the convention that
$\Distance_{G}(U, \emptyset) = \infty$.

To facilitate the exposition, we often use the notion of a \emph{graph object}
$x$ that, depending on the context, can be either a node or an edge, denoting
the set of all such graph objects by $X(G)$ (that serves as a placeholder for
either $V(G)$ or $E(G)$).
For a subset
$S \subseteq X(G)$
of graph objects, let $G(S)$ denote the subgraph induced by $S$ on $G$ and let
$G - S = G(X(G) - S)$.

\paragraph{Configurations.}
Fix a set $\mathcal{O}$ of output values and let $\bot$ be a designated 
symbol that does not belong to $\mathcal{O}$.
Consider a graph
$G \in \mathcal{U}$.
A \emph{node/edge configuration} is a function
$C : X(G) \rightarrow \mathcal{O} \cup \{ \bot \}$
that assigns to each node/edge in $X(G)$ either an output
value in $\mathcal{O}$ or $\bot$.\footnote{%
We emphasize that a node/edge configuration is a purely combinatorial notion.
In particular, a node configuration should not be confused with the global
state of a distributed algorithm running on the graph $G$ (discussed in the
sequel).}\,\footnote{%
We may use the term configuration, omitting the words node/edge, when those
are clear from the context.}
To avoid misunderstandings, we emphasize that in the current paper, we deal
with node configurations (that assign values in
$\mathcal{O} \cup \{ \bot \}$
to the nodes in $V(G)$) and with edge configurations (that assign values in
$\mathcal{O} \cup \{ \bot \}$
to the edges in $E(G)$), but not with ``mixed configurations''.

Let
$\DecidedSet(C) = \{ x \in X(G) \mid C(x) \in \mathcal{O} \}$
denote the set of graph objects that are \emph{decided} under $C$ and
let
$\UnDecidedSet(C) = X(G) - \DecidedSet(C)$
denote the set of graph objects that are \emph{undecided} under $C$.
The node/edge configuration $C$ is said to be \emph{complete} if
$\DecidedSet(C) = X(G)$.
For a graph object
$x \in X(G)$,
let
$C[x]
=
\left\{ C(y) \right\}_{y \in \DecidedSet(C) \cap N_{G}(x)}$
denote the multiset consisting of the output values of $x$'s decided neighbors
under $C$.

\paragraph{Distributed Problems.}
A \emph{distributed node/edge problem} is a
$3$-tuple
$\Problem = \langle
\mathcal{O}, \mathcal{G}, \{ \mathcal{C}_{G} \}_{G \in \mathcal{G}}
\rangle$,
where
$\mathcal{O}$ is a set of output values,
$\mathcal{G} \subseteq \mathcal{U}$
is a family of graphs,
and
$\mathcal{C}_{G} \subseteq \mathcal{O}^{X(G)}$
is a collection of \emph{legal} complete node/edge configurations for graph
$G \in \mathcal{G}$.\footnote{%
Our techniques can be extended to cover also distributed node/edge problems in
which the nodes/edges are hard-coded with an input value, but this goes beyond
the scope of the current paper.}
All distributed node/edge problems considered in the current paper are
\emph{hereditary closed} in the sense that
$G \in \mathcal{G}$
implies that
$G' \in \mathcal{G}$
for every subgraph $G'$ of $G$.

\paragraph{Distributed Algorithms.}
Consider a message passing communication network represented by a graph
$G \in \mathcal{U}_{\Delta}$.
The nodes of $G$ are associated with identical (possibly randomized) state
machines and we refer to the  vector encoding all node states as the network's
\emph{global state}.
These state machines operate in synchronous \emph{rounds} so that in each round
$t \in \Integers_{\geq 0}$,
every node
$v \in V(G)$
executes the following operations:
(1)
$v$ performs local computation and updates its state;
(2)
$v$ sends messages to its neighbors;
and
(3)
$v$ receives the messages sent to it by its neighbors in round $t$.
Throughout, we assume that round $t$ occurs during the time interval
$[t, t + 1)$
so that time $t$ marks the beginning of the round.

We adhere to the conventions of the \emph{port numbering} model of 
distributed
graph algorithms:
Node
$v \in V(G)$
has a port
$i_{u} \in \{ 1, \dots, \Degree_{G}(v) \}$
for each neighbor
$u \in N_{G}(v)$
such that from $v$'s perspective, $u$ is regarded as neighbor $i_{u}$.
The nodes do not have unique identifiers and they are not assumed to know the
graph size (if the graph is finite).
The only piece of global information that the nodes are allowed to hold is the
degree bound $\Delta$.

Consider a distributed node problem
$\Problem = \langle
\mathcal{O}, \mathcal{G}, \{ \mathcal{C}_{G} \}_{G \in \mathcal{G}}
\rangle$
and an algorithm $\Alg$ for $\Problem$.
When $\Alg$ runs on a graph
$G \in \mathcal{G}$,
the state of each node
$v \in V(G)$
includes a designated \emph{output register} (typically among other registers)
denoted by $\OutReg_{v}$.
The output register $\OutReg_{v}$ holds an output value in $\mathcal{O}$ or
$\bot$ if $v$ has no output value (yet).
Let
$\OutReg_{v, t}$
denote the value of register $\OutReg_{v}$ at time $t$.
We refer to the node configuration
$C_{t} : V(G) \rightarrow \mathcal{O} \cup \{ \bot \}$
defined by setting
$C_{t}(v) = \OutReg_{v, t}$
for each node
$v \in V(G)$
as the \emph{configuration associated with $\Alg$} at time $t$.
Node
$v \in V(G)$
is said to be \emph{decided} (resp., \emph{undecided}) under $\Alg$ at time
$t$ if it is decided (resp., undecided) under $C_{t}$.

Consider a distributed edge problem
$\Problem = \langle
\mathcal{O}, \mathcal{G}, \{ \mathcal{C}_{G} \}_{G \in \mathcal{G}}
\rangle$
and an algorithm $\Alg$ for $\Problem$.
When $\Alg$ runs on a graph
$G \in \mathcal{G}$,
the state of each node
$v \in V(G)$
includes a designated \emph{output register} (typically among other registers)
denoted by $\OutReg_{v}(i)$ for each port
$i \in \{ 1, \dots, \Degree_{G}(v) \}$;
to simplify the exposition, we often write $\OutReg_{v}(u)$ instead of 
$\OutReg_{v}(i)$
when port $i$ of $v$ corresponds to
$u \in N_{G}(v)$.
The output register $\OutReg_{v}(u)$ holds an output value in $\mathcal{O}$ or
$\bot$ if $v$ has no output value (yet) for edge
$e = \{ u, v \}$;
edge $e$ is said to be \emph{port-consistent} under \Alg{} if
$\OutReg_{v}(u) = \OutReg_{u}(v)$,
and \emph{port-inconsistent} otherwise.
Let
$\OutReg_{v, t}(u)$
denote the value of register $\OutReg_{v}(u)$ at time $t$.
We refer to the edge configuration
$C_{t} : E(G) \rightarrow \mathcal{O} \cup \{ \bot \}$,
defined by setting
$C_{t}(\{ u, v \}) = \OutReg_{v, t}(u) = \OutReg_{u, t}(v)$
if edge
$\{ u, v \} \in E(G)$
is port-consistent at time $t$ and
$C(\{ u, v \}) = \bot$
otherwise, as the \emph{configuration associated with $\Alg$} at time $t$.
Edge
$e \in E(G)$
is said to be \emph{decided} (resp., \emph{undecided}) under $\Alg$ at time
$t$ if it is decided (resp., undecided) under $C_{t}$.

A distributed algorithm is called \emph{fault free} if it is guaranteed
that its execution commences from a known global state (or a restricted family
thereof) and proceeds according to the state transitions determined by the
algorithm designer.

\paragraph{Phase-Based Algorithms.}
In this paper, we focus on a class of fault free algorithms referred to as
\emph{phase-based}.
The execution of a phase-based algorithm $\Alg$ for a distributed node (resp.,
edge) problem
$\Problem = \langle
\mathcal{O}, \mathcal{G}, \{ \mathcal{C}_{G} \}_{G \in \mathcal{G}}
\rangle$
is divided into \emph{phases} so that every phase consists of $\phi$ rounds,
where
$\phi \in \Integers_{> 0}$
is a parameter of $\Alg$.
The rounds within a phase are referred to as \emph{steps}, indexed by
$j = 0, 1, \dots, \phi - 1$.
The execution of $\Alg$ progresses by running the phases in succession so
that every node executes step
$0 \leq j \leq \phi - 1$
of phase
$i = 0, 1, \dots$
in round
$t = i \phi + j$.

Consider the graph
$G \in \mathcal{G}$
on which the phase-based algorithm $\Alg$ runs.
The operation of $\Alg$ is fully specified by a \emph{phase procedure},
denoted by $\PhaseProcedure_{\Alg}$, that dictates the actions of a node
$v \in V(G)$
in each step
$j = 0, 1, \dots, \phi - 1$
of phase $i$.
A key feature of the phase-based algorithm is that the code of
$\PhaseProcedure_{\Alg}$ for step $j$ is oblivious to the phase number $i$.
As such, $\PhaseProcedure_{\Alg}$ should be seen as a procedure that is
invoked from scratch at the beginning of every phase and runs for $\phi$
rounds.
Moreover, at the beginning of each phase, node $v$ resets all its registers
with the exception of the output register $\OutReg_{v}$ (resp., output
registers $\OutReg_{v}(u)$ for
$u \in N_{G}(v)$),
which means that the only information passed from one phase to the next is the
node (resp., edge) configuration $C$ associated with $\Alg$;
in the context of the new phase, we refer to $C$ as the \emph{initial
configuration} of $\PhaseProcedure_{\Alg}$.
In other words, the usage of all non-output registers of $\Alg$ is confined to
the scope of a single phase.

Our attention in the current paper is restricted to phase-based algorithms in
which nodes are allowed to write into their output registers only during the
last step
$j = \phi - 1$
of the phase, referred to as the \emph{decision step}.
Moreover, the decisions of $\Alg$ are required to be irrevocable in the sense
that $\PhaseProcedure_{\Alg}$ is not allowed to write into output registers that
already hold an output value (rather than $\bot$).
We further assume that
$\phi \geq 3$
(which is without loss of generality as one can always extend the phase with
empty steps).

\paragraph{Self-Stabilization and Fully Adaptive Run-Time.}
An algorithm $\Alg$ for
$\Problem = \langle
\mathcal{O}, \mathcal{G}, \{ \mathcal{C}_{G} \}_{G \in \mathcal{G}}
\rangle$
is \emph{self-stabilizing} if it is guaranteed to reach a legal configuration 
in finite time with probability $1$ from any initial global state.
In the current paper, this notion is captured by a malicious \emph{adversary}
that can modify the content of any register maintained by $\Alg$ (essentially
modifying the nodes' states), including the output registers, the registers
that hold the incoming messages, and any register that may maintain the global
round number $t$ (or a function thereof).
The adversary can also impose dynamic \emph{topology changes}, including the
addition and removal of nodes and edges, as long as the resulting graph
remains in $\mathcal{G}$.
The only piece of information that the adversary cannot modify is the degree
bound $\Delta$, assumed to be hard-coded into the nodes' memory.

To simplify the discussion, we assume that the adversarial manipulations that
occur in round
$t \in \Integers_{\geq 0}$
take place towards the end of the round (say, at time
$t + 1 - \epsilon$),
i.e., after the messages sent in round $t$ have reached their destinations and
before the local computation of round
$t + 1$
begins (this assumption is without loss of generality).

Consider a graph
$G \in \mathcal{G}$
on which $\Alg$ runs.
Node
$v \in V(G)$
is said to be \emph{manipulated} by the adversary if either
(i)
the content of (any of) $v$'s registers is modified;
(ii)
the bijection between $v$'s ports and $N_{G}(v)$ is modified;
or
(iii)
$v$ is added to the graph as a new node.
Notice that condition (ii) includes topological changes involving the edges
incident on $v$ as well as ``rewiring'' of $v$'s ports.

Let $C_{t}$ denote the (node/edge) configuration of \Alg{}'s execution on $G$ at
time $t$.
Consider times
$t^{\circ} < t^{*}$
and an integer
$k > 0$
and suppose that
(1)
configuration
$C_{t^{\circ} - 1} = C_{t^{\circ}}$
is legal and no node is manipulated in round
$t^{\circ} - 1$;
(2)
$k$ nodes (in total) are manipulated during the round interval
$[t^{\circ}, t^{*} - 1]$;
and
(3)
no node is manipulated from round $t^{*}$ onwards.
We say that \Alg{} has \emph{fully adaptive run-time}
$T(\Delta, k)$
for a function
$T : \Integers_{> 0} \times \Integers_{> 0} \rightarrow \Integers_{> 0}$
if there exists a time
$t > t^{*}$
such that
(i)
$C_{t}$ is a legal configuration for the graph resulting from the adversarial
manipulations, i.e., the graph at time $t^{*}$;
(ii)
$C_{t'} = C_{t}$
for every
$t' \geq t$;
and
(iii)
$t \leq t^{*} + T(\Delta, k)$.
If $t$ and $C_{t}$ are random variables determined by the coin tosses of
\Alg{}, then we require that condition (iii) holds in expectation.
It is important to point out that the number $k$ of manipulated nodes is
chosen by the adversary;
the self-stabilizing algorithm does not know $k$ (or any approximation
thereof).

\paragraph{Locally Checkable Labelings.}
A \emph{locally checkable labeling (LCL)} over the output value set
$\mathcal{O}$ is a predicate
$\Predicate :
\mathcal{O} \times \Multisets(\mathcal{O})
\rightarrow
\{ \true, \false \}$.
We restrict our attention to predicates $\Predicate$ that do not admit ``ghost
output values'', namely, output values
$o \in \mathcal{O}$
such that
$\Predicate(o, M) = \false$
for every
$M \in \Multisets(\mathcal{O})$.

Fix an LCL $\Predicate$ and consider a graph
$G \in \mathcal{U}$.
Given a node/edge configuration
$C : X(G) \rightarrow \mathcal{O} \cup \{ \bot \}$,
a decided graph object
$x \in \DecidedSet(C)$
is said to be \emph{content} under $C$ (with respect to $\Predicate$) if
$\Predicate(C(x), C[x]) = \true$,
that is, if $\Predicate$ is evaluated to $\true$ on the output value $C(x)$ of
$v$ and the multiset $C[x]$ of output values of $x$'s decided neighbors;
otherwise, $x$ is said to be \emph{uncontent}.
Let $\ContentSet(C)$ be the set of graph objects that are content under $C$
and let
$\UnContentSet(C) = \DecidedSet(C) - \ContentSet(C)$
be the set of graph objects that are (decided yet) uncontent under $C$.
We emphasize that the distinction between content and uncontent nodes/edges
applies only to decided nodes/edges.
The node/edge configuration $C$ is said to be \emph{content} if
$\ContentSet(C) = \DecidedSet(C)$,
namely, if every decided graph object is also content.

LCL predicates facilitate the definition of an important class of distributed
problems.
A \emph{node-LCL} (resp., \emph{edge-LCL}) is a $3$-tuple
$\Problem = \langle
\mathcal{O}, \mathcal{G}, \Predicate
\rangle$,
where
$\mathcal{O}$ is a set of output values,
$\mathcal{G} \subseteq \mathcal{U}$
is a family of graphs,
and
$\Predicate$
is an LCL over the output value set $\mathcal{O}$.
The distributed node (resp., edge) problem $\Problem$ is defined by
classifying a node (resp., edge) configuration of a given graph
$G \in \mathcal{G}$
as legal for $\Problem$ if and only if it is complete and content (with
respect to $\Predicate$).

\section{Transformer's Overview}
\label{section:transformer-overview}
In this section, we provide an overview of the transformer's operation.
As the transformer for distributed node problems includes some additional
complications on top of that of distributed edge problems, we focus in the
current section on the latter.
Consider an edge-LCL
$\Problem = \langle
\mathcal{O}, \mathcal{G}, \Predicate
\rangle$
to which we wish to apply our transformer.
Assume that
$\mathcal{G} \subseteq \mathcal{U}_{\Delta}$
for a degree bound
$\Delta$
and let
$G \in \mathcal{G}$
be an input graph.
The transformer synthesizes a self-stabilizing algorithm $\SelfStabAlg$
for $\Problem$ from two algorithmic building blocks.
The first and primary algorithmic building block is a phase-based fault free
algorithm $\Alg$ for $\Problem$.
Broadly speaking, $\SelfStabAlg$ runs the phase procedure
$\PhaseProcedure_{\Alg}$ of $\Alg$ on the subgraph of $G$ induced by the
undecided edges
$\{ u, v \} \in E(G)$
and this procedure is responsible for assigning output values from
$\mathcal{O}$ to the corresponding output registers $\OutReg_{u}(v)$ and
$\OutReg_{v}(u)$ (more on that soon).

The second building block is a distributed \emph{detection procedure} for
$\Problem$, denoted hereafter by $\Detect$, that runs indefinitely and
provides a distributed implementation of the LCL predicate $\Predicate$,
determining for each decided edge whether it is content or uncontent.
Upon detecting an uncontent edge
$e = \{ u, v \} \in E(G)$,
the output registers $\OutReg_{u}(v)$ and $\OutReg_{v}(u)$ are reset,
assigning
$\OutReg_{u}(v), \OutReg_{v}(u) \gets \bot$,
thus making $e$ undecided.

When it comes to the synthesis of $\SelfStabAlg$ from the phase procedure
$\PhaseProcedure_{\Alg}$ and the detection procedure $\Detect$, a major
challenge is concerned with integrating the operation of the two procedures,
ensuring that the execution does not get into ``vicious cycles'' of assigning
output values to undecided output registers by the former procedure and then,
resetting the same output registers by the latter one.
To a large extent, this is where the eligibility criteria imposed on
$\Problem$ come into play:
As will be explained later, the eligibility criteria allow us to prove that
starting from a certain (deterministic) ``fault recovery'' time
$t^{r} = t^{*} + \chi$,
where $\chi$ is a parameter of the eligibility criteria, it is guaranteed that
all configurations are content, which means that resets of output registers no
longer occur.
From that time on, the task of converging to a valid global state reduces to
the task of augmenting an (arbitrary) content configuration into a complete
content configuration.

One may hope that the robustness of phase-based algorithms alone ensures that
starting from time $t^{r}$, repeated invocations of $\PhaseProcedure_{\Alg}$
eventually ``take care'' of all undecided edges, leading to a complete (legal)
configuration.
Unfortunately, this (intuitive) argument hides a significant flaw:
The correctness of the phase-based algorithm $\Alg$ requires that
$\PhaseProcedure_{\Alg}$ runs at all nodes in synchrony so that step
$j = 0, 1, \dots, \phi - 1$
of phase
$i = 0, 1, \dots$
is executed by all nodes concurrently in round
$t = i \phi + j$,
where $\phi$ denotes the phase length of $\Alg$.
To satisfy this requirement, the nodes should maintain a \emph{modular clock}
that keeps track of
$j = t \bmod \phi$.
While maintaining such a modular clock in a fault free environment is a trivial
task, the situation becomes trickier in the self-stabilizing realm as the
adversary may manipulate the modular clocks, thus causing the corresponding
nodes to execute their phases ``out of sync'' indefinitely.

We cope with this difficulty by introducing a technique called
\emph{probabilistic phase synchronization} (see
\Sect{}~\ref{section:probabilistic-phase-synchronization}).
In high level terms, this technique replaces the (deterministic) modular clock
by a generic (problem independent) randomized module, referred to as $\PPS$,
that can be viewed as a Markov chain maintained independently by each node
$v \in V(G)$.
This module determines, in a probabilistic fashion, when $v$ invokes
$\PhaseProcedure_{\Alg}$.
By bounding the mixing time of the Markov chain underlying $\PPS$, we
guarantee that each node
$v \in V(G)$
has ample opportunities to execute $\PhaseProcedure_{\Alg}$ in synchrony with
every neighbor
$u \in N_{G}(v)$.
Combined with an additional eligibility criterion ensuring that $\Alg$ admits
a certain type of potential function, we can establish (probabilistically)
sufficient progress per round, ultimately bounding the fully adaptive run-time
of $\SelfStabAlg$.

\section{Probabilistic Phase Synchronization}
\label{section:probabilistic-phase-synchronization}
Fix a node/edge-LCL
$\Problem = \langle
\mathcal{O}, \mathcal{G}, \Predicate
\rangle$
and a fault free algorithm $\Alg$ for $\Problem$ that we wish to transform
into a self-stabilizing algorithm $\SelfStabAlg$.
Assume that $\Alg$ is a phase-based algorithm with a phase of length
$\phi \geq 3$
and a phase procedure $\PhaseProcedure_{\Alg}$.

In the current section, we present the $\PPS$ module whose role is to overcome
the ``out of sync'' challenge (see the discussion in
\Sect{}~\ref{section:transformer-overview}), providing the nodes of the input
graph
$G \in \mathcal{G}$
with a mean to synchronize, in a probabilistic fashion, between the phases
they run.
The $\PPS$ module can be viewed as an ergodic Markov chain with state space
$\mathcal{S}_{\phi} = \{ \HoldSymbol, 0, 1, \dots, \phi - 1 \}$
that each node
$v \in V(G)$
maintains internally, independently of the other nodes, storing its
$\PPS$ state in a designated register denoted by $\Step_{v}$.
Under $\SelfStabAlg$, node $v$ simulates step
$0 \leq j \leq \phi - 1$
of $\PhaseProcedure_{\Alg}$ whenever
$\Step_{v} = j$;
the crux of the transformer is that this simulation is done in conjunction
with, and only with, $v$'s neighbors
$u \in N_{G}(v)$
that are synchronized with $v$ in the sense that
$\Step_{u} = j$
(see \Sect{} \ref{section:edges:implementation} and
\ref{section:nodes:implementation}).

The module includes one procedure, denoted by $\AdvancePPS$, that $v$ invokes
in every round, towards the end of the local computation stage.
The role of $\AdvancePPS$ is to advance the Markov chain and this is the only
procedure that has write access to $\Step_{v}$;
any other access of $\SelfStabAlg$ to this register is read-only.
The properties of $\AdvancePPS$ are summarized in the following observation,
where we take
$\Step_{v, t}$
to denote the value of register $\Step_{v}$ at time $t$ (see
\Fig{}~\ref{figure:transformer:markov-chain} for the Markov chain's transition
chart).

\begin{observation} \label{observation:transformer:step-counter}
For every time
$t \geq t^{*}$
and node
$v \in V$,
we have
\begin{itemize}

\item
$\Pr(\Step_{v, t + 1} = j + 1 \mid \Step_{v, t} = j)
=
1$
for
$0 \leq j < \phi - 1$;

\item
$\Pr(\Step_{v, t + 1} = \HoldSymbol \mid \Step_{v, t} = \phi - 1)
=
1$;
and

\item
$\Pr(\Step_{v, t + 1} = \HoldSymbol \mid \Step_{v, t} = \HoldSymbol)
=
\Pr(\Step_{v, t + 1} = 0 \mid \Step_{v, t} = \HoldSymbol)
=
1 / 2$.
\end{itemize}
This holds independently of any coin toss of $v$ prior to time $t$ and of 
any
coin toss of all other nodes.
\end{observation}

We now turn to show that node $v$ has ample opportunities to simulate each
step
$0 \leq j \leq \phi - 1$
of $\PhaseProcedure_{\Alg}$.
To this end, we up-bound the mixing time of the $\PPS$ Markov chain, proving
that regardless of the current state of $v$'s $\PPS$ module, $v$ starts a
phase sufficiently soon with a sufficiently high probability, independently of
any other node.
This is formally cast in the following lemma, established by showing that the
$\PPS$ ergodic Markov chain belongs to a family of Markov chains that has been
identified and analyzed in \cite{wilmer1999exact};
its proof is deferred to
Appendix~\ref{appendix:proof:lemma:pps:start-phase-probability}.

\begin{lemma} \label{lemma:pps:start-phase-probability}
Fix some time
$t_{0} \geq t^{*}$,
node $v$,
and
$j_{0} \in \mathcal{S}_{\phi}$.
There exists
$\tau = O (\phi^{3})$
such that for every
$t \geq t_{0} + \tau$,
it holds that
$\Pr \left(
\Step_{v, t} = 0 \mid \Step_{v, t_{0}} = j_{0}
\right)
\geq
\frac{1}{2 \phi}$.
This holds independently of any coin toss of $v$ prior to time $t_{0}$ and
of any coin toss of all other nodes.
\end{lemma}

\section{Eligibility Criteria Imposed on the LCL Predicate}
\label{section:LCL-eligibility-criteria}
In this section, we introduce the eligibility criteria imposed on the
predicate $\Predicate$ of a node/edge-LCL
$\Problem = \langle
\mathcal{O}, \mathcal{G}, \Predicate
\rangle$,
recalling that these criteria are common to both edge- and node-LCLs.
We make an extensive use of the following notation:
For an output value
$o \in \mathcal{O}$,
let
\[
\SatMultisets(o)
\, = \,
\left\{
M \in \Multisets(\mathcal{O})
\, \middle| \,
\Predicate(o, M) = \true
\right\}
\]
denote the collection of multisets for which $o$ is evaluated to $\true$ under
$\Predicate$.
Table~\ref{table:LCL-satisfying-multisets-cores} presents these multiset
collections for the four concrete LCL predicates listed in
\Sect{}~\ref{section:introduction:concrete-problems}%
\LongVersion 
\ (i.e., MIS/MM, node/edge
$c$-coloring, maximal node/edge $c$-coloring, and incremental node/edge
$c$-coloring)%
\LongVersionEnd 
;
it is straightforward to verify that the corresponding eight distributed
problems meet the following criterion.

\begin{table}
{\centering
\resizebox{\textwidth}{!}{%
\begin{tabular}{|l|l|l|l|l|}
\hline
LCL &
$o \in \mathcal{O}$ &
$\SatMultisets(o) \subseteq \Multisets(\mathcal{O})$ &
$\Cores(o) \subseteq \SatMultisets(o)$ &
$\psi$
\\
\hline\hline
MIS/MM &
$1$ &
$\displaystyle
\{
M
\mid
M(1) = 0
\}$ &
$\displaystyle
\{
\langle 0, 0 \rangle
\}$ &
\multirow{2}{*}{$1$}
\\
&
$2$ &
$\displaystyle
\{
M
\mid
M(1) \geq 1
\}$ &
$\displaystyle
\{
\langle 1, 0 \rangle
\}$ &
\\
\hline
node/edge  &
$1 \leq i \leq c$ &
$\displaystyle
\{
M
\mid
M(i) = 0
\}$ &
$\displaystyle
\{
\langle 0, \dots, 0 \rangle
\}$ &
$0$
\\
$c$-coloring &
& & &
\\
\hline
max.\ &
$1 \leq i \leq c - 1$ &
$\displaystyle
\{
M
\mid
M(i) = 0
\}$ &
$\displaystyle
\{
\langle 0, \dots, 0 \rangle
\}$ &
\multirow{2}{*}{$1$}
\\
node/edge &
$c$ &
$\displaystyle
\{
M
\mid
M(j) \geq 1 , \, 1 \leq j \leq c - 1
\}$ &
$\displaystyle
\{
\langle 1, \dots, 1, 0 \rangle
\}$ &
\\
$c$-coloring &
& & & \\
\hline
inc.\ &
$1 \leq i \leq c - 1$ &
$\displaystyle
\{
M
\mid
M(i) = 0 \land \sum_{j = 1}^{i - 1} M(j) \geq i - 1
\}$ &
$\displaystyle
\{
\langle x_{1}, \dots, x_{i - 1}, 0, \dots, 0 \rangle
\mid
\sum_{j = 1}^{i - 1} x_{j} = i - 1
\}$ &
\multirow{2}{*}{$c - 1$}
\\
node/edge &
$c$ &
$\displaystyle
\{
M
\mid
\sum_{j = 1}^{c - 1} M(j) \geq c - 1
\}$ &
$\displaystyle
\{
\langle x_{1}, \dots, x_{c - 1}, 0 \rangle
\mid
\sum_{j = 1}^{c - 1} x_{j} = c - 1
\}$ &
\\
$c$-coloring &
& & &
\\
\hline
\end{tabular}
}
\par}
\caption{\label{table:LCL-satisfying-multisets-cores}
The collections $\SatMultisets(o)$ of multisets for which the various output
values $o$ are evaluated to $\true$,
the core collections $\Cores(o)$,
and
the propagation radii $\psi$
for the four concrete LCL predicates listed in
\Sect{}~\ref{section:introduction:concrete-problems} (refer to that section
for the predicate semantics).
The
$\langle \cdots \rangle$
notation represents a multiset interpreted as a vector in $\Integers_{\geq
0}^{\mathcal{O}}$, where
$\mathcal{O} = \{ 1, 2 \}$
for MIS/MM and
$\mathcal{O} = \{ 1, \dots, c \}$
for the other three predicates.}
\end{table}

\begin{criterion}[\emph{gap freeness}]
\label{criterion:common:gap-freeness}
Problem $\Problem$ meets the \emph{gap freeness} criterion if for
every output value
$o \in \mathcal{O}$
and multisets
$M, M', M'' \in \Multisets(\mathcal{O})$
such that
$M \subseteq M' \subseteq M''$,
if
$M, M'' \in \SatMultisets(o)$,
then
$M' \in \SatMultisets(o)$.
\end{criterion}

Assume hereafter that $\Problem$ meets the gap freeness criterion.
The minimal elements of the partially ordered set
$(\SatMultisets(o), \subseteq)$
are referred to as the \emph{cores} of
$o \in \mathcal{O}$,
denoting the collection of $o$'s cores by
\[
\Cores(o)
\, = \,
\left\{
M \in \SatMultisets(o)
\, \middle| \,
M' \subset M \Longrightarrow M' \notin \SatMultisets(o)
\right\} \, .
\]
Table~\ref{table:LCL-satisfying-multisets-cores} presents the core collections
for the four concrete LCL predicates listed in
\Sect{}~\ref{section:introduction}.
We note that $\Cores(o)$ is a (non-empty) antichain of the partially ordered
set
$(\SatMultisets(o), \subseteq)$.\footnote{%
An antichain of a partially ordered set is a subset of mutually incomparable
elements.}

The notion of cores allows us to introduce the (abstract) \emph{supportive
digraph} $D_{\Predicate}$ of the LCL $\Predicate$ whose node set is
$V(D_{\Predicate}) = \mathcal{O}$
and whose edge set $E(D_{\Predicate})$ includes an edge from
$o \in \mathcal{O}$
to
$o' \in \mathcal{O}$
if (and only if) there exists a core
$M \in \Cores(o)$
such that
$o' \in M$.
In other words, the supportive digraph of $\Predicate$ is defined over the
set $\mathcal{O}$ of output values and includes an edge from output value $o$
to output value $o'$ if $o'$ belongs to some (at least one) core of $o$.
Figure~\ref{figure:supportive-digraphs} illustrates the supportive digraphs of
the four concrete LCL predicates listed in
\Sect{}~\ref{section:introduction:concrete-problems}.
The \emph{propagation radius} of predicate $\Predicate$ is defined to be the
length of the longest (directed) path in the supportive digraph
$D_{\Predicate}$ if $D_{\Predicate}$ is acyclic, and $\infty$ otherwise.
Refer to Table~\ref{table:LCL-satisfying-multisets-cores} (see also
Figure~\ref{figure:supportive-digraphs}) for the propagation radii of the four
concrete LCL predicates listed in
\Sect{}~\ref{section:introduction:concrete-problems}.

\begin{criterion}[\emph{$\psi$-bounded propagation}]
\label{criterion:common:bounded-propagation}
Problem $\Problem$ meets the \emph{$\psi$-bounded propagation} criterion
for a parameter
$\psi \in \Integers_{\geq 0}$
if the propagation radius of the predicate $\Predicate$ is at most $\psi$.
\end{criterion}

\begin{remark*}
For an insight on the role of Criteria \ref{criterion:common:gap-freeness} and
\ref{criterion:common:bounded-propagation}, consider a graph $G$ and a node/edge
configuration
$C_{0} : X(G) \rightarrow \mathcal{O} \cup \{ \bot \}$.
Let
$\{ C_{i} : X(G) \rightarrow \mathcal{O} \cup \{ \bot \} \}_{i \geq 0}$
be the configuration sequence generated inductively from $C_{0}$ by setting
\[
C_{i + 1}(x)
\, = \,
\begin{cases}
\bot \, , & x \in \UnContentSet(C_{i}) \\
C_{i}(x) \, , & \text{otherwise}
\end{cases}
\, ;
\]
that is, the configuration
$C_{i + 1}$
is obtained from $C_{i}$ by resetting (to $\bot$) the entry corresponding to
graph object
$x \in X(G)$
if $x$ is (decided and) uncontent in $C_{i}$.
It turns out, that if $\Problem$ meets the gap freeness and $\psi$-bounded
propagation criteria, then
$C_{i + 1} = C_{i}$
for every
$i \geq \psi + 1$
and that this is (existentially) tight up to a constant additive term.
This insight is pivotal for the correctness of our transformer --- see \Sect{}
\ref{section:edges:analysis} and \ref{section:nodes:analysis}, where this
connection is established (implicitly) as part of a stronger claim.
\end{remark*}

\section{The Edge-LCL Transformer}
\label{section:transformer-edges}
This section presents our transformer when applied to an edge-LCL
$\Problem = \langle
\mathcal{O}, \mathcal{G}, \Predicate
\rangle$,
where
$\mathcal{G} \subseteq \mathcal{U}_{\Delta}$
for a degree bound
$\Delta$.
We start, in \Sect{}~\ref{section:edges:algorithmic-eligibility-criteria}, by
introducing the eligibility criteria that $\Problem$ should meet with respect
to the transformer's algorithmic building blocks, namely, the fault free
algorithm $\Alg$ and the detection procedure $\Detect$ (see
\Sect{}~\ref{section:transformer-overview}) assuming that it also meets
Criteria \ref{criterion:common:gap-freeness} and
\ref{criterion:common:bounded-propagation} presented in
\Sect{}~\ref{section:LCL-eligibility-criteria}.
\Sect{}~\ref{section:edges:main-theorem} is dedicated to stating
the formal guarantees of the transformer when applied to (the eligible) problem
$\Problem$, including the dependence of the self-stabilizing algorithm
$\SelfStabAlg$'s fully adaptive run-time and message size on the parameters of
the eligibility criteria.
The transformer's implementation is presented in
\Sect{}~\ref{section:edges:implementation}, providing a precise description of
$\SelfStabAlg$ and how it is synthesized
from $\Alg$ and $\Detect$.
The correctness of $\SelfStabAlg$ is established in
\Sect{}~\ref{section:edges:analysis} together with the promised upper bounds
on its fully adaptive run-time and message size.

\subsection{Algorithmic Eligibility Criteria}
\label{section:edges:algorithmic-eligibility-criteria}

\paragraph{Detection Procedure.}
As discussed in \Sect{}~\ref{section:transformer-overview}, our transformer
employs a distributed detection procedure $\Detect$.
On a graph
$G \in \mathcal{G}$,
procedure $\Detect$ runs indefinitely and does not access any of the registers
maintained by the nodes
$v \in V(G)$
except for the output registers to which $\Detect$ has a read-only access.
The procedure returns a Boolean value denoted by
$\Detect_{v}(u) \in \{ \true, \false \}$
for each output register $\OutReg_{v}(u)$,
$u \in N_{G}(v)$;
let
$\Detect_{v, t}(u)$
denote the value of $\Detect_{v}(u)$ in round $t$.
Informally, the role of $\Detect$ is to return
$\Detect_{v}(u) = \false$
if ``something is wrong'' with the output register $\OutReg_{v}(u)$.

More formally, consider an edge
$e = \{ u, v \} \in E(G)$
and assume that neither $u$ nor $v$ is manipulated in round
$t - 1$.
Procedure $\Detect$ guarantees that
$\Detect_{v, t}(u) = \Detect_{u, t}(v)$
and that
$\Detect_{v, t}(u) = \Detect_{u,t}(v) = \false$
if and only if either
(1)
$\OutReg_{u, t}(v) \neq \OutReg_{v, t}(u)$,
that is, $e$ is port-inconsistent at time $t$;
or
(2)
$\OutReg_{u, t}(v) = \OutReg_{v, t}(u) = o \in \mathcal{O}$
and
$\Predicate(o, M) = \false$,
where
$M \in \Multisets(\mathcal{O})$
is the multiset defined over the output values stored at time $t$ in the
output registers belonging to
$\{ \OutReg_{u}(w) \mid w \in N_{G}(u) - \{ v \} \}
\cup
\{ \OutReg_{v}(w) \mid w \in N_{G}(v) - \{ u \} \}$.

\begin{criterion}[\emph{$\mu_{D}(\Delta)$-detectability}]
\label{criterion:edges:detectability}
Problem $\Problem$ meets the \emph{$\mu_{D}(\Delta)$-detectability} criterion
for a function
$\mu_{D} : \Integers_{> 0} \rightarrow \Integers_{> 0}$
if it admits a detection procedure whose message size is at 
most 
$\mu_{D}(\Delta)$ when it runs on a graph of degree bound $\Delta$.
\end{criterion}

\begin{remark*}
It is straightforward to implement $\Detect$ so that each node
$v \in V(G)$
shares the content of all its output registers
$\OutReg_{v}(\cdot)$
with each one of its neighbors, resulting in
$\mu_{D}(\Delta) = O (\Delta \log |\mathcal{O}|)$.
The challenge in this regard is to come up with more efficient implementations
of $\Detect$ that use smaller messages, which translates to smaller messages
in $\SelfStabAlg$.
\end{remark*}

\paragraph{Restricting the Phase Procedure.}
Assume hereafter that $\Problem$ meets the $\mu_{D}(\Delta)$-detectability
criterion.
In the remainder of this section, we discuss the eligibility criteria imposed
on the fault free algorithm $\Alg$ for $\Problem$.
Consider the graph
$G \in \mathcal{G}$
on which $\Alg$ runs.
Recall that if $\Alg$ is a phase-based algorithm, then it is fully specified
by a phase length parameter $\phi$ and a phase procedure
$\PhaseProcedure_{\Alg}$ invoked at the beginning of every phase.
Moreover, each phase ends with a decision step
$j = \phi - 1$,
which is the only step in which the nodes
$v \in V(G)$
may write into their output registers
$\OutReg_{v}(u)$,
$u \in N_{G}(v)$,
an action that is allowed only when
$\OutReg_{v}(u) = \bot$.

Algorithm $\Alg$ is said be \emph{port-consistent} if it is guaranteed that
the edges remain port-consistent at all times.
Fix some phase of the phase procedure $\PhaseProcedure_{\Alg}$ and let
$D \subseteq E(G)$
be the subset of edges that are decided at the beginning of the phase.
The algorithm is said to be \emph{decision-oblivious} if no messages are
exchanged during the phase over the edges in $D$ and the only step in which
the nodes may take into account the output values associated with the edges in
$D$ is the decision step
$j = \phi - 1$.
In other words, the actions of the nodes (in terms of the values they write
into their local registers) in steps
$j = 0, 1, \dots, \phi - 2$
are identical to their actions in an invocation of $\PhaseProcedure_{\Alg}$ on
the graph
$G(E(G) - D)$
induced by the undecided edges.
This means in particular that while a node
$v \in V(G)$
may base its actions in steps
$j = 0, 1, \dots, \phi - 2$
of $\PhaseProcedure_{\Alg}$ on the (global) degree bound $\Delta$ and on $v$'s
degree in
$G(E(G) - D)$,
its actions in these steps are independent of the number
$\Degree_{G}(v) - \Degree_{G(E(G) - D)}(v)$
of decided edges incident on $v$.
To enable the decision making of a node
$v \in V(G)$
regarding an incident edge
$\{ v, w \} \in E(G) - D$
during the decision step
$j = \phi - 1$,
the decision-oblivious algorithm is required to satisfy the following
property:
the message sent from $w$ to $v$ in step
$\phi - 2$
includes all the information that $v$ may need in (the local computation
stage of) step
$\phi - 1$
regarding the output values associated with the edges in
$\{ e \in D \mid w \in e \}$.

\begin{criterion}[\emph{$(\phi, \mu_{A}(\Delta))$-phase-based}]
\label{criterion:edges:phase-based}
Problem $\Problem$ meets the
\emph{$(\phi, \mu_{A}(\Delta))$-phase-based}
criterion (with respect to $\Alg$) if $\Alg$ is a port-consistent
decision-oblivious phase-based algorithm with a phase of length $\phi$ that
uses messages of size $\mu_{A}(\Delta)$ when it runs on a graph of degree
bound $\Delta$.
\end{criterion}

Assume hereafter that $\Problem$ meets the
$(\phi, \mu_{A}(\Delta))$-phase-based
criterion with respect to $\Alg$.
Given an edge configuration
$C : E(G) \rightarrow \mathcal{O} \cup \{ \bot \}$,
let $\eta$ be the execution of $\PhaseProcedure_{\Alg}$ on the graph $G$ with
initial configuration $C$.
Let
$\PhaseProcedure_{\Alg}(G, C)$
be the edge configuration associated with $\PhaseProcedure_{\Alg}$ when $\eta$
halts, i.e., after $\phi$ rounds.
Notice that if $\PhaseProcedure_{\Alg}$ is randomized, then
$\PhaseProcedure_{\Alg}(G, C)$
is a random variable that depends on the coin tosses of the nodes
$v \in V(G)$
during $\eta$.

\begin{criterion}[\emph{respectful decisions}]
\label{criterion:edges:respectful-decisions}
Problem $\Problem$ meets the \emph{respectful decisions} criterion (with
respect to $\Alg$) if the following two conditions hold with probability $1$
for every graph
$G \in \mathcal{G}$
and edge configuration
$C : E(G) \rightarrow \mathcal{O} \cup \{ \bot \}$: \\
(1)
$\ContentSet(C)
\subseteq
\ContentSet(\PhaseProcedure_{\Alg}(G, C))$;
and \\
(2)
$\UnDecidedSet(C) \cap \DecidedSet(\PhaseProcedure_{\Alg}(G, C))
\subseteq
\ContentSet(\PhaseProcedure_{\Alg}(G, C))$.
\end{criterion}

\begin{remark*}
Criterion~\ref{criterion:edges:respectful-decisions} dictates that the
decisions taken by an invocation of $\PhaseProcedure_{\Alg}$ do not introduce
any (new) uncontent edges:
condition (1) states that edges that were content before the invocation of
$\PhaseProcedure_{\Alg}$ remain content in the new configuration;
condition (2) states that edges that became decided during the execution of
$\PhaseProcedure_{\Alg}$ are content (in the new configuration).
\end{remark*}

\paragraph{Potential Functions.}
Assume hereafter that $\Problem$ meets the respectful decisions criterion with
respect to $\Alg$.
A parameterized integer family
$\left\{ \sigma(M) \right\}_{M \in \Multisets(\mathcal{O})}
\subseteq
\Integers$
is said to be \emph{monotonically non-increasing} if
$M \subseteq M'
\Longrightarrow
\sigma(M) \geq \sigma(M')$
for every
$M, M' \in \Multisets(\mathcal{O})$.
Let $\ContConfGraphs(\Problem)$ denote the collection of pairs of the form
$(G, C)$,
where 
$G \in \mathcal{G}$
and
$C : E(G) \rightarrow \mathcal{O} \cup \{ \bot \}$
is a content edge configuration for $G$.
A function
$\pi : \ContConfGraphs(\Problem) \rightarrow \Integers_{\geq 0}$
is said to be an \emph{edge separable potential function} for $\Problem$ if
there exists a monotonically non-increasing parameterized integer family
$\mathcal{I}
=
\left\{ \sigma(M) \right\}_{M \in \Multisets(\mathcal{O})}
\subset
\Integers_{> 0}$
such that
\[
\pi(G, C)
\, = \,
\sum_{e \in \UnDecidedSet(C)} \sigma(C[e])
\]
for every
$(G, C) \in \ContConfGraphs(\Problem)$.
In this case, we refer to the integers in $\mathcal{I}$ as the
\emph{potential coefficients} of $\pi$ and to $\sigma(\emptyset)$ as the
\emph{top potential coefficient}, observing that $\sigma(\emptyset)$ up-bounds
any other potential coefficient $\sigma(\cdot)$.

\begin{criterion}[\emph{$(\beta, \sigma_{0})$-progress}]
\label{criterion:edges:progress}
Problem $\Problem$ meets the
\emph{$(\beta, \sigma_{0})$-progress}
criterion (with respect to $\Alg$)
for a real
$\beta > 0$
and an integer
$\sigma_{0} > 0$
if $\Problem$ admits an edge separable potential function $\pi$ with top
potential coefficient $\sigma_{0}$ such that for every
$(G, C) \in \ContConfGraphs(\Problem)$,
it is guaranteed that
$\Ex(\pi(G, \PhaseProcedure_{\Alg}(G, C))) \leq (1 - \beta) \cdot \pi(G, C)$.
\end{criterion}

\begin{remark*}
The requirement that the family of potential coefficients is monotonically
non-increasing implies that
$\pi(G, C') \leq \pi(G, C)$
for every graph $G$ and edge configurations $C$ and $C'$ over $G$ satisfying
(1)
$\DecidedSet(C') \supseteq \DecidedSet(C)$;
and
(2)
$C'(e) = C(e)$
for every edge
$e \in \DecidedSet(C)$.
Moreover, since the potential coefficients are strictly positive, if
$\pi(G, C) = 0$,
then $C$ must be complete.
This means that as the execution of $\Alg$ progresses, the potential in
$\Alg$'s associated configuration cannot increase and that once we reach a
configuration with $0$ potential, all edges must be decided.
Criterion~\ref{criterion:edges:progress} ensures that the potential decreases
sufficiently fast, ultimately allowing us to up-bound the run-time of the
self-stabilizing algorithm synthesized from $\Alg$.
\end{remark*}

\subsection{The Main Theorem}
\label{section:edges:main-theorem}
Problem
$\Problem = \langle
\mathcal{O}, \mathcal{G}, \Predicate
\rangle$
is said to be
\emph{$(
\psi,
\mu_{D}(\Delta),
\phi,
\mu_{A}(\Delta),
\beta,
\sigma_{0}
)$-eligible}
if
$\Problem$ meets the
gap freeness
(Criterion~\ref{criterion:common:gap-freeness}),
$\psi$-bounded propagation
(Criterion~\ref{criterion:common:bounded-propagation}),
$\mu_{D}(\Delta)$-detectability
(Criterion~\ref{criterion:edges:detectability}),
$(\phi, \mu_{A}(\Delta))$-phased-based
(Criterion~\ref{criterion:edges:phase-based}),
respectful decisions
(Criterion~\ref{criterion:edges:respectful-decisions}),
and
$(\beta, \sigma_{0})$-progress
(Criterion~\ref{criterion:edges:progress})
criteria.
The guarantees of our transformer are cast in the following theorem.

\begin{theorem} \label{theorem:edges:main}
If an edge-LCL
$\Problem = \langle
\mathcal{O}, \mathcal{G}, \Predicate
\rangle$
is
$(\psi, \mu_{D}(\Delta), \phi, \mu_{A}(\Delta), \beta, \sigma_{0})$-eligible,
then $\Problem$ admits a randomized self-stabilizing algorithm
$\SelfStabAlg$ that uses
messages of size
\[
O \left( \mu_{D}(\Delta) + \mu_{A}(\Delta) + \log \phi \right)
\]
whose fully adaptive run-time is
\[
O \left(
\left( \phi^{5} / \beta \right)
\cdot
\left(
\log (k) + (\psi + \phi) \log (\Delta) + \log (\sigma_{0})
\right)
\right) \, .
\]
Moreover, edges whose distance from any manipulated node is at least
$\psi + \phi + 2$
are guaranteed to maintain their original output value throughout the
execution of $\SelfStabAlg$ (with probability $1$).
\end{theorem}

\subsection{Implementation}
\label{section:edges:implementation}
Suppose that $\Problem$ is
$(\psi, \mu_{D}(\Delta), \phi, \mu_{A}(\Delta), \beta, \sigma_{0})$-eligible
and let $\Detect$ and $\Alg$ be the corresponding detection procedure and
fault free algorithm.
In this section, we describe how the self-stabilizing algorithm
$\SelfStabAlg$, promised in \Thm{}~\ref{section:edges:main-theorem}, is
synthesized from $\Detect$ and from the phase procedure
$\PhaseProcedure_{\Alg}$ associated with $\Alg$.
This synthesis is recounted in Pseudocode~\ref{pseudocode:transformer-edges},
where $\SelfStabAlg$ runs on an input graph
$G \in \mathcal{G}$
and is presented from the perspective of a node
$v \in V(G)$,
denoting the messages sent to (resp., from) $v$
from (resp., to) a neighbor
$u \in N_{G}(v)$
by $\InMsg_{v}(u)$ (resp., $\OutMsg_{v}(u)$).

Each message $m$ sent by $v$ is augmented with a designated field, denoted by
$m.\fieldPPS$, where $v$ records the state of its $\PPS$ module (lines
\ref{line:edges:record-step-begin}--\ref{line:edges:record-step-end}).
Using the $m.\fieldPPS$ fields in its incoming messages, node $v$ can identify
its \emph{$\PPS$-synchronized} neighbors, namely, the nodes
$u \in N_{G}(v)$
that satisfy
$\Step_{u} = \Step_{v}$.
The crux of the transformer is that whenever
$\Step_{v} = 0$,
node $v$ invokes a simulation of a (complete) phase $\Phi$ of
$\PhaseProcedure_{\Alg}$ in conjunction with its $\PPS$-synchronized neighbors
(lines
\ref{line:edges:simulate-phase-procedure-begin}--%
\ref{line:edges:simulate-phase-procedure-end}).

The set of $v$'s $\PPS$-synchronized neighbors is stored in the designated
register $\Peers_{v}$ (which is the only register maintained under
$\SelfStabAlg$ other than $\Step_{v}$ and the registers of
$\PhaseProcedure_{\Alg}$) at the beginning of phase $\Phi$ (lines
\ref{line:edges:compute-peers-begin}--\ref{line:edges:compute-peers-end});
the $\PPS$ module is designed to guarantee that $v$ remains
$\PPS$-synchronized with the nodes in $\Peers_{v}$ throughout the simulation
of $\Phi$ (see \Obs{}~\ref{observation:transformer:step-counter}).
Regardless of the simulation of $\Phi$, node $v$ invokes the detection
procedure $\Detect$ in every round (line~\ref{line:edges:call-detect}) and
resets the output registers
$\OutReg_{v}(u)$
for which
$\Detect_{v}(u) = \false$
(lines \ref{line:edges:detect-begin}--\ref{line:edges:detect-end}).

\begin{algorithm}
\caption{\label{pseudocode:transformer-edges}%
The code (executed in each round) of the self-stabilizing algorithm
$\SelfStabAlg$ for the edge-LCL $\Problem$}
\begin{algorithmic}[1]
\State{call $\Detect$}%
\label{line:edges:call-detect}%
\Comment{assigns values to the (round-temporal) variables
$\Detect_{v}(\cdot)$}
\If{$\Step_{v} = 0$}%
\label{line:edges:compute-peers-begin}
  \State{%
$\Peers_{v} \gets
\{ u \in N_{G}(v) \mid
\OutReg_{v}(u) = \bot
\, \land \,
\InMsg_{v}(u).\fieldPPS = 0
\}$}
\EndIf%
\label{line:edges:compute-peers-end}
\If{$0 \leq \Step_{v} \leq \phi - 1$}%
\label{line:edges:simulate-phase-procedure-begin}
  \State{%
simulate step $\Step_{v}$ of $\PhaseProcedure_{\Alg}$ with the neighbors in
$\Peers_{v}$}%
\label{line:edges:simulate-phase-procedure}%
\EndIf%
\label{line:edges:simulate-phase-procedure-end}
\ForAll{$u \in N_{G}(v)$}%
\label{line:edges:detect-begin}
  \If{$\Detect_{v}(u) = \false$}
    $\OutReg_{v}(u) \gets \bot$%
\label{line:edges:detect}
  \EndIf
\EndFor%
\label{line:edges:detect-end}
\State{call $\AdvancePPS$}%
\Comment{updates the register $\Step_{v}$
(see \Sect{}~\ref{section:probabilistic-phase-synchronization})}
\ForAll{$u \in N_{G}(v)$}%
\label{line:edges:record-step-begin}
  \State{%
$\OutMsg_{v}(u).\fieldPPS \gets \Step_{v}$}
\EndFor%
\label{line:edges:record-step-end}
\end{algorithmic}
\end{algorithm}

\subsection{Analysis}
\label{section:edges:analysis}
In this section, we analyze the self-stabilizing algorithm
$\SelfStabAlg$ developed in \Sect{}~\ref{section:edges:implementation}
and establish \Thm{}~\ref{theorem:edges:main}.
The
$O (\mu_{D}(\Delta) + \mu_{A}(\Delta) + \log \phi)$
message size bound of \Thm{}~\ref{theorem:edges:main} follows directly from
the transformer's design as each message of $\SelfStabAlg$ is composed
of
(i)
a message of the detection procedure $\Detect$ whose size is
$\mu_{D}(\Delta)$;
(ii)
a message of the fault free algorithm $\Alg$ (actually, of the phase procedure
$\PhaseProcedure_{\Alg}$) whose size is
$\mu_{A}(\Delta)$;
and
(iii)
one of the
$\phi + 1$
states of the $\PPS$ module.
Our goal in the remainder of this section is to establish
\Thm{}~\ref{theorem:edges:main}'s fully adaptive run-time bound and the bound
on the distance from the adversarial manipulated nodes to edges that change
their output value (i.e., the theorem's last claim).

\subsubsection{Roadmap}
\label{section:edges:analysis:roadmap}
Recall that the adversarial manipulations may include the addition/removal of
nodes/edges, hence the graph may change during the round interval
$[t^{\circ}, t^{*} - 1]$
(during which the adversarial manipulations take place).
For
$t \geq t^{\circ}$,
let $G_{t}$ be the graph at time $t$ and let
$K_{t} \subseteq V(G_{t})$
be the set of nodes that experience adversarial manipulations during the round
interval
$[t^{\circ}, t - 1]$
and still exist in $G_{t}$.
Let
$C_{t} : E(G_{t}) \rightarrow \mathcal{O} \cup \{ \bot \}$
be the edge configuration associated with $\SelfStabAlg$ at time $t$.
As the nodes that experience topology changes are considered to be
manipulated (see \Sect{}~\ref{section:model-and-preliminaries}), we obtain the
following observation, where we adopt the convention that
$\Distance_{G_{t}}(v, K_{t}) = 0$
for every time
$t > t^{\circ}$
and node
$v \in V(G_{t^{\circ}}) - V(G_{t})$.

\begin{observation}
\label{observation:edges:roadmap:decreasing-distance-to-faults}
The expression
$\Distance_{G_{t}}(v, K_{t})$
is a non-increasing function of $t$ for every node
$v \in V(G_{t^{\circ}})$.
\end{observation}

By the assumption of \Thm{}~\ref{theorem:edges:main}, problem $\Problem$ meets
the $\psi$-bounded propagation criterion
(Criterion~\ref{criterion:common:bounded-propagation}), which means that the
supportive digraph $D_{\Predicate}$ is acyclic and every (directed) path in
$D_{\Predicate}$ is of length at most $\psi$.
A parameter that plays a key role in the transformer's analysis is the
\emph{depth} of an output value
$o \in \mathcal{O}$
(with respect to the predicate $\Predicate$), denoted by
$\DepthSupGraph_{\Predicate}(o)$, defined to be the length of a longest
directed path emerging from $o$ in $D_{\Predicate}$.
Indeed, the journey towards proving \Thm{}~\ref{theorem:edges:main} starts in
\Sect{}~\ref{section:edges:analysis:affected-regions} that is dedicated to
establishing the following proposition.

\begin{proposition}
\label{proposition:edges:roadmap:depth-bounds-spatial-distance}
Consider an edge
$e \in E(G_{t^{\circ}})$
with
$C_{t^{\circ}}(e) = o$
and a time
$t \geq t^{\circ}$.
If
$e \in E(G_{t})$
and
$\Distance_{G_{t}}(e, K_{t})
\geq
\DepthSupGraph_{\Predicate}(o) + \phi + 2$,
then
(1)
$C_{t}(e) = o$;
and
(2)
$e \in \ContentSet(C_{t})$.
\end{proposition}

In what follows, we denote
$G = G_{t^{*}}$
and
$K = K_{t^{*}}$,
observing that the definition of $t^{*}$ implies that
$G = G_{t}$
and
$K = K_{t}$
for every
$t \geq t^{*}$
and that
$|K| \leq k$.
By combining
\Obs{}~\ref{observation:edges:roadmap:decreasing-distance-to-faults} and
\Prop{}~\ref{proposition:edges:roadmap:depth-bounds-spatial-distance}, we
obtain the following corollary that establishes the last claim of
\Thm{}~\ref{theorem:edges:main}.

\begin{corollary} \label{corollary:edges:roadmap:affected-regions}
Consider an edge
$e \in E(G)$
and a time
$t \geq t^{*}$.
If
$\Distance_{G}(e, K) \geq \psi + \phi + 2$,
then
(1)
$e \in E(G_{t^{\circ}})$;
(2)
$C_{t}(e) = C_{t^{\circ}}(e)$;
and
(3)
$e \in \ContentSet(C_{t})$.
\end{corollary}

It remains to establish the fully adaptive run-time bound of
\Thm{}~\ref{theorem:edges:main}.
While \Prop{}~\ref{proposition:edges:roadmap:depth-bounds-spatial-distance}
uses the depth of the output values to provide a ``spatial barrier'' from the
adversarial manipulations, the next proposition, established in
\Sect{}~\ref{section:edges:analysis:fault-recovery}, uses the depth to provide
a ``temporal barrier'' from the adversarial manipulations.

\begin{proposition}
\label{proposition:edges:roadmap:depth-bounds-temporal-distance}
Consider an edge
$e \in E(G)$
and a time
$t$.
If
$t \geq t^{*} + \phi + 2$,
then
(1)
$e$ is port-consistent at time $t$;
and
(2)
if
$C_{t}(e) = o \in \mathcal{O}$
and
$t \geq t^{*} + \DepthSupGraph_{\Predicate}(o) + \phi + 3$,
then
$e \in \ContentSet(C_{t})$.
\end{proposition}

Recalling the guarantees of the detection procedure $\Detect$, the design of
$\SelfStabAlg$ ensures that for every
$t \geq t^{*} + \phi + 2$,
an edge
$e \in \DecidedSet(C_{t})$
becomes undecided in round $t$ if and only if
$e \in \UnContentSet(C_{t})$.
The design of $\SelfStabAlg$ also ensures that if
$e \in \DecidedSet(C_{t}) \cap \DecidedSet(C_{t + 1})$,
then
$C_{t}(e) = C_{t + 1}(e)$.
Setting
\[
t^{r}
\, = \,
t^{*} + \psi + \phi + 3
\, ,
\]
we obtain the following corollary from
\Prop{}~\ref{proposition:edges:roadmap:depth-bounds-temporal-distance}.

\begin{corollary} \label{corollary:edges:roadmap:fault-recovery}
For every
$t \geq t^{r}$,
it holds that
(1)
all edges are port-consistent at time $t$;
(2)
the edge configuration $C_{t}$ is content;
and
(3)
if
$e \in \DecidedSet(C_{t})$,
then
$C_{t + 1}(e) = C_{t}(e)$.
\end{corollary}

Owing to \Cor{}~\ref{corollary:edges:roadmap:fault-recovery}, our remaining
task is to prove that starting from time $t^{r}$, it does not take
$\SelfStabAlg$ too long until it reaches a complete (content) configuration.
To this end, we recall that by the assumption of
\Thm{}~\ref{theorem:edges:main}, problem $\Problem$ meets the
$(\beta, \sigma_{0})$-progress
criterion (Criterion~\ref{criterion:edges:progress}) and recruit the edge
separable potential function $\pi$ promised in this criterion.
Since the number of edges in $G$ at distance smaller than
$\psi + \phi + 2$
from $K$ is up-bounded by
$O (|K| \cdot (\Delta - 1)^{\psi + \phi + 1} \cdot \Delta)
\leq
O (k \cdot \Delta^{\psi + \phi + 2})$,
we can employ \Cor{}~\ref{corollary:edges:roadmap:affected-regions} to
conclude that
$|\UnDecidedSet(C_{t^{r}})|
\leq
O (k \cdot \Delta^{\psi + \phi + 2})$.
\Cor{}~\ref{corollary:edges:roadmap:bound-initial-potential} follows from the
definition of an edge separable potential function as the top potential
coefficient of $\pi$ is $\sigma_{0}$.

\begin{corollary} \label{corollary:edges:roadmap:bound-initial-potential}
The potential in $C_{t^{r}}$ satisfies
$\pi(G, C_{t^{r}})
\leq
O \left( \sigma_{0} \cdot k \cdot \Delta^{\psi + \phi + 2} \right)$.
\end{corollary}

To complete the proof, we show that from time $t^{r}$ onwards, the potential
decreases fast.
Recalling the definition of an edge separable potential function, we deduce
from \Cor{}~\ref{corollary:edges:roadmap:fault-recovery} that
$\pi(G, C_{t})$
is a non-increasing function of
$t \geq t^{r}$.
Moreover, if
$\pi(G, C_{t}) = 0$
for some
$t \geq t^{r}$,
then $C_{t'}$ is a legal configuration for every
$t' \geq t$.
Therefore, the task of establishing the run-time bound reduces to that of
up-bounding the first time
$t \geq t^{r}$
at which
$\pi(G, C_{t}) = 0$.
To this end, we take
$\tau = O (\phi^{3})$
to be the parameter promised in \Lem{}~\ref{lemma:pps:start-phase-probability}
and establish the following proposition in
\Sect{}~\ref{section:edges:analysis:progress}.

\begin{proposition} \label{proposition:edges:roadmap:progress}
Fix some
$t \geq t^{r}$
and the global state of $\SelfStabAlg$ at time $t$.
Then,
\[
\Ex(\pi(G, C_{t + \tau + \phi}))
\, \leq \,
\left( 1 - \frac{\beta}{4 \phi^{2}} \right)
\cdot
\pi(G, C_{t}) \, .
\]
\end{proposition}

To complete the run-time analysis, we define the random variables
$P_{i} = \pi(G, C_{t^{r} + i (\tau + \phi)})$
for
$i \in \Integers_{\geq 0}$.
Taking
$\chi = 1 - \beta / (4 \phi^{2})$,
\Prop{}~\ref{proposition:edges:roadmap:progress} ensures that
$\Ex(P_{i} \mid P_{i - 1}) \leq \chi \cdot P_{i - 1}$,
hence
\[
\Ex(P_{i})
\, = \,
\Ex(\Ex(P_{i} \mid P_{i - 1}))
\, \leq \,
\chi \cdot \Ex(P_{i - 1})
\, \leq \,
\chi^{i} \cdot \Ex(P_{0})
\, \leq \,
\chi^{i} \cdot O (\sigma_{0} \cdot k \cdot \Delta^{\psi + \phi + 2})
\, ,
\]
where
the third transition is by induction on $i$ and the last transition is due to
\Cor{}~\ref{corollary:edges:roadmap:bound-initial-potential}.
Since
$\chi = 1 - \Omega (\beta / \phi^{2})$
and since
$P_{i} < 1
\Longrightarrow
P_{i} = 0$,
we deduce by standard arguments that
\[
\Ex \left( \min \{ i \geq 0 \mid P_{i} = 0 \} \right)
\, \leq \,
O \left(
\left( \phi^{2} / \beta \right)
\cdot
\left(
\log (\sigma_{0}) + \log (k) + (\psi + \phi) \log (\Delta)
\right)
\right) \, .
\]
\Thm{}~\ref{theorem:edges:main} follows by recalling that
$\tau + \phi = O (\phi^{3})$
and that
$t^{r} = t^{*} + O (\psi + \phi)$.

The remainder of \Sect{}~\ref{section:edges:analysis} is dedicated to proving
\Prop{} \ref{proposition:edges:roadmap:depth-bounds-spatial-distance},
\ref{proposition:edges:roadmap:depth-bounds-temporal-distance}, and
\ref{proposition:edges:roadmap:progress};
this is done in \Sect{} \ref{section:edges:analysis:affected-regions},
\ref{section:edges:analysis:fault-recovery}, and
\ref{section:edges:analysis:progress}, respectively.
The proofs of the three propositions rely on two important ``service lemmas''
that are stated and established in
\Sect{}~\ref{section:edges:analysis:service-lemmas}.

\subsubsection{Service Lemmas}
\label{section:edges:analysis:service-lemmas}
In this section, we establish two important technical lemmas that play a
crucial role in the remainder of
\Sect{}~\ref{section:edges:analysis}, starting with
the following one.

\begin{lemma} \label{lemma:edges:service:conditions-port-inconsistency}
Fix some time
$t \geq t^{\circ}$
and an edge
$\{ u, v \} \in E(G_{t})$.
Assume that $u$ and $v$ are free of adversarial manipulations in rounds
$t - 2$
and
$t - 1$
and that
$\{ u, v \}$
is port-inconsistent at time $t$.
Then,
$\OutReg_{u, t - 1}(v) = \OutReg_{v, t - 1}(u) = \bot$
and if
$\OutReg_{u, t}(v) \neq \bot$,
then
$\Step_{u, t} = \HoldSymbol$.
\end{lemma}
\begin{proof}
Recall that $\Detect$ guarantees that
$\Detect_{u, t - 1}(v) = \Detect_{v, t - 1}(u)$
and that if
$\{ u, v \}$
is port-inconsistent at time
$t - 1$,
then
$\Detect_{u, t - 1}(v) = \Detect_{v, t - 1}(u) = \false$.
This results in resetting
$\OutReg_{u}(v), \OutReg_{v}(u) \gets \bot$
in round
$t - 1$
(line~\ref{line:edges:detect} of
Pseudocode~\ref{pseudocode:transformer-edges}),
making
$\{ u, v \}$
port-consistent at time $t$, in contradiction to the assumption.

Since an output value may be written into the output register
$\OutReg_{u}(v)$ (resp., $\OutReg_{v}(u)$) in round
$t - 1$
only if
$\OutReg_{u, t - 1}(v) = \bot$
(resp.,
$\OutReg_{v, t - 1}(u) = \bot$)
and since
$\Detect_{u, t - 1}(v) = \Detect_{v, t - 1}(u)$,
it follows that if
$\{ u, v \}$
is decided at time
$t - 1$,
then
$\{ u, v \}$
cannot be port-inconsistent at time $t$. 
Therefore, it must be the case that
$\OutReg_{u, t - 1}(v) = \OutReg_{v, t - 1}(u) = \bot$.

Finally, another necessary condition for an output value to be written into
$\OutReg_{u}(v)$ in round
$t - 1$
is that
$\Step_{u, t - 1} = \phi - 1$.
By \Obs{}~\ref{observation:transformer:step-counter}, this implies that
$\Step_{u, t} = \HoldSymbol$,
thus yielding the assertion.
\end{proof}

The next lemma requires the following notation and terminology.
Fix some time
$t \geq t^{\circ}$.
A node
$v \in V(G_{t})$
is said to be \emph{clean} at time $t$ if $v$ is not manipulated
during the round interval
$[t - \phi - 1, t)$.
Let
$U^{c}_{t} \subseteq V(G_{t})$
be the set of nodes that are clean at time $t$ and let
$H^{c}_{t} = G_{t}(U^{c}_{t})$,
observing that $H^{c}_{t}$ is a node induced subgraph of $G_{t'}$ for every
$t - \phi - 1 \leq t' \leq t$.
A node
$v \in V(G_{t})$
is said to be \emph{deeply-clean} at time $t$ if
$\{ w \in V(G_{t}) \mid \Distance_{G_{t}}(v, w) \leq \phi \}
\subseteq
U^{c}_{t}$,
that is, if all nodes in the ball of radius $\phi$ around $v$ are clean.
Let
$U^{dc}_{t} \subseteq V(G_{t})$
be the set of nodes that are deeply-clean at time $t$ and let
$H^{dc}_{t} = G_{t}(U^{dc}_{t})$.
Finally, for a step
$0 \leq j \leq \phi - 1$
of the phase procedure $\PhaseProcedure_{\Alg}$,
let
\[
S_{t}^{j}
\, = \,
\left\{ v \in V(G_{t}) \mid \Step_{v, t} = j \right\}
\, .
\]

\begin{lemma} \label{lemma:edges:service:complete-phase-simulation}
Consider some time
$t \geq t^{\circ}$
and define
\[
Q
\, = \,
\left\{
\{ u, v \} \in E(H^{c}_{t}) \cap \UnDecidedSet(C_{t})
\mid
u, v \in S_{t}^{\phi - 1}
\right\}
\quad \text{and} \quad
R
\, = \,
\left\{
\OutReg_{u}(v) \mid \{ u, v \} \in E(H^{dc}_{t}) \cap Q
\right\}
\, .
\]
Let $\eta$ be the (fault free) execution of $\PhaseProcedure_{\Alg}$ when
invoked on the graph
$H^{c}_{t}(Q \cup (\DecidedSet(C_{t}) \cap E(H^{c}_{t})))$
with the restriction of $C_{t}$ to
$Q \cup (\DecidedSet(C_{t}) \cap E(H^{c}_{t}))$
as the initial configuration.
Then, the assignment of output values to the output registers in $R$ performed
by $\SelfStabAlg$ in round $t$ obeys the same probability distribution as the
assignment of output values to the output registers in $R$ performed by $\eta$
in its decision step.
\end{lemma}
\begin{proof}
For a node
$v \in U^{c}_{t}$
and a step
$0 \leq j \leq \phi - 1$,
\Obs{}~\ref{observation:transformer:step-counter} ensures that
$v \in S_{t}^{\phi - 1}$
if and only if
$v \in S_{t - \phi + 1 + j}^{j}$.
This means that the nodes
$v \in U^{c}_{t}$
that simulate step
$\phi - 1$
of a phase $\Phi$ of $\PhaseProcedure_{\Alg}$ in round $t$ are exactly the
nodes
$v \in U^{c}_{t}$
that simulate step
$0 \leq j \leq \phi - 1$
of $\Phi$ in round
$t - \phi + 1 + j$
(see Pseudocode~\ref{pseudocode:transformer-edges}).
Moreover, since $\PhaseProcedure_{\Alg}$ resets all its non-output registers
at the beginning of $\Phi$, it follows by the definition of the deeply-clean
nodes that if
$v \in U^{dc}_{t}$,
then throughout the simulation of $\Phi$, node $v$ does not receive any
adversarially manipulated information.

\Lem{}~\ref{lemma:edges:service:conditions-port-inconsistency} ensures that the
edges in $Q$ are port-consistent at time
$t - \phi + 1 + j$
for every
$0 \leq j \leq \phi - 1$.
As the nodes
$v \in U^{c}_{t} \cap S_{t}^{\phi - 1}$
do not assign output values to their output registers $\OutReg_{v}(\cdot)$ in
step $j$ of phase $\Phi$ for any
$0 \leq j \leq \phi - 2$
(recall that step
$j = \phi - 1$
is the decision step), this means that
$Q = \left\{
\{ u, v \} \in E(H^{c}_{t}) \cap \UnDecidedSet(C_{t - \phi + 1})
\mid
u, v \in S_{t - \phi + 1}^{0}
\right\}$.
\Lem{}~\ref{lemma:edges:service:conditions-port-inconsistency} also ensures
that if
$v \in U^{dc}_{t} \cap S_{t}^{\phi - 1}$
and
$\OutReg_{v, t}(w) \in \mathcal{O}$
for an incident edge
$e = \{ v, w \} \in E(G_{t})$,
then $e$ is port-consistent, and hence decided,  at time $t$, which means that
$e \in \DecidedSet(C_{t}) \cap E(H^{c}_{t})$.
The assertion follows by recalling that $\Alg$ is a decision-oblivious
phase-based algorithm (see Criterion~\ref{criterion:edges:phase-based}).
\end{proof}

\begin{remark*}
Notice that the conditions of
\Lem{}~\ref{lemma:edges:service:complete-phase-simulation} are carefully
tailored to ensure that the assignment of output values to the output
registers in $R$ performed by $\SelfStabAlg$ in round $t$ indeed corresponds
to an actual (fault free) invocation of $\PhaseProcedure_{\Alg}$.
If these conditions are not satisfied, then the aforementioned assignment may
be the result of the simulation of a $\PhaseProcedure_{\Alg}$ phase that is
corrupted by the adversarial manipulations and hence, does not necessarily
correspond to any actual (fault free) invocation of $\PhaseProcedure_{\Alg}$.
\end{remark*}

\subsubsection{Affected Regions}
\label{section:edges:analysis:affected-regions}
Consider an edge
$e \in E(G_{t^{\circ}})$,
where
$C_{t^{\circ}}(e) = o$,
and a time
$t \geq t^{\circ}$
and assume that
$e \in E(G_{t})$
with
$\Distance_{G_{t}}(e, K_{t})
\geq
\DepthSupGraph_{\Predicate}(o) + \phi + 2$.
Our goal in this section is to prove
\Prop{}~\ref{proposition:edges:roadmap:depth-bounds-spatial-distance}, stating
that
(1)
$C_{t}(e) = o$;
and
(2)
$e \in \ContentSet(C_{t})$.
We augment this assertion with the additional claim that
(3)
all edges adjacent to $e$ in $G_{t}$ are port-consistent at time $t$.

Claims (1)--(3) are proved by induction on $t$.
The base case of
$t = t^{\circ}$
holds by the definition of $t^{\circ}$, ensuring that $C_{t^{\circ}}$ is a
content complete configuration.
So assume that claims (1)--(3) hold for time
$t \geq t^{\circ}$
and consider time
$t + 1$.
Employing \Obs{}~\ref{observation:edges:roadmap:decreasing-distance-to-faults},
the assumption that
$e \in E(G_{t + 1})$
with
$\Distance_{G_{t + 1}}(e, K_{t + 1})
\geq
\DepthSupGraph_{\Predicate}(o) + \phi + 2$
yields the following observation.

\begin{observation}
\label{observation:edges:affected-regions:distances-from-faults}
For every time
$t^{\circ} \leq t' \leq t + 1$,
it holds that
$e \in E(G_{t'})$
and
$N_{G_{t'}(e)} = N_{G_{t^{\circ}}}(e)$.
Moreover,
$\Distance_{G_{t'}}(e, K_{t'})
\geq
\DepthSupGraph_{\Predicate}(o) + \phi + 2$
and
$\Distance_{G_{t'}}(f, K_{t'})
\geq
\DepthSupGraph_{\Predicate}(o) + \phi + 1$
for every edge
$f \in N_{G_{t^{\circ}}}(e)$.
\end{observation}

Owing to
\Obs{}~\ref{observation:edges:affected-regions:distances-from-faults}, we
denote
$A = N_{G_{t^{\circ}}}(e)$
and notice that
$A = N_{G_{t'}}(e)$
for all
$t^{\circ} \leq t' \leq t + 1$.
\Obs{}~\ref{observation:edges:affected-regions:distances-from-faults} also
allows us to use the inductive hypothesis to conclude that
(1)
$C_{t}(e) = o$;
(2)
$e \in \ContentSet(C_{t})$;
and
(3)
all edges in $A$ are port-consistent at time $t$.
Taking $u$ and $v$ to be the endpoints of $e$, the guarantees of $\Detect$
imply that the output registers $\OutReg_{u}(v)$ and $\OutReg_{v}(u)$ are not
reset in round $t$ (see line~\ref{line:edges:detect} in
Pseudocode~\ref{pseudocode:transformer-edges}), thus
$C_{t + 1}(e) = C_{t}(e) = o$
establishing claim (1) of the inductive step.

En route to establishing claims (2) and (3) of the inductive step, let
$B = \{ w \in V(G_{t})
\mid
\Distance_{G_{t}}(w, e) \leq \phi + 1 \}$
be the ball of radius
$\phi + 1$
around $e$ in $G_{t}$
and
let
$H = G_{t}(B)$.
\Obs{}~\ref{observation:edges:affected-regions:distances-from-faults} ensures
that the nodes in $B$ are clean at time $t$ and that the endpoints of the
edges in $A$ are deeply-clean at time $t$ (refer to
\Sect{}~\ref{section:edges:analysis:service-lemmas} for the definitions of
clean and deeply-clean nodes).
This allows us to apply
\Lem{}~\ref{lemma:edges:service:complete-phase-simulation} and obtain the
following corollary.

\begin{corollary}
\label{corollary:edges:affected-regions:complete-phase-simulation}
Define
\[
Q
\, = \,
\left\{
\{ u, v \} \in E(H) \cap \UnDecidedSet(C_{t})
\mid
u, v \in S_{t}^{\phi - 1}
\right\}
\quad \text{and} \quad
R
\, = \,
\left\{
\OutReg_{u}(v) \mid \{ u, v \} \in A \cap Q
\right\}
\, .
\]
Let $\eta$ be the (fault free) execution of $\PhaseProcedure_{\Alg}$ when
invoked on the graph
$H(Q \cup (\DecidedSet(C_{t}) \cap E(H)))$
with the restriction of $C_{t}$ to
$Q \cup (\DecidedSet(C_{t}) \cap E(H))$
as the initial configuration.
Then, the assignment of output values to the output registers in $R$ performed
by $\SelfStabAlg$ in round $t$ obeys the same probability distribution as the
assignment of output values to the output registers in $R$ performed by $\eta$
in its decision step.
\end{corollary}

Next, we partition the set $A$ of edges adjacent to $e$ into the subsets
\[
A_{<}
\, = \,
\left\{
f \in A
\mid
\DepthSupGraph_{\Predicate}(C_{t^{\circ}}(f))
<
\DepthSupGraph_{\Predicate}(o)
\right\}
\quad \text{and} \quad
A_{\geq}
\, = \,
\left\{
f \in A
\mid
\DepthSupGraph_{\Predicate}(C_{t^{\circ}}(f))
\geq
\DepthSupGraph_{\Predicate}(o)
\right\}
\]
and argue about each subset separately, staring with the former.

\begin{lemma}
\label{lemma:edges:affected-regions:adjacent-edges-with-smaller-depth}
For each edge
$f \in A_{<}$,
it holds that
$C_{t + 1}(f) = C_{t}(f) = C_{t^{\circ}}(f)$.
\end{lemma}
\begin{proof}
\Obs{}~\ref{observation:edges:affected-regions:distances-from-faults} ensures
that
$\Distance_{G_{t}}(f, K_{t})
\geq
\DepthSupGraph_{\Predicate}(o) + \phi + 1$.
By the definition of $A_{<}$, we conclude that
$\Distance_{G_{t}}(f, K_{t})
\geq
\DepthSupGraph_{\Predicate}(C_{t^{\circ}}(f)) + \phi + 2$.
This allows us to apply the inductive hypothesis and deduce that
(1)
$C_{t}(f) = C_{t^{\circ}}(f)$;
(2)
$f \in \ContentSet(C_{t})$;
and
(3)
all edges adjacent to $f$ in $G_{t}$ are port-consistent at time $t$.
Taking $u$ and $v$ to be the endpoints of $f$, the guarantees of
$\Detect$ imply that the output registers $\OutReg_{u}(v)$ and
$\OutReg_{v}(u)$ are not reset in round $t$ (see line~\ref{line:edges:detect}
in Pseudocode~\ref{pseudocode:transformer-edges}), yielding the assertion.
\end{proof}

\Lem{}~\ref{lemma:edges:affected-regions:adjacent-edges-with-smaller-depth}
ensures that the edges in $A_{<}$ are port-consistent at time
$t + 1$.
To see that the edges in $A_{\geq}$ are also port-consistent at time
$t + 1$,
we further partition this edge set into the subsets
\[
A_{\geq}^{\DecidedSet}
\, = \,
A_{\geq} \cap \DecidedSet(C_{t})
\quad \text{and} \quad
A_{\geq}^{\UnDecidedSet}
\, = \,
A_{\geq} \cap \UnDecidedSet(C_{t})
\, ,
\]
observing that
$A_{\geq}^{\UnDecidedSet} = A \cap \UnDecidedSet(C_{t})$
due to
\Lem{}~\ref{lemma:edges:affected-regions:adjacent-edges-with-smaller-depth}.
\Lem{}~\ref{lemma:edges:service:conditions-port-inconsistency} guarantees that
the edges in $A_{\geq}^{\DecidedSet}$ are port-consistent at time
$t + 1$.
Recalling that $\Alg$ is a port-consistent phase-based algorithm (see
Criterion~\ref{criterion:edges:phase-based}), the fact that the edges in
$A_{\geq}^{\UnDecidedSet}$ are port-consistent at time
$t + 1$
follows from
\Cor{}~\ref{corollary:edges:affected-regions:complete-phase-simulation}, thus
establishing claim (3) of the inductive step.

It remains to establish claim (2) of the inductive step.
To this end, let $M_{<}$ be the multiset over
$\mathcal{O}$ defined as
\[
M_{<}
=
\{ C_{t^{\circ}}(f) \}_{f \in A_{<}}
\, .
\]
As $C_{t^{\circ}}$ is a content complete configuration,
we know that
$C_{t^{\circ}}[e] \in \SatMultisets(o)$,
which means that there exists a core
$M^{*} \in \Cores(o) \subseteq \SatMultisets(o)$
such that
$M^{*} \subseteq C_{t^{\circ}}[e]$
(recall the definition of cores from
\Sect{}~\ref{section:LCL-eligibility-criteria}).
By the definition of the supportive digraph $D_{\Predicate}$ (see
\Sect{}~\ref{section:LCL-eligibility-criteria}), we conclude that
$\DepthSupGraph_{\Predicate}(o') < \DepthSupGraph_{\Predicate}(o)$
for every
$o' \in M^{*}$,
hence
$M^{*} \subseteq M_{<}$.

Let
$M_{\geq}^{\DecidedSet, t}$,
$M_{\geq}^{\DecidedSet, t + 1}$,
and
$M_{\geq}^{\UnDecidedSet, t + 1}$
be the multisets over $\mathcal{O}$ defined as
\[
M_{\geq}^{\DecidedSet, t}
=
\{ C_{t}(f) \}_{f \in A_{\geq}^{\DecidedSet}}
\, , \;
M_{\geq}^{\DecidedSet, t + 1}
=
\{
C_{t + 1}(f)
\}_{f \in A_{\geq}^{\DecidedSet} \cap \DecidedSet(C_{t + 1})}
\, , \; \text{and} \;
M_{\geq}^{\UnDecidedSet, t + 1}
=
\{
C_{t + 1}(f)
\}_{f \in A_{\geq}^{\UnDecidedSet} \cap \DecidedSet(C_{t + 1})}
\, ,
\]
observing that
$M_{\geq}^{\DecidedSet, t + 1} \subseteq M_{\geq}^{\DecidedSet, t}$.
Since problem $\Problem$ meets the respectful decisions criterion
(Criterion~\ref{criterion:edges:respectful-decisions}) and since the edges in
$A_{\geq}^{\UnDecidedSet} \cap \DecidedSet(C_{t + 1})$
(i.e., the edges that realize
$M_{\geq}^{\UnDecidedSet, t + 1}$)
are exactly the edges adjacent to $e$ that obtain an output value in round
$t$, we can apply
\Cor{}~\ref{corollary:edges:affected-regions:complete-phase-simulation} once
more to deduce that
$M_{<} + M_{\geq}^{\DecidedSet, t} + M_{\geq}^{\UnDecidedSet, t + 1}
\in
\SatMultisets(o)$.\footnote{%
Given two multisets
$M, M' \in \Multisets(S)$,
the multiset
$M + M' \in \Multisets(S)$
is defined by setting
$(M + M')(s) = M(s) + M'(s)$
for each
$s \in S$.}
As $\Problem$ meets the gap freeness criterion
(Criterion~\ref{criterion:common:gap-freeness}), we conclude that
$M_{<} + M_{\geq}^{\DecidedSet, t + 1} + M_{\geq}^{\UnDecidedSet, t + 1}
\in
\SatMultisets(o)$
by observing that
\[
M^{*}
\, \subseteq \,
M_{<}
\, \subseteq \,
M_{<} + M_{\geq}^{\DecidedSet, t + 1} + M_{\geq}^{\UnDecidedSet, t + 1}
\, \subseteq \,
M_{<} + M_{\geq}^{\DecidedSet, t} + M_{\geq}^{\UnDecidedSet, t + 1}
\, .
\]
This establishes claim (2) of the inductive step as
$M_{<} + M_{\geq}^{\DecidedSet, t + 1} + M_{\geq}^{\UnDecidedSet, t + 1}
=
C_{t + 1}[e]$.

\subsubsection{Fault Recovery}
\label{section:edges:analysis:fault-recovery}
In this section, we establish
\Prop{}~\ref{proposition:edges:roadmap:depth-bounds-temporal-distance}.
To this end, let
$\hat{C}_{t} : E(G) \rightarrow \mathcal{O} \cup \{ \bot \}$,
$t \geq t^{*}$,
be the edge configuration defined by setting
\[
\hat{C}_{t}(e)
\, = \,
\begin{cases}
C_{t + 1}(e) \, , & e \in \UnDecidedSet(C_{t}) \cap \DecidedSet(C_{t + 1}) \\
C_{t}(e) \, , & \text{otherwise}
\end{cases}
\, ,
\]
where we observe that
$\UnDecidedSet(C_{t}) \cap \DecidedSet(C_{t + 1})$
is the subset of edges that become decided (as a result of the simulation of
$\PhaseProcedure_{\Alg}$'s decision step
$j = \phi - 1$
in line~\ref{line:edges:simulate-phase-procedure} of
Pseudocode~\ref{pseudocode:transformer-edges}) in round $t$.
The definition of $\hat{C}_{t}$ allows us to state the following basic
observation.

\begin{observation} \label{observation:edges:fault-recovery:early-relations}
For every
$t \geq t^{*} + 1$,
it holds that
\\
(1)
$\DecidedSet(\hat{C}_{t}) = \DecidedSet(C_{t}) \cup \DecidedSet(C_{t + 1})$;
\\
(2)
$\hat{C}_{t}(e) = C_{t + 1}(e) = \hat{C}_{t + 1}(e)$
for every edge
$e \in \DecidedSet(C_{t + 1})$;
and
\\
(3)
$C_{t}(e) = C_{t + 1}(e)$
for every edge
$e \in \DecidedSet(C_{t}) \cap \DecidedSet(C_{t + 1})$.
\end{observation}

Next, recall the definition of deeply-clean nodes from
\Sect{}~\ref{section:edges:analysis:service-lemmas} and notice that all
nodes in $V(G)$ are deeply clean at any time
$t \geq t^{*} + \phi + 1$.
Therefore, we can apply
\Lem{}~\ref{lemma:edges:service:complete-phase-simulation} to obtain the
following corollary.

\begin{corollary}
\label{corollary:edges:fault-recovery:complete-phase-simulation}
Consider some time
$t \geq t^{*} + \phi + 1$
and define
\[
Q
\, = \,
\left\{
\{ u, v \} \in \UnDecidedSet(C_{t})
\mid
u, v \in S_{t}^{\phi - 1}
\right\}
\quad \text{and} \quad
R
\, = \,
\left\{
\OutReg_{u}(v) \mid \{ u, v \} \in Q
\right\}
\, .
\]
Let $\eta$ be the (fault free) execution of $\PhaseProcedure_{\Alg}$ when
invoked on the graph
$G(Q \cup \DecidedSet(C_{t}))$
with the restriction of $C_{t}$ to
$Q \cup \DecidedSet(C_{t})$
as the initial configuration.
Then, the assignment of output values into the output registers in $R$
performed by $\SelfStabAlg$ in round $t$ obeys the same probability
distribution as the assignment of output values into the output registers in
$R$ performed by $\eta$ in its decision step.
\end{corollary}

We are now ready to establish the following lemma.

\begin{lemma} \label{lemma:edges:fault-recovery:late-relations}
For every
$t \geq t^{*} + \phi + 2$,
it holds that
\\
(1)
all edges are port-consistent at time $t$;
\\
(2)
$\UnContentSet(C_{t})
=
\DecidedSet(C_{t}) \cap \UnDecidedSet(C_{t + 1})
=
\DecidedSet(\hat{C}_{t}) \cap \UnDecidedSet(C_{t + 1})$;
\\
(3)
$\ContentSet(C_{t}) \subseteq \ContentSet(\hat{C}_{t})$;
and
\\
(4)
$\UnDecidedSet(C_{t}) \cap \DecidedSet(\hat{C}_{t}) \subseteq
\ContentSet(\hat{C}_{t})$.
\end{lemma}
\begin{proof}
Claim (1) is a consequence of
\Cor{}~\ref{corollary:edges:fault-recovery:complete-phase-simulation} as
$\Alg$ is port-consistent.
Claims (3) and (4) are also a consequence of
\Cor{}~\ref{corollary:edges:fault-recovery:complete-phase-simulation} due to
Criterion~\ref{criterion:edges:respectful-decisions}.
To see that claim (2) holds, notice that if all edges are port-consistent at
time $t$ (as promised by claim (1)), then the guarantees of $\Detect$ imply
that
$\Detect_{u, t}(v) = \Detect_{v, t}(u) = \false$
if and only if
$\{ u, v \} \in \UnContentSet(C_{t})$
for every edge
$\{ u, v \} \in \DecidedSet(C_{t})$.
The assertion follows by the design of $\SelfStabAlg$.
\end{proof}

Owing to \Obs{}~\ref{observation:edges:fault-recovery:early-relations}(3) and
\Lem{}~\ref{lemma:edges:fault-recovery:late-relations}(1), the task of
establishing
\Prop{}~\ref{proposition:edges:roadmap:depth-bounds-temporal-distance} reduces
to that of proving the following lemma.

\begin{lemma} \label{lemma:edges:fault-recovery:directed-path}
For every
$t \geq t^{*} + \phi + 2$,
if
$e \in \UnContentSet(C_{t})$,
then the supportive digraph
$D_{\Predicate}$
associated with the LCL predicate $\Predicate$ admits a directed path $P$ of
length
$t - (t^{*} + \phi + 2)$
emerging from $C_{t}(e)$.
\end{lemma}
\begin{proof}
We establish the assertion by induction on $t$.
The base case of
$t = t^{*} + \phi + 2$
is trivial as the digraph $D_{\Predicate}$ admits a path of length $0$
emerging from $o$ for every
$o \in \mathcal{O}$
including
$o = C_{t}(e)$.

Assume that the assertion holds for
$t - 1 \geq t^{*} + \phi + 2$
and consider an edge
$e \in \UnContentSet(C_{t})$.
We start by proving that
$e \in \ContentSet(\hat{C}_{t - 1})$.
As
$e \in \UnContentSet(C_{t}) \subseteq \DecidedSet(C_{t})$,
\Lem{}~\ref{lemma:edges:fault-recovery:late-relations}(2) ensures that
$e \notin \UnContentSet(C_{t - 1})$.
If
$e \in \UnDecidedSet(C_{t - 1})$,
then since
$e \in \DecidedSet(C_{t})$,
we know that
$e \in \UnDecidedSet(C_{t - 1}) \cap \DecidedSet(\hat{C}_{t - 1})$
by \Obs{}~\ref{observation:edges:fault-recovery:early-relations}(1).
Thus,
$e \in \ContentSet(\hat{C}_{t - 1})$
by \Lem{}~\ref{lemma:edges:fault-recovery:late-relations}(4).
Otherwise, it holds that
$e \in \ContentSet(C_{t - 1})$
and we can apply \Lem{}~\ref{lemma:edges:fault-recovery:late-relations}(3) to
conclude that
$e \in \ContentSet(\hat{C}_{t - 1})$.

Let
$o = \hat{C}_{t - 1}(e) \in \mathcal{O}$
be the output value of $e$ under $\hat{C}_{t - 1}$.
By \Obs{}~\ref{observation:edges:fault-recovery:early-relations}(2), we know
that
$C_{t}(e) = o$
as well.
Since
$e \in \ContentSet(\hat{C}_{t - 1})$,
it follows that
$\hat{C}_{t - 1}[e] \in \SatMultisets(o)$.
Let
$\hat{M} \in \Cores(o)$
be a core for $o$ such that
$\hat{M} \subseteq \hat{C}_{t - 1}[e]$
(recall that cores are defined in
\Sect{}~\ref{section:LCL-eligibility-criteria}).
We argue that
$\hat{M} \neq \emptyset$.
Indeed, \Obs{}~\ref{observation:edges:fault-recovery:early-relations}(1, 2)
implies that
$C_{t}[e] \subseteq \hat{C}_{t - 1}[e]$,
hence if
$\hat{M} = \emptyset$,
then
$C_{t}[e] \in \SatMultisets(o)$
due to Criterion~\ref{criterion:common:gap-freeness},
which contradicts the assumption that
$e \in \UnContentSet(C_{t})$.

Let
$\hat{A} \subseteq N_{G}(e)$
be a subset of $e$'s neighbors whose decisions under
$\hat{C}_{t - 1}$
realize $\hat{M}$.
For each
$e' \in \hat{A}$,
either
(1)
$e' \in \DecidedSet(C_{t})$
and
$C_{t}(e') = \hat{C}_{t - 1}(e')$
by \Obs{}~\ref{observation:edges:fault-recovery:early-relations}(2);
or
(2)
$e' \in \UnDecidedSet(C_{t})$.
Recalling that
$C_{t}[e] \subseteq \hat{C}_{t - 1}[e]$,
if
$C_{t}(e') = \hat{C}_{t - 1}(e')$
for every
$e' \in \hat{A}$,
then
$\hat{M} \subseteq C_{t}[e] \subseteq \hat{C}_{t - 1}[e]$
implying that
$C_{t}[e] \in \SatMultisets(o)$
due to Criterion~\ref{criterion:common:gap-freeness}.
But this means that
$e \in \ContentSet(C_{t})$,
in contradiction to the assumption that
$e \in \UnContentSet(C_{t})$.

Therefore, there must exist an edge
$e' \in \hat{A}$
such that
$e' \in \UnDecidedSet(C_{t})$.
As
$e' \in \DecidedSet(\hat{C}_{t - 1})$,
\Lem{}~\ref{lemma:edges:fault-recovery:late-relations}(2) ensures that
$e' \in \UnContentSet(C_{t - 1})$.
By the inductive hypothesis, the supportive digraph $D_{\Predicate}$ admits a
directed path $P'$
of length
$t - 1 - (t^{*} + \phi + 2)$
emerging from
$o' = C_{t - 1}(e') \in \mathcal{O}$.
Recalling that
$o' \in \hat{M} \in \Cores(o)$,
we conclude that
$(o, o') \in E(D_{\Predicate})$.
The assertion follows as
$(o, o') \circ P'$
is a directed path of length
$t - (t^{*} + \phi + 2)$
emerging from $o$ in $D_{\Predicate}$.
\end{proof}

\subsubsection{Progress}
\label{section:edges:analysis:progress}
This section is dedicated to proving
\Prop{}~\ref{proposition:edges:roadmap:progress}, showing that
$\Ex(\pi(G, C_{t + \tau + \phi}))
\leq
\left( \frac{1 - \beta}{4 \phi^{2}} \right)
\cdot
\pi(G, C_{t})$
for every time
$t \geq t^{r}$,
where $\tau$ is the parameter promised in
\Lem{}~\ref{lemma:pps:start-phase-probability} and
$\pi : \ContConfGraphs(\Problem) \rightarrow \Integers_{\geq 0}$
is the edge separable potential function promised in
Criterion~\ref{criterion:edges:progress}.
We do so by analyzing the following $4$-stage artificial process:
\begin{enumerate}

\item[(I)]
The adversary determines the global state of $\SelfStabAlg$ at time $t$.
In particular, the adversary determines the configuration $C_{t}$ and the
value
$\Step_{v, t}$
of the $\PPS$ state at time $t$ for each node
$v \in V(G)$.

\item[(II)]
Nature determines the random node subset
$S = S_{t + \tau}^{0}$
by selecting each node
$v \in V(G)$
to be included in $S$, independently, with probability $p_{v}$, where
$p_{v} = \Pr(\Step_{v, t + \tau} = 0 \mid \Step_{v, t})$.

\item[(III)]
The node subset $S$ is revealed to the adversary who then determines the coin
tosses of all nodes for the round interval
$[t, t + \tau)$
subject to 
$S_{t + \tau}^{0} = S$;
and the coin tosses of the nodes in
$V(G) - S$
for the round interval
$[t + \tau, t + \tau + \phi)$.
In particular, the adversary determines the configuration
$C_{t + \tau}$
and the value of
$C_{t + \tau + \phi}(e)$
for each edge
$e \in \UnDecidedSet(C_{t + \tau})$
such that
$e \cap S = \emptyset$.

\item[(IV)]
Nature tosses the coins of the nodes in $S$ for the round interval
$[t + \tau, t + \tau + \phi)$.

\end{enumerate}
If \Prop{}~\ref{proposition:edges:roadmap:progress} holds for this artificial
process, then it also holds for the actual execution of $\SelfStabAlg$ during
the round interval
$[t, t + \tau + \phi)$
as in the latter, the role of the adversary is assumed by (the less malicious)
nature.

Let
$\mathcal{I}
=
\left\{ \sigma(M) \right\}_{M \in \Multisets(\mathcal{O})}
\subset
\Integers_{> 0}$
be $\pi$'s family of potential coefficients.
Fix some time
$t \leq t' \leq t + \tau + \phi$.
For and an edge
$e \in E(G)$,
let
\[
\lambda_{t'}(e)
\, = \,
\begin{cases}
\sigma(C_{t'}[e]) \, , & e \in \UnDecidedSet(C_{t'}) \\
0 \, , & \text{otherwise}
\end{cases}
\, ,
\]
observing that as $\mathcal{I}$ is monotonically non-increasing,
\Cor{}~\ref{corollary:edges:roadmap:fault-recovery} implies that
$\lambda_{t'}(e)$ is a monotonically non-increasing function of $t'$.
This notation is extended to edge subsets
$F \subseteq E(G)$
by defining
$\lambda_{t'}(F)
=
\sum_{e \in F} \lambda_{t'}(e)$,
which allows us to express the potential in $G$ with respect to $C_{t}$ as
\begin{equation} \label{equation:edges:progress:express-potential-with-lambda}
\pi(G, C_{t'})
\, = \,
\lambda_{t'}(E(G))
\, .
\end{equation}
Moreover,
$\lambda_{t'}(F_{1} \cup F_{2})
=
\lambda_{t'}(F_{1}) + \lambda_{t'}(F_{2})$
for every disjoint edge subsets
$F_{1}, F_{2} \subseteq E(G)$.

Let
$F^{S} = \{ \{ u, v \} \in E(G) \mid u, v \in S \}$.
Since \Lem~\ref{lemma:pps:start-phase-probability} ensures that
$p_{v} \geq 1 / (2 \phi)$
for each node
$v \in V$,
it follows that
$\Pr(e \in F^{S}) \geq 1 / (2 \phi)^{2} = 1 / (4 \phi^{2})$
for each edge
$e \in E(G)$,
thus
\begin{equation} \label{equation:edges:progress:large-share}
\Ex \left( \lambda_{t}(F^{S}) \right)
\, \geq \,
\frac{1}{4 \phi^{2}} \cdot \lambda_{t}(E(G))
\, ,
\end{equation}
where the expectation is over the coin tosses of stage (II).

For a time
$t \leq t' \leq t + \tau + \phi$,
let
$H_{t'} = G(F^{S} \cup \DecidedSet(C_{t'}))$
and let
$C^{H}_{t'} : E(H_{t'}) \rightarrow \mathcal{O} \cup \{ \bot \}$
be the restriction of $C_{t'}$ to the edges in $E(H_{t'})$.
By definition, we know that
$\UnDecidedSet(C^{H}_{t'}) \subseteq F^{S}$
and if
$e \in F^{S}$,
then
$C^{H}_{t'}[e] = C_{t'}[e]$,
thus
\[
\pi(H_{t'}, C^{H}_{t'})
\, = \,
\lambda_{t'}(F^{S})
\, .
\]
Since only the nodes
$v \in S$
may write output values into their output registers $\OutReg_{v}(\cdot)$ in
round
$t + \tau + \phi - 1$,
it follows that
$\DecidedSet(C_{t + \tau + \phi - 1}) - F^{S}
=
\DecidedSet(C_{t + \tau + \phi}) - F^{S}$,
hence
$H_{t + \tau + \phi - 1} = H_{t + \tau + \phi}$.
As problem $\Problem$ meets the
$(\beta, \sigma_{0})$-progress
criterion (Criterion~\ref{criterion:edges:progress}), we can employ
\Cor{}~\ref{corollary:edges:fault-recovery:complete-phase-simulation} to
conclude that
\[
\Ex \left( \pi(H_{t + \tau + \phi}, C^{H}_{t + \tau + \phi}) \right)
\, \leq \,
(1 - \beta) \cdot \pi(H_{t + \tau + \phi - 1}, C^{H}_{t + \tau + \phi - 1})
\, ,
\]
and therefore,
\begin{equation} \label{equation:edges:progress:phase-progress}
\Ex \left(
\lambda_{t + \tau + \phi}(F^{S})
\right)
\, \leq \,
(1 - \beta) \cdot \left(
\lambda_{t + \tau + \phi - 1}(F^{S})
\right)
\, ,
\end{equation}
where the expectation is over the coin tosses of stage (IV).

Put together, we conclude that
\begin{align*}
\Ex \left( \pi(G, C_{t + \tau + \phi}) \right)
\, = \, &
\Ex \left(
\lambda_{t + \tau + \phi}(E(G))
\right)
\\
= \, &
\Ex \left(
\lambda_{t + \tau + \phi}(F^{S})
\right)
+
\Ex \left(
\lambda_{t + \tau + \phi}(E(G) - F^{S})
\right)
\\
\leq \, &
(1 - \beta) \cdot \Ex \left(
\lambda_{t + \tau + \phi - 1}(F^{S})
\right)
+
\Ex \left(
\lambda_{t + \tau + \phi}(E(G) - F^{S})
\right)
\\
\leq \, &
(1 - \beta) \cdot \Ex \left(
\lambda_{t}(F^{S})
\right)
+
\Ex \left(
\lambda_{t}(E(G) - F^{S})
\right)
\\
= \, &
(1 - \beta) \cdot \Ex \left(
\lambda_{t}(F^{S})
\right)
+
\lambda_{t}(E(G))
-
\Ex \left(
\lambda_{t}(F^{S})
\right)
\\
= \, &
\lambda_{t}(E(G))
-
\beta \cdot \Ex \left(
\lambda_{t}(F^{S})
\right)
\\
\leq \, &
\lambda_{t}(E(G))
-
\frac{\beta}{4 \phi^{2}} \cdot \left(
\lambda_{t}(E(G))
\right)
\, = \,
\left( 1 - \frac{\beta}{4 \phi^{2}} \right) \cdot \pi(G, C_{t})
\, ,
\end{align*}
where
the first, third, seventh, and last transitions follow from
(\ref{equation:edges:progress:express-potential-with-lambda}),
(\ref{equation:edges:progress:phase-progress}),
(\ref{equation:edges:progress:large-share}),
and
(\ref{equation:edges:progress:express-potential-with-lambda}),
respectively, thus establishing
\Prop{}~\ref{proposition:edges:roadmap:progress}.

\section{The Node-LCL Transformer}
\label{section:transformer-nodes}
This section presents our transformer when applied to a node-LCL
$\Problem = \langle
\mathcal{O}, \mathcal{G}, \Predicate
\rangle$,
where
$\mathcal{G} \subseteq \mathcal{U}_{\Delta}$
for a degree bound
$\Delta$.
We start, in \Sect{}~\ref{section:nodes:algorithmic-eligibility-criteria}, by
introducing the eligibility criteria that $\Problem$ should meet with respect
to the transformer's algorithmic building blocks, namely, the fault free
algorithm $\Alg$ and the detection procedure $\Detect$ assuming that it also
meets Criteria \ref{criterion:common:gap-freeness} and
\ref{criterion:common:bounded-propagation} presented in
\Sect{}~\ref{section:LCL-eligibility-criteria}.
\Sect{}~\ref{section:nodes:main-theorem} is dedicated to stating
the formal guarantees of the transformer when applied to (the eligible) problem
$\Problem$, including the dependence of the self-stabilizing algorithm
$\SelfStabAlg$'s fully adaptive run-time and message size on the parameters of
the eligibility criteria.
The transformer's implementation is presented in
\Sect{}~\ref{section:nodes:implementation}, providing a precise description of
$\SelfStabAlg$ and how it is synthesized
from $\Alg$ and $\Detect$.
The correctness of $\SelfStabAlg$ is established in
\Sect{}~\ref{section:nodes:analysis} together with the promised upper bounds
on its fully adaptive run-time and message size.
We start with a short discussion of the similarities and differences between
the node- and edge-LCL transformers.

\paragraph{Comparison with the Edge-LCL Transformer.}
There are many common themes between the edge-LCL transformer presented in
\Sect{}~\ref{section:transformer-edges} and the node-LCL transformer presented
in the current section.
Nevertheless, there are several significant differences that prevent a unified
presentation.
One such difference is cast in the definition of a decision-oblivious
phase-based algorithm that is slightly more restrictive in the context of
node-LCLs (see Criterion~\ref{criterion:nodes:phase-based} and its preceding
discussion).
Node-LCLs also need a different type of separable potential functions that use
two families of potential coefficients (see
Criterion~\ref{criterion:nodes:progress}).
Moreover, when dealing with node-LCLs, the nodes themselves may become
decided, essentially withdrawing from active participation in the invocations
of the phase procedure, which introduces some additional complications for the
transformer's implementation (see \Sect{}~\ref{section:nodes:implementation}).
In particular, while the edge-LCL transformer does not have to ``interfere''
with the messages sent by the simulated fault free algorithm (an utterly
black-box simulation), the node-LCL transformer ``takes control'' over a
certain type of messages, confiscating them from the simulated phase
procedure (see \Sect{}~\ref{section:nodes:implementation}).

For the sake of a smooth exposition, the node-LCL transformer is presented in
the remainder of this section in a self-contained manner.
The reader is not assumed to be fluent with the notation, terminology, and
concepts introduced in \Sect{}~\ref{section:transformer-edges}.

\subsection{Algorithmic Eligibility Criteria}
\label{section:nodes:algorithmic-eligibility-criteria}

\paragraph{Detection Procedure.}
Our transformer employs a distributed detection procedure $\Detect$.
On a graph
$G \in \mathcal{G}$,
procedure $\Detect$ runs indefinitely and does not access any of the registers
maintained by the nodes
$v \in V(G)$
except for the output register $\OutReg_{v}$ to which $\Detect$ has a
read-only access.
The procedure returns a Boolean value denoted by
$\Detect_{v} \in \{ \true, \false \}$;
let
$\Detect_{v, t}$
denote the value of $\Detect_{v}$ in round $t$.
Informally, the role of $\Detect$ is to return
$\Detect_{v} = \false$
if ``something is wrong'' with the output register $\OutReg_{v}$.

More formally, consider a node
$v \in V(G)$
and assume that neither $v$ nor any of its neighbors in $G$ is manipulated in
round
$t - 1$.
Procedure $\Detect$ guarantees that
$\Detect_{v, t} = \false$
if and only if
$\Predicate(o, M) = \false$,
where
$M \in \Multisets(\mathcal{O})$
is the multiset defined over the output values stored at time $t$ in the
output registers belonging to
$\{ \OutReg_{u} \mid u \in N_{G}(v) \}$.

\begin{criterion}[\emph{$\mu_{D}(\Delta)$-detectability}]
\label{criterion:nodes:detectability}
Problem $\Problem$ meets the \emph{$\mu_{D}(\Delta)$-detectability} criterion
for a function
$\mu_{D} : \Integers_{> 0} \rightarrow \Integers_{> 0}$
if it admits a detection procedure whose message size is at 
most 
$\mu_{D}(\Delta)$ when it runs on a graph of degree bound $\Delta$.
\end{criterion}

\begin{remark*}
It is straightforward to implement $\Detect$ so that each node
$v \in V(G)$
shares the content of its output register $\OutReg_{v}$
with each one of its neighbors, resulting in
$\mu_{D}(\Delta) = O (\log |\mathcal{O}|)$.
In contrast to distributed edge problems, where implementing improved
(non-trivial) detection procedures may lead to an overall message size
improvement, in the context of distributed node problems, such improved
implementations do not lead to an overall improvement as the nodes have to
share the content of their output register with their neighbors regardless of
the detection procedure.
\end{remark*}

\paragraph{Restricting the Phase Procedure.}
Assume hereafter that $\Problem$ meets the $\mu_{D}(\Delta)$-detectability
criterion.
In the remainder of this section, we discuss the eligibility criteria imposed
on the fault free algorithm $\Alg$ for $\Problem$.
Consider the graph
$G \in \mathcal{G}$
on which $\Alg$ runs.
Recall that if $\Alg$ is a phase-based algorithm, then it is fully specified
by a phase length parameter $\phi$ and a phase procedure
$\PhaseProcedure_{\Alg}$ invoked at the beginning of every phase.
Moreover, each phase ends with a decision step
$j = \phi - 1$,
which is the only step in which the nodes
$v \in V(G)$
may write into their output registers
$\OutReg_{v}$,
an action that is allowed only if
$\OutReg_{v} = \bot$,
i.e., if $v$ is still undecided.

Recall further that when a new phase begins, all registers of a node
$v \in V(G)$
with the exception of the output register $\OutReg_{v}$ are reset.
This means, in particular, that $v$ loses any track of its neighbors' status
including which of them are still undecided and which already hold an
(irrevocable) output value.
To overcome this technical difficulty, we assume hereafter that step
$j = 0$
of the phase is a designated ``announcement step'' in which each node
$v \in V(G)$
informs its neighbors whether or not
$\OutReg_{v} = \bot$
(essentially sending a $1$-bit message).
The actual actions of the phase procedure $\PhaseProcedure_{\Alg}$ then start
in step
$j = 1$
after $v$ already identified its undecided neighbors.

Fix some phase of the phase procedure $\PhaseProcedure_{\Alg}$ and let
$D \subseteq V(G)$
be the subset of nodes that are decided at the beginning of the phase.
The algorithm is said to be \emph{decision-oblivious} if the only step in
which a node
$v \in V(G) - D$
may take into account the output values associated with the nodes in $D$ is
the decision step
$j = \phi - 1$.
In other words, the actions of the nodes in
$V(G) - D$
(in terms of the values they write into their local registers) in steps
$j = 1, \dots, \phi - 2$
are identical to their actions in an invocation of $\PhaseProcedure_{\Alg}$ on
the graph
$G(V(G) - D)$
induced by the undecided nodes.
This means in particular that while a node
$v \in V(G) - D$
may base its actions in steps
$j = 1, \dots, \phi - 2$
of $\PhaseProcedure_{\Alg}$ on the (global) degree bound $\Delta$ and on $v$'s
degree in
$G(V(G) - D)$,
its actions in these steps are independent of the number
$\Degree_{G}(v) - \Degree_{G(V(G) - D)}(v)$
of decided nodes adjacent to $v$.

To enable the decision making of a node
$v \in V(G)$
during the decision step
$j = \phi - 1$,
the decision-oblivious algorithm is required to satisfy the following
property:
in step
$\phi - 2$,
each node
$u \in D \cap N_{G}(v)$
sends to $v$ a message that contains all the information that $v$ may need in
(the local computation stage of) step
$\phi - 1$
regarding the output value
$o \in \mathcal{O}$
stored in $u$'s output register
$\OutReg_{u}$.
To this end, we assume that $u$ writes the string $g_{\Alg}(o)$ into a
designated field
$m.\fieldOut$
appended to the messages $m$ that $u$ sends in step 
$\phi - 2$,
where
$g_{\Alg} : \mathcal{O} \rightarrow \{ 0, 1 \}^{*}$
is an abstract \emph{output encoding} function determined by the algorithm
designer.\footnote{%
In most node-LCLs, the natural choice for an output encoding function is the
identity function
$g_{\Alg}(o) = o$
(this is also the output encoding used for the concrete algorithms in the
current paper).
Nevertheless, we support the more general output encoding abstraction since it
opens the gate for algorithms whose message size is asymptotically smaller
than
$\log |\mathcal{O}|$.}
The $m.\fieldOut$ fields of the incoming messages received by $v$ in step
$\phi - 2$,
allow $v$ to account for the output values of its decided neighbors in the
decisions $v$ makes during the decision step
$\phi - 1$.
It is required though that $v$ regards this information as a multiset
$\{ g_{\Alg}(\OutReg_{u}) \}_{u \in N_{G}(v) \cap D}$.

We note that the role of a decided node
$u \in D$
in $\PhaseProcedure_{\Alg}$ is restricted to step
$j = 0$,
in which $u$ informs its neighbors that it is decided,
and step
$j = \phi - 2$,
in which $u$ shares $g_{\Alg}(\OutReg_{u})$ with its neighbors.
In all other steps of $\PhaseProcedure_{\Alg}$, node $u$ does not exchange any
information with its neighbors.

\begin{criterion}[\emph{$(\phi, \mu_{A}(\Delta))$-phase-based}]
\label{criterion:nodes:phase-based}
Problem $\Problem$ meets the
\emph{$(\phi, \mu_{A}(\Delta))$-phase-based}
criterion (with respect to $\Alg$) if $\Alg$ is a decision-oblivious
phase-based algorithm with a phase of length $\phi$ that uses messages of size
$\mu_{A}(\Delta)$ when it runs on a graph of degree bound $\Delta$.
\end{criterion}

Assume hereafter that $\Problem$ meets the
$(\phi, \mu_{A}(\Delta))$-phase-based
criterion with respect to $\Alg$.
Given a node configuration
$C : V(G) \rightarrow \mathcal{O} \cup \{ \bot \}$,
let $\eta$ be the execution of $\PhaseProcedure_{\Alg}$ on the graph $G$ with
initial configuration $C$.
Let
$\PhaseProcedure_{\Alg}(G, C)$
be the node configuration associated with $\PhaseProcedure_{\Alg}$ when $\eta$
halts, i.e., after $\phi$ rounds.
Notice that if $\PhaseProcedure_{\Alg}$ is randomized, then
$\PhaseProcedure_{\Alg}(G, C)$
is a random variable that depends on the coin tosses of the nodes
$v \in V(G)$
during $\eta$.

\begin{criterion}[\emph{respectful decisions}]
\label{criterion:nodes:respectful-decisions}
Problem $\Problem$ meets the \emph{respectful decisions} criterion (with
respect to $\Alg$) if the following two conditions hold with probability $1$
for every graph
$G \in \mathcal{G}$
and node configuration
$C : V(G) \rightarrow \mathcal{O} \cup \{ \bot \}$: \\
(1)
$\ContentSet(C)
\subseteq
\ContentSet(\PhaseProcedure_{\Alg}(G, C))$;
and \\
(2)
$\UnDecidedSet(C) \cap \DecidedSet(\PhaseProcedure_{\Alg}(G, C))
\subseteq
\ContentSet(\PhaseProcedure_{\Alg}(G, C))$.
\end{criterion}

\begin{remark*}
Criterion~\ref{criterion:nodes:respectful-decisions} dictates that the
decisions taken by an invocation of $\PhaseProcedure_{\Alg}$ do not introduce
any (new) uncontent nodes:
condition (1) states that nodes that were content before the invocation of
$\PhaseProcedure_{\Alg}$ remain content in the new configuration;
condition (2) states that nodes that became decided during the execution of
$\PhaseProcedure_{\Alg}$ are content (in the new configuration).
\end{remark*}

\paragraph{Potential Functions.}
Assume hereafter that $\Problem$ meets the respectful decisions criterion with
respect to $\Alg$.
Let $\ContConfGraphs(\Problem)$ denote the collection of pairs of the form
$(G, C)$,
where 
$G \in \mathcal{G}$
and
$C : V(G) \rightarrow \mathcal{O} \cup \{ \bot \}$
is a content node configuration for $G$.
A function
$\pi : \ContConfGraphs(\Problem) \rightarrow \Integers_{\geq 0}$
is said to be a \emph{node separable potential function} for $\Problem$ if
there exist two monotonically non-increasing parameterized integer families
$\mathcal{I}^{\sigma}
=
\left\{ \sigma(M) \right\}_{M \in \Multisets(\mathcal{O})}
\subset
\Integers_{> 0}$
and
$\mathcal{I}^{\rho}
=
\left\{ \rho(M) \right\}_{M \in \Multisets(\mathcal{O})}
\subset
\Integers_{\geq 0}$
(see \Sect{}~\ref{section:edges:algorithmic-eligibility-criteria} for the
definition of monotonically non-increasing parameterized integer families)
such that
\[
\pi(G, C)
\, = \,
\sum_{v \in \UnDecidedSet(C)}
\sigma(C[v])
+
\rho(C[v]) \cdot \left| N_{G}(v) \cap \UnDecidedSet(C) \right|
\]
for every
$(G, C) \in \ContConfGraphs(\Problem)$.
In this case, we refer to the integers in $\mathcal{I}^{\sigma}$ (resp.,
$\mathcal{I}^{\rho}$) as the \emph{potential $\sigma$-coefficients} (resp.,
\emph{potential $\rho$-coefficients}) of $\pi$ and to $\sigma(\emptyset)$
(resp., $\rho(\emptyset)$) as the \emph{top potential $\sigma$-coefficient}
(resp., \emph{top potential $\rho$-coefficient}), observing that
$\sigma(\emptyset)$ (resp., $\rho(\emptyset)$) up-bounds any other potential
$\sigma$-coefficient $\sigma(\cdot)$ (resp., potential
$\rho$-coefficient $\rho(\cdot)$).

\begin{criterion}[\emph{$(\beta, \sigma_{0}, \rho_{0})$-progress}]
\label{criterion:nodes:progress}
Problem $\Problem$ meets the
\emph{$(\beta, \sigma_{0}, \rho_{0})$-progress}
criterion (with respect to $\Alg$)
for a real
$\beta > 0$
and integers
$\sigma_{0} > 0$
and
$\rho_{0} \geq 0$
if $\Problem$ admits a node separable potential function $\pi$ with top
potential $\sigma$-coefficient $\sigma_{0}$ and top potential
$\rho$-coefficient $\rho_{0}$ such that for every
$(G, C) \in \ContConfGraphs(\Problem)$,
it is guaranteed that
$\Ex(\pi(G, \PhaseProcedure_{\Alg}(G, C))) \leq (1 - \beta) \cdot \pi(G, C)$.
\end{criterion}

\begin{remark*}
The requirement that the two families of potential coefficients are
monotonically non-increasing implies that
$\pi(G, C') \leq \pi(G, C)$
for every graph $G$ and node configurations $C$ and $C'$ over $G$ satisfying
(1)
$\DecidedSet(C') \supseteq \DecidedSet(C)$;
and
(2)
$C'(v) = C(v)$
for every node
$v \in \DecidedSet(C)$.
Moreover, since the potential $\sigma$-coefficients are strictly positive, if
$\pi(G, C) = 0$,
then $C$ must be complete.
This means that as the execution of $\Alg$ progresses, the potential in
$\Alg$'s associated configuration cannot increase and that once we reach a
configuration with $0$ potential, all nodes must be decided.
Criterion~\ref{criterion:nodes:progress} ensures that the potential decreases
sufficiently fast, ultimately allowing us to up-bound the run-time of the
self-stabilizing algorithm synthesized from $\Alg$.
\end{remark*}

\subsection{The Main Theorem}
\label{section:nodes:main-theorem}
Problem
$\Problem = \langle
\mathcal{O}, \mathcal{G}, \Predicate
\rangle$
is said to be
\emph{$(
\psi,
\mu_{D}(\Delta),
\phi,
\mu_{A}(\Delta),
\beta,
\sigma_{0},
\rho_{0}
)$-eligible}
if
$\Problem$ meets the
gap freeness
(Criterion~\ref{criterion:common:gap-freeness}),
$\psi$-bounded propagation
(Criterion~\ref{criterion:common:bounded-propagation}),
$\mu_{D}(\Delta)$-detectability
(Criterion~\ref{criterion:nodes:detectability}),
$(\phi, \mu_{A}(\Delta))$-phased-based
(Criterion~\ref{criterion:nodes:phase-based}),
respectful decisions
(Criterion~\ref{criterion:nodes:respectful-decisions}),
and
$(\beta, \sigma_{0}, \rho_{0})$-progress
(Criterion~\ref{criterion:nodes:progress})
criteria.
The guarantees of our transformer are cast in the following theorem.

\begin{theorem} \label{theorem:nodes:main}
If a node-LCL
$\Problem = \langle
\mathcal{O}, \mathcal{G}, \Predicate
\rangle$
is
$(\psi,
\mu_{D}(\Delta),
\phi,
\mu_{A}(\Delta),
\beta,
\sigma_{0},
\rho_{0})$-eligible,
then $\Problem$ admits a randomized self-stabilizing algorithm
$\SelfStabAlg$ that uses
messages of size
\[
O \left(
\mu_{D}(\Delta) + \mu_{A}(\Delta) + \log \phi
\right)
\]
whose fully adaptive run-time is
\[
O \left(
\left( \phi^{5} / \beta \right)
\cdot
\left(
\log (k) + (\psi + \phi) \log (\Delta) + \log (\sigma_{0} + \rho_{0})
\right)
\right) \, .
\]
Moreover, nodes whose distance from any manipulated node is at least
$\psi + \phi + 2$
are guaranteed to maintain their original output value throughout the
execution of $\SelfStabAlg$ (with probability $1$).
\end{theorem}

\subsection{Implementation}
\label{section:nodes:implementation}
Suppose that $\Problem$ is
$(\psi, \mu_{D}(\Delta), \phi, \mu_{A}(\Delta), \beta, \sigma_{0},
\rho_{0})$-eligible and let $\Detect$ and $\Alg$ be the corresponding
detection procedure and fault free algorithm.
In this section, we describe how the self-stabilizing algorithm
$\SelfStabAlg$, promised in \Thm{}~\ref{section:nodes:main-theorem}, is
synthesized from $\Detect$ and from the phase procedure
$\PhaseProcedure_{\Alg}$ associated with $\Alg$.
This synthesis is recounted in Algorithm~\ref{pseudocode:transformer-nodes},
where $\SelfStabAlg$ runs on an input graph
$G \in \mathcal{G}$
and is presented from the perspective of a node
$v \in V(G)$,
denoting the messages sent to (resp., from) $v$
from (resp., to) a neighbor
$u \in N_{G}(v)$
by $\InMsg_{v}(u)$ (resp., $\OutMsg_{v}(u)$).

Each message $m$ sent by $v$ is augmented with a designated field, denoted by
$m.\fieldPPS$, where $v$ records the state of its $\PPS$ module
(line~\ref{line:nodes:record-step}).
Using the $m.\fieldPPS$ fields in its incoming messages, node $v$ can identify
its \emph{$\PPS$-synchronized} neighbors, namely, the nodes
$u \in N_{G}(v)$
that satisfy
$\Step_{u} = \Step_{v}$.
The crux of the transformer is that whenever
$\Step_{v} = 0$,
node $v$ invokes a simulation of a (complete) phase $\Phi$ of
$\PhaseProcedure_{\Alg}$ in conjunction with its $\PPS$-synchronized neighbors
(lines
\ref{line:nodes:simulate-phase-procedure-begin}--%
\ref{line:nodes:simulate-phase-procedure-end}).
The set of $v$'s $\PPS$-synchronized neighbors is stored in the designated
register $\Peers_{v}$ at the beginning of phase $\Phi$ (lines
\ref{line:nodes:compute-peers-begin}--\ref{line:nodes:compute-peers-end});
the $\PPS$ module is designed to guarantee that $v$ remains
$\PPS$-synchronized with the nodes in $\Peers_{v}$ throughout the simulation
of $\Phi$ (see \Obs{}~\ref{observation:transformer:step-counter}).

During the simulation of $\Phi$, we distinguish between steps
$0 \leq j \leq \phi - 2$,
in which $v$ simulates $\Phi$ in conjunction with its $\PPS$-synchronized
neighbors (line~\ref{line:nodes:simulate-early-steps}),
and step
$j = \phi - 1$,
in which $v$ simulates $\Phi$ in conjunction with its $\PPS$-synchronized
neighbors, but also collecting the output encoding
$g_{\Alg}(\OutReg_{u})$
of its decided neighbors
$u \in N_{G}(v)$
that are not necessarily $\PPS$-synchronized with $v$
(line~\ref{line:nodes:simulate-decision-step}).
To enable that, the transformer ``takes over'' the $\fieldOut$ fields of the
messages from the simulated procedure $\PhaseProcedure_{\Alg}$:
in $\SelfStabAlg$, each message $\InMsg_{v}(u)$ incoming from a decided
neighbor
$u \in N_{G}(v)$
carries the value of
$g_{\Alg}(\OutReg_{u})$ in the $\InMsg_{v}(u).\fieldOut$ field, just in case
$v$ is approaching a decision step.
Conversely, if $v$ itself is decided, then $v$ assigns
$\OutMsg_{v}(\cdot).\fieldOut \gets g_{\Alg}(\OutReg_{v})$
to each outgoing message (line~\ref{line:nodes:record-output-encoding}).

Regardless of the simulation of $\Phi$, node $v$ invokes the detection
procedure $\Detect$ in every round (line~\ref{line:nodes:call-detect}) and
resets its output register
$\OutReg_{v}$
if
$\Detect_{v} = \false$
(line~\ref{line:nodes:reset-output-register}).
If
$\Detect_{v} = \false$,
then $v$ also turns on the designated flag $\Wait_{v}$
(line~\ref{line:nodes:turn-wait-on}), whose role is to prevent $v$ from
participating in a simulation of a phase that already started.
This flag is reset whenever
$\Step_{v} = \HoldSymbol$
(line~\ref{line:nodes:turn-wait-off}),
thus allowing $v$ to participate in the simulation of the next phase.

\begin{algorithm}
\caption{\label{pseudocode:transformer-nodes}%
The code (executed in each round) of the self-stabilizing algorithm
$\SelfStabAlg$ for the node-LCL $\Problem$}
\begin{algorithmic}[1]
\State{call $\Detect$}%
\label{line:nodes:call-detect}%
\Comment{assigns a value to the (round-temporal) variable $\Detect_{v}$}
\If{$\Step_{v} = 0$}%
\label{line:nodes:compute-peers-begin}
  \State{%
$\Peers_{v} \gets
\{ u \in N_{G}(v)
\mid
\InMsg_{v}(u).\fieldPPS = 0
\}$}%
\label{line:nodes:simulate-early-steps}
\EndIf%
\label{line:nodes:compute-peers-end}
\If{%
$0 \leq \Step_{v} \leq \phi - 2$
and
$\Wait_{v} = \false$}%
\label{line:nodes:simulate-phase-procedure-begin}
  \State{%
simulate step $\Step_{v}$ of $\PhaseProcedure_{\Alg}$ with the neighbors in
$\Peers_{v}$}%
\label{line:nodes:simulate-decision-step}
\EndIf%
\label{line:nodes:simulate-phase-procedure}
\If{%
$\Step_{v} = \phi - 1$
and
$\Wait_{v} = \false$}%
\Comment{decision step}
  \State{%
simulate step
$\phi - 1$
of $\PhaseProcedure_{\Alg}$ with the neighbors in
$\Peers_{v} \cup \{ u \in N_{G}(v) \mid \InMsg_{v}(u).\fieldOut \neq \bot \}$}
\EndIf%
\label{line:nodes:simulate-phase-procedure-end}
\If{$\Detect_{v} = \false$}%
\label{line:nodes:detect}
  \State{$\OutReg_{v} \gets \bot$}%
\label{line:nodes:reset-output-register}
  \State{$\Wait_{v} \gets \true$}%
\label{line:nodes:turn-wait-on}
\EndIf
\If{$\Step_{v} = \HoldSymbol$}
  \State{$\Wait_{v} \gets \false$}%
\label{line:nodes:turn-wait-off}
\EndIf
\State{call $\AdvancePPS$}%
\Comment{updates the register $\Step_{v}$
(\Sect{}~\ref{section:probabilistic-phase-synchronization})}
\ForAll{$u \in N_{G}(v)$}
  \State{$\OutMsg_{v}(u).\fieldPPS \gets \Step_{v}$}%
\label{line:nodes:record-step}
  \If{$\OutReg_{v} \neq \bot$}
    \State{$\OutMsg_{v}(u).\fieldOut \gets g_{\Alg}(\OutReg_{v})$}%
\label{line:nodes:record-output-encoding}
  \Else
    \State{$\OutMsg_{v}(u).\fieldOut \gets \bot$}
  \EndIf
\EndFor
\end{algorithmic}
\end{algorithm}

\subsection{Analysis}
\label{section:nodes:analysis}
In this section, we analyze the self-stabilizing algorithm
$\SelfStabAlg$ developed in \Sect{}~\ref{section:nodes:implementation}
and establish \Thm{}~\ref{theorem:nodes:main}.
The
$O (\mu_{D}(\Delta) + \mu_{A}(\Delta) + \log \phi)$
message size bound of \Thm{}~\ref{theorem:nodes:main} follows directly from
the transformer's design as each message of $\SelfStabAlg$ is composed
of
(i)
a message of the detection procedure $\Detect$ whose size is
$\mu_{D}(\Delta)$;
(ii)
a message of the fault free algorithm $\Alg$ (actually, of the phase procedure
$\PhaseProcedure_{\Alg}$) whose size is
$\mu_{A}(\Delta)$;
(iii)
one of the
$\phi + 1$
states of the $\PPS$ module;
and
(iv)
possibly, an output encoding taken from
$\{ g_{\Alg}(o) \mid o \in \mathcal{O} \}$
that does not require more than
$\mu_{A}(\Delta)$ bits (see
\Sect{}~\ref{section:nodes:algorithmic-eligibility-criteria}). 
Our goal in the remainder of this section is to establish
\Thm{}~\ref{theorem:nodes:main}'s fully adaptive run-time bound and the bound
on the distance from the adversarial manipulated nodes to nodes that change
their output value (i.e., the theorem's last claim).

\begin{remark*}
In contrast to the analysis presented in \Sect{}~\ref{section:edges:analysis}
for the product of our edge-LCL transformer, in the current section, we do not
have to be concerned with the issue of port-inconsistency, which makes the
analysis somewhat simpler in several places.
On the other hand, since node separable potential functions are slightly more
complicated than edge separable potential functions, the part of the analysis
in which we exploit Criterion~\ref{criterion:nodes:progress} is a bit more
involved than its edge-LCL counterpart.
\end{remark*}

\subsubsection{Roadmap}
\label{section:nodes:analysis:roadmap}
Recall that the adversarial manipulations may include the addition/removal of
nodes/edges, hence the graph may change during the round interval
$[t^{\circ}, t^{*} - 1]$
(during which the adversarial manipulations take place).
For
$t \geq t^{\circ}$,
let $G_{t}$ be the graph at time $t$ and let
$K_{t} \subseteq V(G_{t})$
be the set of nodes that experience adversarial manipulations during the round
interval
$[t^{\circ}, t - 1]$
and still exist in $G_{t}$.
Let
$C_{t} : V(G_{t}) \rightarrow \mathcal{O} \cup \{ \bot \}$
be the node configuration associated with $\SelfStabAlg$ at time $t$.
The journey towards proving \Thm{}~\ref{theorem:nodes:main} starts in
\Sect{}~\ref{section:nodes:analysis:affected-regions} that is dedicated to
establishing the following proposition.

\begin{proposition}
\label{proposition:nodes:roadmap:depth-bounds-spatial-distance}
Consider a node
$v \in V(G_{t^{\circ}})$
with
$C_{t^{\circ}}(v) = o$
and a time
$t \geq t^{\circ}$.
If
$v \in V(G_{t})$
and
$\Distance_{G_{t}}(v, K_{t})
\geq
\DepthSupGraph_{\Predicate}(o) + \phi + 2$,
then
(1)
$C_{t}(v) = o$;
and
(2)
$v \in \ContentSet(C_{t})$.
\end{proposition}

In what follows, we denote
$G = G_{t^{*}}$
and
$K = K_{t^{*}}$,
observing that the definition of $t^{*}$ implies that
$G = G_{t}$
and
$K = K_{t}$
for every
$t \geq t^{*}$
and that
$|K| \leq k$.
By combining
\Obs{}~\ref{observation:edges:roadmap:decreasing-distance-to-faults} and
\Prop{}~\ref{proposition:nodes:roadmap:depth-bounds-spatial-distance}, we
obtain the following corollary that establishes the last claim of
\Thm{}~\ref{theorem:nodes:main}.

\begin{corollary} \label{corollary:nodes:roadmap:affected-regions}
Consider a node
$v \in V(G)$
and a time
$t \geq t^{*}$.
If
$\Distance_{G}(v, K) \geq \psi + \phi + 2$,
then
(1)
$v \in V(G_{t^{\circ}})$;
(2)
$C_{t}(v) = C_{t^{\circ}}(v)$;
and
(3)
$v \in \ContentSet(C_{t})$.
\end{corollary}

It remains to establish the fully adaptive run-time bound of
\Thm{}~\ref{theorem:nodes:main}.
While \Prop{}~\ref{proposition:nodes:roadmap:depth-bounds-spatial-distance}
uses the depth of the output values to provide a ``spatial barrier'' from the
adversarial manipulations, the next proposition, established in
\Sect{}~\ref{section:nodes:analysis:fault-recovery}, uses the depth to provide
a ``temporal barrier'' from the adversarial manipulations.

\begin{proposition}
\label{proposition:nodes:roadmap:depth-bounds-temporal-distance}
Consider a node
$v \in V(G)$
and a time
$t$.
If
$C_{t}(v) = o \in \mathcal{O}$
and
$t \geq t^{*} + \DepthSupGraph_{\Predicate}(o) + \phi + 2$,
then
$v \in \ContentSet(C_{t})$.
\end{proposition}

Recalling the guarantees of the detection procedure $\Detect$, the design of
$\SelfStabAlg$ ensures that for every
$t \geq t^{*} + 1$,
a node
$v \in \DecidedSet(C_{t})$
becomes undecided in round $t$ if and only if
$v \in \UnContentSet(C_{t})$.
The design of $\SelfStabAlg$ also ensures that if
$v \in \DecidedSet(C_{t}) \cap \DecidedSet(C_{t + 1})$,
then
$C_{t}(v) = C_{t + 1}(v)$.
Setting
\[
t^{r}
\, = \,
t^{*} + \psi + \phi + 2
\, ,
\]
we obtain the following corollary from
\Prop{}~\ref{proposition:nodes:roadmap:depth-bounds-temporal-distance}.

\begin{corollary} \label{corollary:nodes:roadmap:fault-recovery}
For every
$t \geq t^{r}$,
it holds that
(1)
the node configuration $C_{t}$ is content;
and
(2)
if
$v \in \DecidedSet(C_{t})$,
then
$C_{t + 1}(v) = C_{t}(v)$.
\end{corollary}

Owing to \Cor{}~\ref{corollary:nodes:roadmap:fault-recovery}, our remaining
task is to prove that starting from time $t^{r}$, it does not take
$\SelfStabAlg$ too long until it reaches a complete (content) configuration.
To this end, we recall that by the assumption of
\Thm{}~\ref{theorem:nodes:main}, problem $\Problem$ meets the
$(\beta, \sigma_{0}, \rho_{0})$-progress
criterion (Criterion~\ref{criterion:nodes:progress}) and
recruit the node separable potential function $\pi$ promised in this
criterion.
Since the number of nodes in $G$ at distance smaller than
$\psi + \phi + 2$
from $K$ is up-bounded by
$O (|K| \cdot (\Delta - 1)^{\psi + \phi + 1})
\leq
O (k \cdot \Delta^{\psi + \phi + 1})$,
we can employ \Cor{}~\ref{corollary:nodes:roadmap:affected-regions} to
conclude that
$|\UnDecidedSet(C_{t^{r}})|
\leq
O (k \cdot \Delta^{\psi + \phi + 1})$.
\Cor{}~\ref{corollary:nodes:roadmap:bound-initial-potential} follows from the
definition of a node separable potential function as the top potential
$\sigma$-coefficient and top potential $\rho$-coefficient of $\pi$ are
$\sigma_{0}$ and $\rho_{0}$, respectively.

\begin{corollary} \label{corollary:nodes:roadmap:bound-initial-potential}
The potential in $C_{t^{r}}$ satisfies
$\pi(G, C_{t^{r}})
\leq
O \left(
(\sigma_{0} + \Delta \cdot \rho_{0}) \cdot k \cdot \Delta^{\psi + \phi + 1}
\right)$.
\end{corollary}

To complete the proof, we show that from time $t^{r}$ onwards, the potential
decreases fast.
Recalling the definition of a node separable potential function, we deduce
from \Cor{}~\ref{corollary:nodes:roadmap:fault-recovery} that
$\pi(G, C_{t})$
is a non-increasing function of
$t \geq t^{r}$.
Moreover, if
$\pi(G, C_{t}) = 0$
for some
$t \geq t^{r}$,
then $C_{t'}$ is a legal configuration for every
$t' \geq t$.
Therefore, the task of establishing the run-time bound reduces to that of
up-bounding the first time
$t \geq t^{r}$
at which
$\pi(G, C_{t}) = 0$.
To this end, we take
$\tau = O (\phi^{3})$
to be the parameter promised in \Lem{}~\ref{lemma:pps:start-phase-probability}
and establish the following proposition in
\Sect{}~\ref{section:nodes:analysis:progress}.

\begin{proposition} \label{proposition:nodes:roadmap:progress}
Fix some
$t \geq t^{r}$
and the global state of $\SelfStabAlg$ at time $t$.
Then,
\[
\Ex(\pi(G, C_{t + \tau + \phi}))
\, \leq \,
\left( 1 - \frac{\beta}{4 \phi^{2}} \right)
\cdot
\pi(G, C_{t}) \, .
\]
\end{proposition}

To complete the run-time analysis, we define the random variables
$P_{i} = \pi(G, C_{t^{r} + i (\tau + \phi)})$
for
$i \in \Integers_{\geq 0}$.
Taking
$\chi = 1 - \beta / (4 \phi^{2})$,
\Prop{}~\ref{proposition:nodes:roadmap:progress} ensures that
$\Ex(P_{i} \mid P_{i - 1}) \leq \chi \cdot P_{i - 1}$,
hence
\[
\Ex(P_{i})
\, = \,
\Ex(\Ex(P_{i} \mid P_{i - 1}))
\, \leq \,
\chi \cdot \Ex(P_{i - 1})
\, \leq \,
\chi^{i} \cdot \Ex(P_{0})
\, \leq \,
\chi^{i} \cdot
O ((\sigma_{0} + \Delta \cdot \rho_{0}) \cdot k \cdot \Delta^{\psi + \phi + 1})
\, ,
\]
where
the third transition is by induction on $i$ and the last transition is due to
\Cor{}~\ref{corollary:nodes:roadmap:bound-initial-potential}.
Since
$\chi = 1 - \Omega (\beta / \phi^{2})$
and since
$P_{i} < 1
\Longrightarrow
P_{i} = 0$,
we deduce by standard arguments that
\[
\Ex \left( \min \{ i \geq 0 \mid P_{i} = 0 \} \right)
\, \leq \,
O \left(
\left( \phi^{2} / \beta \right)
\cdot
\left(
\log (\sigma_{0} + \rho_{0}) + \log (k) + (\psi + \phi) \log (\Delta)
\right)
\right) \, .
\]
\Thm{}~\ref{theorem:nodes:main} follows by recalling that
$\tau + \phi = O (\phi^{3})$
and that
$t^{r} = t^{*} + O (\psi + \phi)$.

The remainder of this section is dedicated to proving \Prop{}
\ref{proposition:nodes:roadmap:depth-bounds-spatial-distance},
\ref{proposition:nodes:roadmap:depth-bounds-temporal-distance}, and
\ref{proposition:nodes:roadmap:progress};
this is done in \Sect{} \ref{section:nodes:analysis:affected-regions},
\ref{section:nodes:analysis:fault-recovery}, and
\ref{section:nodes:analysis:progress}, respectively.
The proofs of the three propositions rely on an important ``service lemma''
that is stated and established in
\Sect{}~\ref{section:nodes:analysis:service-lemma}.

\subsubsection{Service Lemma}
\label{section:nodes:analysis:service-lemma}
In this section, we establish an important technical lemma that plays a
crucial role in the remainder of \Sect{}~\ref{section:nodes:analysis}.
The lemma requires the following notation and terminology.
Fix some time
$t \geq t^{\circ}$.
A node
$v \in V(G_{t})$
is said to be \emph{clean} at time $t$ if $v$ is not manipulated
during the round interval
$[t - \phi, t)$.
Let
$U^{c}_{t} \subseteq V(G_{t})$
be the set of nodes that are clean at time $t$, observing that
$U^{c}_{t} \subseteq V(G_{t'})$
for every
$t - \phi \leq t' \leq t$.
A node
$v \in V(G_{t})$
is said to be \emph{deeply-clean} at time $t$ if
$\{ w \in V(G_{t}) \mid \Distance_{G_{t}}(v, w) \leq \phi \}
\subseteq
U^{c}_{t}$,
that is, if all nodes in the ball of radius $\phi$
around $v$ are clean.
Let
$U^{dc}_{t} \subseteq V(G_{t})$
be the set of nodes that are deeply-clean at time $t$.
Finally, for a step
$0 \leq j \leq \phi - 1$
of the phase procedure $\PhaseProcedure_{\Alg}$,
let
\[
S_{t}^{j}
\, = \,
\left\{ v \in V(G_{t}) \mid \Step_{v, t} = j \right\}
\, .
\]

\begin{lemma} \label{lemma:nodes:service:complete-phase-simulation}
Consider some time
$t \geq t^{\circ}$
and define
\[
Q
\, = \,
\UnDecidedSet(C_{t}) \cap U^{c}_{t} \cap S_{t}^{\phi - 1}
\quad \text{and} \quad
R
\, = \,
\left\{
\OutReg_{v} \mid v \in U^{dc}_{t} \cap Q
\right\}
\, .
\]
Let $\eta$ be the (fault free) execution of $\PhaseProcedure_{\Alg}$ when
invoked on the graph
$G_{t}(Q \cup (\DecidedSet(C_{t}) \cap U^{c}_{t}))$
with the restriction of $C_{t}$ to
$Q \cup (\DecidedSet(C_{t}) \cap U^{c}_{t})$
as the initial configuration.
Then, the assignment of output values to the output registers in $R$ performed
by $\SelfStabAlg$ in round $t$ obeys the same probability distribution as the
assignment of output values to the output registers in $R$ performed by $\eta$
in its decision step.
\end{lemma}
\begin{proof}
For a node
$v \in U^{c}_{t}$
and a step
$0 \leq j \leq \phi - 1$,
\Obs{}~\ref{observation:transformer:step-counter} ensures that
$v \in S_{t}^{\phi - 1}$
if and only if
$v \in S_{t - \phi + 1 + j}^{j}$.
This means that the nodes
$v \in U^{c}_{t}$
that simulate step
$\phi - 1$
of a phase $\Phi$ of $\PhaseProcedure_{\Alg}$ in round $t$ are exactly the
nodes
$v \in U^{c}_{t}$
that simulate step
$0 \leq j \leq \phi - 1$
of $\Phi$ in round
$t - \phi + 1 + j$
(see Pseudocode~\ref{pseudocode:transformer-nodes}).
Moreover, since $\PhaseProcedure_{\Alg}$ resets all its non-output registers
at the beginning of $\Phi$, it follows by the definition of the deeply-clean
nodes that if
$v \in U^{dc}_{t}$,
then throughout the simulation of $\Phi$, node $v$ does not receive any
adversarially manipulated information.
The assertion follows by recalling that $\Alg$ is a decision-oblivious
phase-based algorithm (see Criterion~\ref{criterion:nodes:phase-based}).
\end{proof}

\begin{remark*}
Notice that the conditions of
\Lem{}~\ref{lemma:nodes:service:complete-phase-simulation} are carefully
tailored to ensure that the assignment of output values to the output
registers in $R$ performed by $\SelfStabAlg$ in round $t$ indeed corresponds
to an actual (fault free) invocation of $\PhaseProcedure_{\Alg}$.
If these conditions are not satisfied, then the aforementioned assignment may
be the result of the simulation of a $\PhaseProcedure_{\Alg}$ phase that is
corrupted by the adversarial manipulations and hence, does not necessarily
correspond to any actual (fault free) invocation of $\PhaseProcedure_{\Alg}$.
\end{remark*}

\subsubsection{Affected Regions}
\label{section:nodes:analysis:affected-regions}
Consider a node
$v \in V(G_{t^{\circ}})$,
where
$C_{t^{\circ}}(v) = o$,
and a time
$t \geq t^{\circ}$
and assume that
$v \in V(G_{t})$
with
$\Distance_{G_{t}}(v, K_{t})
\geq
\DepthSupGraph_{\Predicate}(o) + \phi + 2$.
Our goal in this section is to prove
\Prop{}~\ref{proposition:nodes:roadmap:depth-bounds-spatial-distance}, stating
that
(1)
$C_{t}(v) = o$;
and
(2)
$v \in \ContentSet(C_{t})$.

Claims (1) and (2) are proved by induction on $t$.
The base case of
$t = t^{\circ}$
holds by the definition of $t^{\circ}$, ensuring that $C_{t^{\circ}}$ is a
content complete configuration.
So assume that claims (1) and (2) hold for time
$t \geq t^{\circ}$
and consider time
$t + 1$.
Employing \Obs{}~\ref{observation:edges:roadmap:decreasing-distance-to-faults},
the assumption that
$v \in V(G_{t + 1})$
with
$\Distance_{G_{t + 1}}(v, K_{t + 1})
\geq
\DepthSupGraph_{\Predicate}(o) + \phi + 2$
yields the following observation.

\begin{observation}
\label{observation:nodes:affected-regions:distances-from-faults}
For every time
$t^{\circ} \leq t' \leq t + 1$,
it holds that
$v \in V(G_{t'})$
and
$N_{G_{t'}(v)} = N_{G_{t^{\circ}}}(v)$.
Moreover,
$\Distance_{G_{t'}}(v, K_{t'})
\geq
\DepthSupGraph_{\Predicate}(o) + \phi + 2$
and
$\Distance_{G_{t'}}(u, K_{t'})
\geq
\DepthSupGraph_{\Predicate}(o) + \phi + 1$
for every node
$u \in N_{G_{t^{\circ}}}(v)$.
\end{observation}

Owing to
\Obs{}~\ref{observation:nodes:affected-regions:distances-from-faults}, we
denote
$A = N_{G_{t^{\circ}}}(v)$
and notice that
$A = N_{G_{t'}}(v)$
for all
$t^{\circ} \leq t' \leq t + 1$.
\Obs{}~\ref{observation:nodes:affected-regions:distances-from-faults} also
allows us to use the inductive hypothesis to conclude that
(1)
$C_{t}(v) = o$;
and
(2)
$v \in \ContentSet(C_{t})$.
The guarantees of $\Detect$ imply that the output register $\OutReg_{v}$ is not
reset in round $t$ (see line~\ref{line:nodes:detect} in
Pseudocode~\ref{pseudocode:transformer-nodes}), thus
$C_{t + 1}(v) = C_{t}(v) = o$
establishing claim (1) of the inductive step.

En route to establishing claim (2) of the inductive step,
let
$B = \{ w \in V(G_{t})
\mid
\Distance_{G_{t}}(w, v) \leq \phi + 1 \}$
be the ball of radius
$\phi + 1$
around $v$ in $G_{t}$.
\Obs{}~\ref{observation:nodes:affected-regions:distances-from-faults} ensures
that the nodes in $B$ are clean at time $t$ and that the nodes in
$\{ v \} \cup A$
are deeply-clean at time $t$ (refer to
\Sect{}~\ref{section:nodes:analysis:service-lemma} for the definitions of
clean and deeply-clean nodes).
This allows us to apply
\Lem{}~\ref{lemma:nodes:service:complete-phase-simulation} and obtain the
following corollary.

\begin{corollary}
\label{corollary:nodes:affected-regions:complete-phase-simulation}
Define
\[
Q
\, = \,
\UnDecidedSet(C_{t}) \cap B \cap S_{t}^{\phi - 1}
\quad \text{and} \quad
R
\, = \,
\left\{
\OutReg_{u} \mid u \in (\{ v \} \cup A) \cap Q
\right\}
\, .
\]
Let $\eta$ be the (fault free) execution of $\PhaseProcedure_{\Alg}$ when
invoked on the graph
$G_{t}(Q \cup (\DecidedSet(C_{t}) \cap B))$
with the restriction of $C_{t}$ to
$Q \cup (\DecidedSet(C_{t}) \cap B)$
as the initial configuration.
Then, the assignment of output values to the output registers in $R$ performed
by $\SelfStabAlg$ in round $t$ obeys the same probability distribution as the
assignment of output values to the output registers in $R$ performed by $\eta$
in its decision step.
\end{corollary}

Next, we partition the set $A$ of nodes adjacent to $v$ into the subsets
\[
A_{<}
\, = \,
\left\{
u \in A
\mid
\DepthSupGraph_{\Predicate}(C_{t^{\circ}}(u))
<
\DepthSupGraph_{\Predicate}(o)
\right\}
\quad \text{and} \quad
A_{\geq}
\, = \,
\left\{
u \in A
\mid
\DepthSupGraph_{\Predicate}(C_{t^{\circ}}(u))
\geq
\DepthSupGraph_{\Predicate}(o)
\right\}
\]
and argue about each subset separately, staring with the former.

\begin{lemma}
\label{lemma:nodes:affected-regions:adjacent-nodes-with-smaller-depth}
For each node
$u \in A_{<}$,
it holds that
$C_{t + 1}(u) = C_{t}(u) = C_{t^{\circ}}(u)$.
\end{lemma}
\begin{proof}
\Obs{}~\ref{observation:nodes:affected-regions:distances-from-faults} ensures
that
$\Distance_{G_{t}}(u, K_{t})
\geq
\DepthSupGraph_{\Predicate}(o) + \phi + 1$.
By the definition of $A_{<}$, we conclude that
$\Distance_{G_{t}}(u, K_{t})
\geq
\DepthSupGraph_{\Predicate}(C_{t^{\circ}}(u)) + \phi + 2$.
This allows us to apply the inductive hypothesis and deduce that
(1)
$C_{t}(u) = C_{t^{\circ}}(u)$;
and
(2)
$u \in \ContentSet(C_{t})$.
The guarantees of $\Detect$ imply that the output register $\OutReg_{u}$ is
not reset in round $t$ (see line~\ref{line:nodes:detect} in
Pseudocode~\ref{pseudocode:transformer-nodes}), yielding the assertion.
\end{proof}

Let $M_{<}$ be the multiset over
$\mathcal{O}$ defined as
\[
M_{<}
=
\{ C_{t^{\circ}}(u) \}_{u \in A_{<}}
\, .
\]
As $C_{t^{\circ}}$ is a content complete configuration,
we know that
$C_{t^{\circ}}[v] \in \SatMultisets(o)$,
which means that there exists a core
$M^{*} \in \Cores(o) \subseteq \SatMultisets(o)$
such that
$M^{*} \subseteq C_{t^{\circ}}[v]$
(recall the definition of cores from
\Sect{}~\ref{section:LCL-eligibility-criteria}).
By the definition of the supportive digraph $D_{\Predicate}$ (see
\Sect{}~\ref{section:LCL-eligibility-criteria}), we conclude that
$\DepthSupGraph_{\Predicate}(o') < \DepthSupGraph_{\Predicate}(o)$
for every
$o' \in M^{*}$,
hence
$M^{*} \subseteq M_{<}$.

Next, we partition the node set $A_{\geq}$ into the subsets
\[
A_{\geq}^{\DecidedSet}
\, = \,
A_{\geq} \cap \DecidedSet(C_{t})
\quad \text{and} \quad
A_{\geq}^{\UnDecidedSet}
\, = \,
A_{\geq} \cap \UnDecidedSet(C_{t})
\, ,
\]
observing that
$A_{\geq}^{\UnDecidedSet} = A \cap \UnDecidedSet(C_{t})$
due to
\Lem{}~\ref{lemma:nodes:affected-regions:adjacent-nodes-with-smaller-depth}.
Let
$M_{\geq}^{\DecidedSet, t}$,
$M_{\geq}^{\DecidedSet, t + 1}$,
and
$M_{\geq}^{\UnDecidedSet, t + 1}$
be the multisets over $\mathcal{O}$ defined as
\[
M_{\geq}^{\DecidedSet, t}
=
\{ C_{t}(u) \}_{u \in A_{\geq}^{\DecidedSet}}
\, , \;
M_{\geq}^{\DecidedSet, t + 1}
=
\{
C_{t + 1}(u)
\}_{u \in A_{\geq}^{\DecidedSet} \cap \DecidedSet(C_{t + 1})}
\, , \; \text{and} \;
M_{\geq}^{\UnDecidedSet, t + 1}
=
\{
C_{t + 1}(u)
\}_{u \in A_{\geq}^{\UnDecidedSet} \cap \DecidedSet(C_{t + 1})}
\, ,
\]
observing that
$M_{\geq}^{\DecidedSet, t + 1} \subseteq M_{\geq}^{\DecidedSet, t}$.
Since problem $\Problem$ meets the respectful decisions criterion
(Criterion~\ref{criterion:nodes:respectful-decisions}) and since the nodes in
$A_{\geq}^{\UnDecidedSet} \cap \DecidedSet(C_{t + 1})$
(i.e., the nodes that realize
$M_{\geq}^{\UnDecidedSet, t + 1}$)
are exactly the nodes adjacent to $v$ that obtain an output value in round
$t$, we can apply
\Cor{}~\ref{corollary:nodes:affected-regions:complete-phase-simulation} to
deduce that
$M_{<} + M_{\geq}^{\DecidedSet, t} + M_{\geq}^{\UnDecidedSet, t + 1}
\in
\SatMultisets(o)$.
As $\Problem$ meets the gap freeness criterion
(Criterion~\ref{criterion:common:gap-freeness}), we conclude that
$M_{<} + M_{\geq}^{\DecidedSet, t + 1} + M_{\geq}^{\UnDecidedSet, t + 1}
\in
\SatMultisets(o)$
by observing that
\[
M^{*}
\, \subseteq \,
M_{<}
\, \subseteq \,
M_{<} + M_{\geq}^{\DecidedSet, t + 1} + M_{\geq}^{\UnDecidedSet, t + 1}
\, \subseteq \,
M_{<} + M_{\geq}^{\DecidedSet, t} + M_{\geq}^{\UnDecidedSet, t + 1}
\, .
\]
This establishes claim (2) of the inductive step as
$M_{<} + M_{\geq}^{\DecidedSet, t + 1} + M_{\geq}^{\UnDecidedSet, t + 1}
=
C_{t + 1}[e]$.

\subsubsection{Fault Recovery}
\label{section:nodes:analysis:fault-recovery}
In this section, we establish
\Prop{}~\ref{proposition:nodes:roadmap:depth-bounds-temporal-distance}.
To this end, let
$\hat{C}_{t} : V(G) \rightarrow \mathcal{O} \cup \{ \bot \}$,
$t \geq t^{*}$,
be the node configuration defined by setting
\[
\hat{C}_{t}(v)
\, = \,
\begin{cases}
C_{t + 1}(v) \, , & v \in \UnDecidedSet(C_{t}) \cap \DecidedSet(C_{t + 1}) \\
C_{t}(v) \, , & \text{otherwise}
\end{cases}
\, ,
\]
where we observe that
$\UnDecidedSet(C_{t}) \cap \DecidedSet(C_{t + 1})$
is the subset of nodes that become decided (as a result of the simulation of
$\PhaseProcedure_{\Alg}$'s decision step
$j = \phi - 1$
in line~\ref{line:nodes:simulate-phase-procedure} of
Pseudocode~\ref{pseudocode:transformer-nodes}) in round $t$.
The definition of $\hat{C}_{t}$ allows us to state the following basic
observation.

\begin{observation} \label{observation:nodes:fault-recovery:early-relations}
For every
$t \geq t^{*} + 1$,
it holds that
\\
(1)
$\DecidedSet(\hat{C}_{t}) = \DecidedSet(C_{t}) \cup \DecidedSet(C_{t + 1})$;
\\
(2)
$\hat{C}_{t}(v) = C_{t + 1}(v) = \hat{C}_{t + 1}(v)$
for every node
$v \in \DecidedSet(C_{t + 1})$;
and
\\
(3)
$C_{t}(v) = C_{t + 1}(v)$
for every node
$v \in \DecidedSet(C_{t}) \cap \DecidedSet(C_{t + 1})$.
\end{observation}

Next, recall the definition of deeply-clean nodes from
\Sect{}~\ref{section:nodes:analysis:service-lemma} and notice that all
nodes in $V(G)$ are deeply clean at any time
$t \geq t^{*} + \phi$.
Therefore, we can apply
\Lem{}~\ref{lemma:nodes:service:complete-phase-simulation} to obtain the
following corollary.

\begin{corollary}
\label{corollary:nodes:fault-recovery:complete-phase-simulation}
Consider some time
$t \geq t^{*} + \phi$
and define
\[
Q
\, = \,
\UnDecidedSet(C_{t}) \cap S_{t}^{\phi - 1}
\quad \text{and} \quad
R
\, = \,
\left\{
\OutReg_{v} \mid v \in Q
\right\}
\, .
\]
Let $\eta$ be the (fault free) execution of $\PhaseProcedure_{\Alg}$ when
invoked on the graph
$G(Q \cup \DecidedSet(C_{t}))$
with the restriction of $C_{t}$ to
$Q \cup \DecidedSet(C_{t})$
as the initial configuration.
Then, the assignment of output values into the output registers in $R$
performed by $\SelfStabAlg$ in round $t$ obeys the same probability
distribution as the assignment of output values into the output registers in
$R$ performed by $\eta$ in its decision step.
\end{corollary}

We are now ready to establish the following lemma.

\begin{lemma} \label{lemma:nodes:fault-recovery:late-relations}
For every
$t \geq t^{*} + \phi + 1$,
it holds that
\\
(1)
$\UnContentSet(C_{t})
=
\DecidedSet(C_{t}) \cap \UnDecidedSet(C_{t + 1})
=
\DecidedSet(\hat{C}_{t}) \cap \UnDecidedSet(C_{t + 1})$;
\\
(2)
$\ContentSet(C_{t}) \subseteq \ContentSet(\hat{C}_{t})$;
and
\\
(3)
$\UnDecidedSet(C_{t}) \cap \DecidedSet(\hat{C}_{t}) \subseteq
\ContentSet(\hat{C}_{t})$.
\end{lemma}
\begin{proof}
Claims (2) and (3) are a consequence of
\Cor{}~\ref{corollary:nodes:fault-recovery:complete-phase-simulation} due to
Criterion~\ref{criterion:nodes:respectful-decisions}.
To see that claim (1) holds, notice that the guarantees of $\Detect$ imply
that
$\Detect_{v, t} = \false$
if and only if
$v \in \UnContentSet(C_{t})$
for every node
$v \in \DecidedSet(C_{t})$.
The assertion follows by the design of $\SelfStabAlg$.
\end{proof}

Owing to \Obs{}~\ref{observation:nodes:fault-recovery:early-relations}(3), the
task of establishing
\Prop{}~\ref{proposition:nodes:roadmap:depth-bounds-temporal-distance} reduces
to that of proving the following lemma.

\begin{lemma} \label{lemma:nodes:fault-recovery:directed-path}
For every
$t \geq t^{*} + \phi + 1$,
if
$v \in \UnContentSet(C_{t})$,
then the supportive digraph
$D_{\Predicate}$
associated with the LCL predicate $\Predicate$ admits a directed path $P$ of
length
$t - (t^{*} + \phi + 1)$
emerging from $C_{t}(v)$.
\end{lemma}
\begin{proof}
We establish the assertion by induction on $t$.
The base case of
$t = t^{*} + \phi + 1$
is trivial as the digraph $D_{\Predicate}$ admits a path of length $0$
emerging from $o$ for every
$o \in \mathcal{O}$
including
$o = C_{t}(e)$.

Assume that the assertion holds for
$t - 1 \geq t^{*} + \phi + 1$
and consider a node
$v \in \UnContentSet(C_{t})$.
We start by proving that
$v \in \ContentSet(\hat{C}_{t - 1})$.
As
$v \in \UnContentSet(C_{t}) \subseteq \DecidedSet(C_{t})$,
\Lem{}~\ref{lemma:nodes:fault-recovery:late-relations}(1) ensures that
$v \notin \UnContentSet(C_{t - 1})$.
If
$v \in \UnDecidedSet(C_{t - 1})$,
then since
$v \in \DecidedSet(C_{t})$,
we know that
$v \in \UnDecidedSet(C_{t - 1}) \cap \DecidedSet(\hat{C}_{t - 1})$
by \Obs{}~\ref{observation:nodes:fault-recovery:early-relations}(1).
Thus,
$v \in \ContentSet(\hat{C}_{t - 1})$
by \Lem{}~\ref{lemma:nodes:fault-recovery:late-relations}(3).
Otherwise, it holds that
$v \in \ContentSet(C_{t - 1})$
and we can apply \Lem{}~\ref{lemma:nodes:fault-recovery:late-relations}(2) to
conclude that
$v \in \ContentSet(\hat{C}_{t - 1})$.

Let
$o = \hat{C}_{t - 1}(v) \in \mathcal{O}$
be the output value of $v$ under $\hat{C}_{t - 1}$.
By \Obs{}~\ref{observation:nodes:fault-recovery:early-relations}(2), we know
that
$C_{t}(v) = o$
as well.
Since
$v \in \ContentSet(\hat{C}_{t - 1})$,
it follows that
$\hat{C}_{t - 1}[v] \in \SatMultisets(o)$.
Let
$\hat{M} \in \Cores(o)$
be a core for $o$ such that
$\hat{M} \subseteq \hat{C}_{t - 1}[v]$
(recall that cores are defined in
\Sect{}~\ref{section:LCL-eligibility-criteria}).
We argue that
$\hat{M} \neq \emptyset$.
Indeed, \Obs{}~\ref{observation:nodes:fault-recovery:early-relations}(1, 2)
implies that
$C_{t}[v] \subseteq \hat{C}_{t - 1}[v]$,
hence if
$\hat{M} = \emptyset$,
then
$C_{t}[v] \in \SatMultisets(o)$
due to Criterion~\ref{criterion:common:gap-freeness},
which contradicts the assumption that
$v \in \UnContentSet(C_{t})$.

Let
$\hat{A} \subseteq N_{G}(v)$
be a subset of $v$'s neighbors whose decisions under
$\hat{C}_{t - 1}$
realize $\hat{M}$.
For each
$v' \in \hat{A}$,
either
(1)
$v' \in \DecidedSet(C_{t})$
and
$C_{t}(v') = \hat{C}_{t - 1}(v')$
by \Obs{}~\ref{observation:nodes:fault-recovery:early-relations}(2);
or
(2)
$v' \in \UnDecidedSet(C_{t})$.
Recalling that
$C_{t}[v] \subseteq \hat{C}_{t - 1}[v]$,
if
$C_{t}(v') = \hat{C}_{t - 1}(v')$
for every
$v' \in \hat{A}$,
then
$\hat{M} \subseteq C_{t}[v] \subseteq \hat{C}_{t - 1}[v]$
implying that
$C_{t}[v] \in \SatMultisets(o)$
due to Criterion~\ref{criterion:common:gap-freeness}.
But this means that
$v \in \ContentSet(C_{t})$,
in contradiction to the assumption that
$v \in \UnContentSet(C_{t})$.

Therefore, there must exist a node
$v' \in \hat{A}$
such that
$v' \in \UnDecidedSet(C_{t})$.
As
$v' \in \DecidedSet(\hat{C}_{t - 1})$,
\Lem{}~\ref{lemma:nodes:fault-recovery:late-relations}(1) ensures that
$v' \in \UnContentSet(C_{t - 1})$.
By the inductive hypothesis, the supportive digraph $D_{\Predicate}$ admits a
directed path $P'$
of length
$t - 1 - (t^{*} + \phi + 1)$
emerging from
$o' = C_{t - 1}(v') \in \mathcal{O}$.
Recalling that
$o' \in \hat{M} \in \Cores(o)$,
we conclude that
$(o, o') \in E(D_{\Predicate})$.
The assertion follows as
$(o, o') \circ P'$
is a directed path of length
$t - (t^{*} + \phi + 1)$
emerging from $o$ in $D_{\Predicate}$.
\end{proof}

\subsubsection{Progress}
\label{section:nodes:analysis:progress}
This section is dedicated to proving
\Prop{}~\ref{proposition:nodes:roadmap:progress}, showing that
$\Ex(\pi(G, C_{t + \tau + \phi}))
\leq
\left( \frac{1 - \beta}{4 \phi^{2}} \right)
\cdot
\pi(G, C_{t})$
for every time
$t \geq t^{r}$,
where $\tau$ is the parameter promised in
\Lem{}~\ref{lemma:pps:start-phase-probability} and
$\pi : \ContConfGraphs(\Problem) \rightarrow \Integers_{\geq 0}$
is the node separable potential function promised in
Criterion~\ref{criterion:nodes:progress}.
We do so by analyzing the following $4$-stage artificial process:
\begin{enumerate}

\item[(I)]
The adversary determines the global state of $\SelfStabAlg$ at time $t$.
In particular, the adversary determines the configuration $C_{t}$ and the
value
$\Step_{v, t}$
of the $\PPS$ state at time $t$ for each node
$v \in V(G)$.

\item[(II)]
Nature determines the random node subset
$S = S_{t + \tau}^{0}$
by selecting each node
$v \in V(G)$
to be included in $S$, independently, with probability $p_{v}$, where
$p_{v} = \Pr(\Step_{v, t + \tau} = 0 \mid \Step_{v, t})$. 

\item[(III)]
The node subset $S$ is revealed to the adversary who then determines the coin
tosses of all nodes for the round interval
$[t, t + \tau)$
subject to 
$S_{t + \tau}^{0} = S$;
and the coin tosses of the nodes in
$V(G) - S$
for the round interval
$[t + \tau, t + \tau + \phi)$.
In particular, the adversary determines the configuration
$C_{t + \tau}$
and the value of
$C_{t + \tau + \phi}(v)$
for each node
$v \in \UnDecidedSet(C_{t + \tau}) \cap (\bigcup_{u \in S} N_{G}(u) - S)$.

\item[(IV)]
Nature tosses the coins of the nodes in $S$ for the round interval
$[t + \tau, t + \tau + \phi)$.

\end{enumerate}
If \Prop{}~\ref{proposition:nodes:roadmap:progress} holds for this artificial
process, then it also holds for the actual execution of $\SelfStabAlg$ during
the round interval
$[t, t + \tau + \phi)$
as in the latter, the role of the adversary is assumed by (the less malicious)
nature.

Let
$\mathcal{I}^{\sigma}
=
\left\{ \sigma(M) \right\}_{M \in \Multisets(\mathcal{O})}
\subset
\Integers_{> 0}$
and
$\mathcal{I}^{\rho}
=
\left\{ \rho(M) \right\}_{M \in \Multisets(\mathcal{O})}
\subset
\Integers_{\geq 0}$
be $\pi$'s families of potential $\sigma$-coefficients and potential
$\rho$-coefficient, respectively.
Fix some time
$t \leq t' \leq t + \tau + \phi$.
For a node
$u \in V(G)$
and an edge
$e = \{ v, w\} \in E(G)$,
let
\[
\lambda^{\sigma}_{t'}(u)
\, = \,
\begin{cases}
\sigma(C_{t'}[u]) \, , & u \in \UnDecidedSet(C_{t'}) \\
0 \, , & \text{otherwise}
\end{cases}
\quad \text{and} \quad
\lambda^{\rho}_{t'}(e)
\, = \,
\begin{cases}
\rho(C_{t'}[v]) + \rho(C_{t'}[w]) \, , & v, w \in \UnDecidedSet(C_{t'}) \\
0 \, , & \text{otherwise}
\end{cases}
\, ,
\]
observing that as $\mathcal{I}^{\sigma}$ and $\mathcal{I}^{\rho}$ are
monotonically non-increasing,
\Cor{}~\ref{corollary:nodes:roadmap:fault-recovery} implies that both
$\lambda^{\sigma}_{t'}(u)$ and $\lambda^{\rho}_{t'}(e)$ are monotonically
non-increasing functions of $t'$.
This notation is extended to node subsets
$U \subseteq V(G)$
and edge subsets
$F \subseteq E(G)$
by defining
$\lambda^{\sigma}_{t'}(U)
=
\sum_{v \in U} \lambda^{\sigma}_{t'}(v)$
and
$\lambda^{\rho}_{t'}(F)
=
\sum_{e \in F} \lambda^{\rho}_{t'}(e)$,
which allows us to express the potential in $G$ with respect to $C_{t}$ as
\begin{equation} \label{equation:nodes:progress:express-potential-with-lambda}
\pi(G, C_{t'})
\, = \,
\lambda^{\sigma}_{t'}(V(G)) + \lambda^{\rho}_{t'}(E(G))
\, .
\end{equation}
Moreover,
$\lambda^{\sigma}_{t'}(U_{1} \cup U_{2})
=
\lambda^{\sigma}_{t'}(U_{1}) + \lambda^{\sigma}_{t'}(U_{2})$
for every disjoint node subsets
$U_{1}, U_{2} \subseteq V(G)$
and
$\lambda^{\rho}_{t'}(F_{1} \cup F_{2})
=
\lambda^{\rho}_{t'}(F_{1}) + \lambda^{\rho}_{t'}(F_{2})$
for every disjoint edge subsets
$F_{1}, F_{2} \subseteq E(G)$.

Let
$F^{S} = \{ \{ u, v \} \in E(G) \mid u, v \in S \}$.
Since \Lem~\ref{lemma:pps:start-phase-probability} ensures that
$p_{v} \geq 1 / (2 \phi)$
for each node
$v \in V$,
it follows that
$\Pr(v \in S) \geq 1 / (2 \phi)$
for each node
$v \in V(G)$
and
$\Pr(e \in F^{S}) \geq 1 / (2 \phi)^{2} = 1 / (4 \phi^{2})$
for each edge
$e \in E(G)$,
thus
\begin{equation} \label{equation:nodes:progress:large-share}
\Ex \left( \lambda^{\sigma}_{t}(S) + \lambda^{\rho}_{t}(F^{S}) \right)
\, \geq \,
\frac{1}{2 \phi} \cdot \lambda^{\sigma}_{t}(V(G))
+
\frac{1}{4 \phi^{2}} \cdot \lambda^{\rho}_{t}(E(G))
\, \geq \,
\frac{1}{4 \phi^{2}} \cdot \left(
\lambda^{\sigma}_{t}(V(G)) + \lambda^{\rho}_{t}(E(G))
\right)
\, ,
\end{equation}
where the expectation is over the coin tosses of stage (II).

For a time
$t \leq t' \leq t + \tau + \phi$,
let
$H_{t'} = G(S \cup \DecidedSet(C_{t'}))$
and let
$C^{H}_{t'} : V(H_{t'}) \rightarrow \mathcal{O} \cup \{ \bot \}$
be the restriction of $C_{t'}$ to the nodes in $V(H_{t'})$.
By definition, we know that
$\UnDecidedSet(C^{H}_{t'}) \subseteq S$
and if
$v \in S$,
then
$C^{H}_{t'}[v] = C_{t'}[v]$,
thus
\[
\pi(H_{t'}, C^{H}_{t'})
\, = \,
\lambda^{\sigma}_{t'}(S) + \lambda^{\rho}_{t'}(F^{S})
\, .
\]
Since only the nodes
$v \in S$
may write output values into their output registers $\OutReg_{v}$ in round
$t + \tau + \phi - 1$,
it follows that
$\DecidedSet(C_{t + \tau + \phi - 1}) - S
=
\DecidedSet(C_{t + \tau + \phi}) - S$,
hence
$H_{t + \tau + \phi - 1} = H_{t + \tau + \phi}$.
As problem $\Problem$ meets the
$(\beta, \sigma_{0}, \rho_{0})$-progress
criterion (Criterion~\ref{criterion:nodes:progress}), we can employ
\Cor{}~\ref{corollary:nodes:fault-recovery:complete-phase-simulation} to
conclude that
\[
\Ex \left( \pi(H_{t + \tau + \phi}, C^{H}_{t + \tau + \phi}) \right)
\, \leq \,
(1 - \beta) \cdot \pi(H_{t + \tau + \phi - 1}, C^{H}_{t + \tau + \phi - 1})
\, ,
\]
and therefore,
\begin{equation} \label{equation:nodes:progress:phase-progress}
\Ex \left(
\lambda^{\sigma}_{t + \tau + \phi}(S)
+
\lambda^{\rho}_{t + \tau + \phi}(F^{S})
\right)
\, \leq \,
(1 - \beta) \cdot \left(
\lambda^{\sigma}_{t + \tau + \phi - 1}(S)
+
\lambda^{\rho}_{t + \tau + \phi - 1}(F^{S})
\right)
\, ,
\end{equation}
where the expectation is over the coin tosses of stage (IV).

Put together, we conclude that
\begin{align*}
\Ex \left( \pi(G, C_{t + \tau + \phi}) \right)
\, = \, &
\Ex \left(
\lambda^{\sigma}_{t + \tau + \phi}(V(G))
+
\lambda^{\rho}_{t + \tau + \phi}(E(G))
\right)
\\
= \, &
\Ex \left(
\lambda^{\sigma}_{t + \tau + \phi}(S)
+
\lambda^{\rho}_{t + \tau + \phi}(F^{S})
\right)
+
\Ex \left(
\lambda^{\sigma}_{t + \tau + \phi}(V(G) - S)
+
\lambda^{\rho}_{t + \tau + \phi}(E(G) - F^{S})
\right)
\\
\leq \, &
(1 - \beta) \cdot \Ex \left(
\lambda^{\sigma}_{t + \tau + \phi - 1}(S)
+
\lambda^{\rho}_{t + \tau + \phi - 1}(F^{S})
\right)
\\
&
+
\Ex \left(
\lambda^{\sigma}_{t + \tau + \phi}(V(G) - S)
+
\lambda^{\rho}_{t + \tau + \phi}(E(G) - F^{S})
\right)
\\
\leq \, &
(1 - \beta) \cdot \Ex \left(
\lambda^{\sigma}_{t}(S)
+
\lambda^{\rho}_{t}(F^{S})
\right)
+
\Ex \left(
\lambda^{\sigma}_{t}(V(G) - S)
+
\lambda^{\rho}_{t}(E(G) - F^{S})
\right)
\\
= \, &
(1 - \beta) \cdot \Ex \left(
\lambda^{\sigma}_{t}(S) + \lambda^{\rho}_{t}(F^{S})
\right)
+
\lambda^{\sigma}_{t}(V(G)) + \lambda^{\rho}_{t}(E(G))
-
\Ex \left(
\lambda^{\sigma}_{t}(S) + \lambda^{\rho}_{t}(F^{S})
\right)
\\
= \, &
\lambda^{\sigma}_{t}(V(G))
+
\lambda^{\rho}_{t}(E(G))
-
\beta \cdot \Ex \left(
\lambda^{\sigma}_{t}(S)
+
\lambda^{\rho}_{t}(F^{S})
\right)
\\
\leq \, &
\lambda^{\sigma}_{t}(V(G))
+
\lambda^{\rho}_{t}(E(G))
-
\frac{\beta}{4 \phi^{2}} \cdot \left(
\lambda^{\sigma}_{t}(V(G))
+
\lambda^{\rho}_{t}(E(G))
\right) \\
= \, &
\left( 1 - \frac{\beta}{4 \phi^{2}} \right) \cdot \pi(G, C_{t})
\, ,
\end{align*}
where
the first, third, seventh, and last transitions follow from
(\ref{equation:nodes:progress:express-potential-with-lambda}),
(\ref{equation:nodes:progress:phase-progress}),
(\ref{equation:nodes:progress:large-share}),
and
(\ref{equation:nodes:progress:express-potential-with-lambda}),
respectively, thus establishing
\Prop{}~\ref{proposition:nodes:roadmap:progress}.

\section{Simulating a Node-LCL Algorithm on the Line Graph}
\label{section:simulation-line-graph}
In this section we explain how to simulate a self-stabilizing algorithm 
produced form our transformer for distributed node problems on the line graph.
This simulation provides another way to develop a self-stabilizing algorithm
for edge problems which is usually less efficient.

Simulating an algorithm \Alg{} for a distributed node problem on the line
graph $L(G)$ of a graph $G$ is a simple task in a fault free environment with
unique node IDs:
It suffices to appoint an endpoint
$x(e) \in e$
for each edge
$e \in E(G) = V(L(G))$
that takes on the responsibility for simulating the computation associated
with $e$ throughout the execution (e.g., the endpoint with a smaller ID);
since
$\Distance_{G}(x(e), x(e')) \leq 2$
for every
$e, e' \in E(G)$
such that
$\{ e, e' \} \in E(L(G))$,
it suffices to simulate each round of $\Alg$ on $L(G)$ with $2$ rounds in $G$,
e.g., allocating the even rounds for simulating the actions of $\Alg$ on
$L(G)$ and the odd rounds for gathering the information from nodes that are
$2$ hops away.

Things become more involved when one wishes to simulate an anonymous
self-stabilizing algorithm \SelfStabAlg{} on $L(G)$.
First, we need to ensure that faulty nodes are safely detected.
Second, we no longer have a natural symmetry breaking rule that determines the
endpoint $x(e)$ responsible for simulating the computation associated with an
edge
$e \in E(G)$.
Third, in a self-stabilizing setting, the adversary can manipulate the nodes'
clocks, thus preventing them from partitioning the rounds into even and odd in
synchrony.

To overcome these difficulties, we shall design a self-stabilizing algorithm,
denoted by $\SelfStabAlg'$, that runs on $G$ and simulates a run of
$\SelfStabAlg$ on $L(G)$.
To this end, we exploit the structure of the self-stabilizing algorithms
developed in \Sect{}~\ref{section:nodes:implementation}, recalling
that $\SelfStabAlg$ is composed of a detection procedure \Detect{}, a
phase procedure \PhaseProcedure{}, and the \PPS{} module.
We design $\SelfStabAlg'$ by adapting each component of
\SelfStabAlg{} separately, denoting the components that run on $G$ and
simulate the operation of \Detect{}, $\PhaseProcedure_{\Alg}$, and \PPS{} on 
$L(G)$
by $\Detect'$, $\PhaseProcedure_{\Alg}'$, and $\PPS'$, respectively.

To simulate a run of \Detect{} on $L(G)$ by the simulating procedure
$\Detect'$, consider an edge
$e = \{ u, v \} \in E(G)$.
The output register $\OutReg_{e}$ of $e$ is represented under $\Detect'$ by the
two output registers $\OutReg_{u}(v)$ and $\OutReg_{v}(u)$.
We augment the messages exchanged between $u$ and $v$ under $\Detect'$ with
the content of the simulated output register $\OutReg_{e}$, thus allowing the
two nodes to verify that edge $e$ is port-consistent.
Beyond that piece of information, nodes $u$ and $v$ simply share with each
other under $\Detect'$ the messages that $\Detect$ sends when simulated on
$L(G)$ (typically, these messages simply contain the content of all their
output registers).
If a port-inconsistency is detected or if $\Detect$ returns $\false$, then
node $u$ (resp., $v$) resets
$\OutReg_{u}(v) \gets \bot$
(resp.,
$\OutReg_{v}(u) \gets \bot$).

Next, we turn to explain how an invocation of the phase procedure
$\PhaseProcedure_{\Alg}$ on a subgraph of $L(G)$ is simulated by procedure
$\PhaseProcedure_{\Alg}'$ running on $G$.
Assuming that each phase of \PhaseProcedure{} consists of $\phi$ steps, a
phase of $\PhaseProcedure_{\Alg}'$ is stretch over
$2 \phi$
steps.
Given an edge
$e \in E(G)$,
module $\PPS'$ is responsible for (probabilistically) starting new phases of
$\PhaseProcedure_{\Alg}'$ and for advancing the steps once a phase starts so 
that both endpoints of $e$ agree on the current step
$j \in \{ 0, 1, \dots, 2 \phi - 1 \}$ 
of edge $e$.
Module $\PPS'$ is also responsible for breaking the symmetry between the
endpoints $x(e)$ and $y(e)$ of edge
$e = \{ x(e), y(e) \}$
so that $x(e)$ is responsible for simulating the execution of
$\PhaseProcedure_{\Alg}$ on $e$ in the current phase and $y(e)$ assists $x(e)$ 
in collecting information from nodes that are $2$ hops away from $x(e)$.
Specifically, each step
$j = 0, 1, \dots, \phi - 1$
of $\PhaseProcedure_{\Alg}$ is simulated by steps
$2 j$
and
$2 j + 1$
of $\PhaseProcedure_{\Alg}'$, where in step
$2 j$,
node $y(e)$ passes to $x(e)$ the messages sent under $\PhaseProcedure_{\Alg}$ 
from edges
$e' \in E(G)$
of the form
$e' = \{ y(e) = y(e'), x(e') \}$
to edge $e$, and in step
$2 j + 1$,
node $x(e)$ updates the registers of $\PhaseProcedure_{\Alg}$ at $e$, maintained
under $\PhaseProcedure_{\Alg}'$ at $x(e)$.

The simulation of the decision step
$j = \phi - 1$
of $\PhaseProcedure_{\Alg}$ is slightly different, recalling that this is the 
only step in which $\PhaseProcedure_{\Alg}$ is allowed to write an output value
$o \in \mathcal{O}$
into the output register
$\OutReg_{e}$ of (the undecided) edge
$e \in E(G)$,
implemented under $\PhaseProcedure_{\Alg}'$ by writing $o$ into the output 
registers
$\OutReg_{x(e)}(y(e))$ and $\OutReg_{y(e)}(x(e))$.
To ensure that the output value $o$ is written into $\OutReg_{x(e)}(y(e))$ and
$\OutReg_{y(e)}(x(e))$ concurrently, node $x(e)$ employs step
$2 \phi - 2$
of $\PhaseProcedure_{\Alg}'$ to inform $y(e)$ of $o$.
Then, in step
$2 \phi - 1$,
serving as the decision step of $\PhaseProcedure_{\Alg}'$,
nodes $x(e)$ and $y(e)$ set
$\OutReg_{x(e)}(y(e)) \gets o$
and
$\OutReg_{y(e)}(x(e)) \gets o$.

The last action is conditioned on the output values of the decided edges
$e' \in N_{G}(e)$
that are revealed to $x(e)$ and $y(e)$ at the beginning of step
$2 \phi - 1$.
Here, we exploit the trivial (yet key) observation that if
$e' \in N_{G}(e)$,
then
$|e' \cap \{ x(e), y(e) \}| = 1$,
hence if $e'$ is already decided at the end of step
$2 \phi - 2$,
then both $x(e)$ and $y(e)$ know about it at the beginning of step
$2 \phi - 1$.

Finally, we explain how module $\PPS'$ simulates module \PPS{} (see
\Sect{}~\ref{section:probabilistic-phase-synchronization}) that 
runs in an edge
$e = \{ u, v \} \in E(G)$
and determines when a new phase of $\PhaseProcedure_{\Alg}'$ begins (as well as
advancing the steps within a phase).
Module $\PPS'$ is also responsible for breaking the symmetry between $u$ and
$v$, appointing one of them to $x(e)$ and the other to $y(e)$, when a new
phase begins.

To implement module $\PPS'$, nodes $u$ and $v$ maintain the register
$\Step_{e}$ in conjunction.
Specifically, node $u$ (resp., $v$) maintains a local copy of $\Step_{e}$,
denoted by $\Step_{u}(v)$ (resp., $\Step_{v}(u)$).
In each round, nodes $u$ and $v$ exchange the values of $\Step_{u}(v)$ and
$\Step_{v}(u)$;
if
$\Step_{u}(v) \neq \Step_{v}(u)$,
then nodes $u$ and $v$ reset $\PPS'$ by setting
$\Step_{u}(v) \gets \HoldSymbol$
and
$\Step_{v}(u) \gets \HoldSymbol$,
respectively.

Assuming that
$\Step_{u}(v) = \Step_{v}(u)$,
nodes $u$ and $v$ update $\Step_{e}$ in accordance with the policy of \PPS{}
(see \Sect{}~\ref{section:probabilistic-phase-synchronization}).
If
$\Step_{e} = j \in \{ 0, 1, \dots, 2 \phi - 1 \}$,
then this update is deterministic, hence it can be performed by the two nodes
with no further interaction.
Otherwise
($\Step_{e} = \HoldSymbol$),
nodes $u$ and $v$ toss an (unbiased) coin $r_{e}$ to determine the next
state of $\Step_{e}$.
To this end, node $u$ (resp., $v$) tosses a coin denoted by $r_{u}(v)$ (resp.,
$r_{v}(u)$) and shares its value with $v$ (resp., $u$).
The two nodes then set
$r_{e} \gets r_{u}(v) \oplus r_{v}(u)$
and update $\Step_{e}$ accordingly.

To assign the roles of $x(e)$ and $y(e)$ to the endpoints of edge $e$ in a
given phase, we use a similar trick:
Whenever
$\Step_{e} = \HoldSymbol$,
node $u$ (resp., $v$) tosses yet another (unbiased) coin, denoted by
$\hat{r}_{u}(v)$ (resp., $\hat{r}_{v}(u)$) and shares its value with $v$
(resp., ($u$).
If $\Step_{e}$ advances in the subsequent round from
$\Step_{e} = \HoldSymbol$
to
$\Step_{e} = 0$,
then edge $e$ actually starts a phase of $\PhaseProcedure_{\Alg}$ if and only if
$\hat{r}_{u}(v) \neq \hat{r}_{v}(u)$,
in which case the node whose coin is $1$ (resp., $0$) assumes the role of
$x(e)$ (resp., $y(e)$) for the duration of the current phase.
Notice that with this additional condition, the lower bound promised in
\Lem{}~\ref{lemma:pps:start-phase-probability} on the 
probability to start a phase under $\PPS'$ decreases by (no more than) a 
constant factor.
Recalling that a degree bound of $\Delta$ for $G$ implies a degree bound of
$2 \Delta - 2$
for $L(G)$, the following theorem is derived from
\Thm{}~\ref{theorem:nodes:main}.

\begin{theorem} \label{theorem:transformer:simulation-line-graph}
Consider a node-LCL
$\Problem = \langle
\mathcal{O}, \mathcal{G}, \Predicate
\rangle$
that is
$(\psi,
\mu_{D}(\Delta),
\phi,
\mu_{A}(\Delta),
\beta,
\sigma_{0},
\rho_{0})$-eligible
and let
$\mathcal{H} = \{ G \in \mathcal{U} \mid L(G) \in \mathcal{G} \}$.
If $\mathcal{H}$ is hereditary closed,
then, the edge-LCL
$\Problem' = \langle
\mathcal{O}, \mathcal{H}, \Predicate
\rangle$
admits a randomized self-stabilizing algorithm $\SelfStabAlg'$ that
uses messages of size
\[
O \left(
\Delta (\mu_{D}(2 \Delta - 2) + \mu_{A}(2 \Delta - 2)) + \log \phi
\right)
\]
whose fully adaptive run time is
\[
O \left(
\left( \phi^{5} / \beta \right)
\cdot
\left(
\log (k) + (\psi + \phi) \log (\Delta) +
\log (\sigma_{0} + \rho_{0})
\right)
\right) \, .
\]
Moreover, edges whose distance from any manipulated node is at least
$\psi + 2 \phi + 2$
are guaranteed to maintain their original output value throughout the
execution of $\SelfStabAlg'$ (with probability $1$).
\end{theorem}

\section{Concrete Problems --- Edge-LCLs}
\label{section:concrete-problems-edges}
In this section, we develop self-stabilizing algorithms for the four
concrete edge-LCLs listed in
\Sect{}~\ref{section:introduction:concrete-problems} and establish the fully
adaptive run-time and message size bounds promised in (the bottom part of)
Table~\ref{table:transformer:concrete-problems}, with the exception of the
improved MM algorithm developed separately in
\Sect{}~\ref{section:improved-MM}.
To do so, we present the fault free algorithms and detection procedures
provided to our transformer and prove that the corresponding problems meet the
necessary eligibility criteria.

\subsection{Incremental Edge $c$-Coloring}
\label{section:concrete-edges:incremental}
This section is dedicated to showing that the incremental edge $c$-coloring
problem meets
the
$(O (\log c))$-detectability criterion
(Criterion~\ref{criterion:edges:detectability}),
the
$(3, O (c + \log\log \Delta))$-phased-based criterion
(Criterion~\ref{criterion:edges:phase-based}),
the respectful decisions criterion
(Criterion~\ref{criterion:edges:respectful-decisions}),
and
the
$(\Omega (1 / c), 3 c)$-progress criterion
(Criterion~\ref{criterion:edges:progress}).
\Thm{}~\ref{theorem:concrete-edges:incremental-direct} follows from
\Thm{}~\ref{theorem:edges:main} as we
have already established in \Sect{}~\ref{section:LCL-eligibility-criteria}
that the problem meets the
gap freeness criterion (Criterion~\ref{criterion:common:gap-freeness})
and
the
$(c - 1)$-bounded
propagation criterion (Criterion~\ref{criterion:common:bounded-propagation}).

\begin{theorem} \label{theorem:concrete-edges:incremental-direct}
The incremental edge $c$-coloring problem admits a randomized self-stabilizing
algorithm that uses messages of size
$O (c + \log\log \Delta)$
whose fully adaptive run-time is
$O (c \log (k) + c^{2} \log (\Delta))$.
\end{theorem}

\subsubsection{Implementing the Detection Procedure}
\label{section:concrete-edges:incremental:detection}
Consider a graph
$G \in \mathcal{U}$.
We implement the detection procedure $\Detect$ for incremental $c$-coloring
with
messages of size
$\mu_{D}(\Delta) = O (\log c)$.
To this end, a node
$v \in V(G)$
shares the content of
$\OutReg_{v}(u)$
with each neighbor
$u \in N_{G}(v)$.
In addition, if
$\OutReg_{v}(u) = i \in \{ 1, \dots, c \}$,
then $v$ also shares with $u$ the size of the set
$\left\{
w \in N_{G}(v)
\mid
\OutReg_{v}(w) \neq \bot
\land
\OutReg_{v}(w) < i
\right\}$.
Clearly, this information allows $u$ and $v$ to determine if either
(I)
edge
$\{ u, v \}$
is port-inconsistent;
or
(II)
edge
$\{ u, v \}$
is uncontent.
Therefore, the incremental edge $c$-coloring problem meets the
$(O (\log c))$-detectability
criterion (Criterion~\ref{criterion:edges:detectability}) as promised.

\subsubsection{Implementing the Fault Free Algorithm}
\label{section:concrete-edges:incremental:algorithm}
The (fault free) algorithm $\Alg$ developed in the current section uses
messages of size
$O (c + \log\log \Delta)$
and follows a mode of operation similar to that of the classic (fault free) MM
algorithm of Israeli and Itai~\cite{IsraeliI1986fast}.
Being a phase-based algorithm, $\Alg$ is recounted through its phase procedure
$\PhaseProcedure_{\Alg}$ that works with $3$-step phases.

Let $G$ be the graph on which $\PhaseProcedure_{\Alg}$ is invoked and let
$C : E(G) \rightarrow \{ 1, \dots, c \} \cup \{ \bot \}$
be the configuration associated with $\PhaseProcedure_{\Alg}$ at the beginning
of the execution.
Let
$H = G(\UnDecidedSet(C))$
be the subgraph induced on $G$ by the set of edges that are undecided in $C$.

Consider a node
$v \in V(H)$.
In step $0$, node $v$ randomly marks itself as \emph{active} or 
\emph{passive} with equal probability.
If $v$ is active, then
(1)
$v$ picks a neighbor
$u \in N_{H}(v)$
uniformly at random and sends a \emph{proposal} message to $u$ that encodes
the value of
$\LogDeg(v) = \lfloor \log \Degree_{H}(v) \rfloor$,
using
$O (\log\log \Delta)$
bits.

In step $1$, if $v$ is passive and $v$ received at least one proposal, then
(1)
$v$ selects a neighbor
$u \in N_{H}(v)$
from which a proposal arrived with probability proportional to
$2^{\LogDeg(u)}$
--- notice that this is a $2$-approximation of $\Degree_{H}(u)$;
(2)
$v$ marks the selected node $u$ as \emph{accepted};
and
(3)
$v$ sends an \emph{acceptance} message to $u$.
Moreover, preparing for the decision step, regardless of whether $v$ is active
or passive, $v$ shares the vector
$b_{v} \in \{ 0, 1 \}^{c - 1}$
with each neighbor
$w \in N_{H}(v)$,
where $b_{v}$ is defined by setting
$b_{v}(i) = 1$
if and only if $v$ admits a (decided) incident edge
$\{ v, v' \} \in E(G)$
(at least one)
such that
$C(\{ v, v' \}) = i$.

Consider the decision step (step $2$).
For each incident edge
$f = \{ v, w \} \in E(H)$,
node $v$ uses the vector $b_{w}$ received from node $w$ to check if
$C[f] \supseteq \{ 1, \dots c - 1 \}$,
that is,
if
$i \in C[f]$
for every
$1 \leq i \leq c - 1$.
If this is the case, then $v$ sets
$\OutReg_{v}(w) \gets c$.
Following that, assuming that edge
$e = \{ u, v \} \in E(H)$
was not decided yet (i.e., $\OutReg_{v}(u)$ still holds $\bot$), node $v$ sets
$\OutReg_{v}(u) \gets i_{\min}$,
where
$i_{\min} = \min \{ 1 \leq i \leq c - 1 \mid i \notin C[e] \}$,
if either
(1)
$v$ is active and $u$ is marked as accepted;
or
(2)
$v$ is passive and it received an acceptance message from $u$.

This completes the description of the phase procedure
$\PhaseProcedure_{\Alg}$.
Let
$C' : E(G) \rightarrow \{ 1, \dots, c \} \cup \{ \bot \}$
be the configuration associated with $\PhaseProcedure_{\Alg}$ when the
execution terminates and notice that this is a random variable depending on
the coin tosses of the procedure in steps $0$ and $1$.
The fact that incremental edge $c$-coloring meets
the
$(3, O (c + \log\log \Delta))$-phase
based criterion (Criterion~\ref{criterion:edges:phase-based})
and
the respectful decisions criterion
(Criterion~\ref{criterion:edges:respectful-decisions}) with respect to $\Alg$
follows from \Obs{}~\ref{observation:concrete-edges:incremental:correctness}.

\sloppy
\begin{observation}
\label{observation:concrete-edges:incremental:correctness}
Algorithm $\Alg$ is a port-consistent decision-oblivious phase-based
algorithm.
Moreover, for each edge
$e \in E(G)$,
if
$e \in \UnDecidedSet(C) \cap \DecidedSet(C')$,
then \\
(1)
$C'(e) = i \in \{ 1, \dots, c - 1 \}$
only if
$i \notin C'[e]$
and
$j \in C[e]$
for every
$1 \leq j \leq i - 1$;
and \\
(2)
$C'(e) = c$
if and only if
$i \in C[e]$
for every
$1 \leq i \leq c - 1$.
\end{observation}
\par\fussy

\paragraph{Edge Separable Potential Function.}
It remains to show that incremental edge $c$-coloring meets the
$(\Omega (1 / c), 3 c)$-progress
criterion (Criterion~\ref{criterion:edges:progress}) with respect to $\Alg$.
To this end, we introduce an edge separable potential function $\pi$ for
incremental edge $c$-coloring with top potential coefficient
$\sigma_{0} = 3 c$
and prove that
\begin{equation} \label{equation:concrete-edges:incremental:progress}
\Ex \left( \pi(G, C') \right)
\, \leq \,
(1 - \beta) \cdot \pi(G, C)
\end{equation}
for
$\beta = \Omega (1 / c)$,
where the expectation is over $C'$.
The family
$\mathcal{I}
=
\{ \sigma(M) \}_{M \in \Multisets(\{ 1, \dots, c \})}$
of potential coefficients for $\pi$ is defined by setting
\[
\sigma(M)
\, = \,
3 c
-
\sum_{i = 1}^{c - 1} \min \{ M(i), 2 \}
\, .
\]
By definition, this family is monotonically non-increasing.
Moreover, the construction of $\mathcal{I}$ ensures that
\[
c + 2 \leq \sigma(M) \leq 3 c
\]
for every
$M \in \Multisets(\{ 1, \dots, c \})$,
hence it is a legitimate family of potential coefficients with
$\sigma_{0} = 3 c$.

En route to establishing
(\ref{equation:concrete-edges:incremental:progress}), we define the
notation
\[
\lambda(e)
\, = \,
\begin{cases}
\sigma(C[e]) \, , & e \in \UnDecidedSet(C) \\
0 \, , & \text{otherwise}
\end{cases}
\quad \text{and} \quad
\lambda'(e)
\, = \,
\begin{cases}
\sigma(C'[e]) \, , & e \in \UnDecidedSet(C') \\
0 \, , & \text{otherwise}
\end{cases}
\]
for each edge
$e \in E(G)$.
This notation is extended to edge subsets
$F \subseteq E(G)$
by defining
$\sigma(F) = \sum_{e \in F}\sigma(e)$
and
$\sigma'(F) = \sum_{e \in F}\sigma'(e)$,
observing that
$\pi(G, C) = \lambda(E(G)) = \lambda(E(H))$
and
$\pi(G, C') = \lambda'(E(G)) = \lambda'(E(H))$.

A node
$v \in V(H)$
is said to be \emph{good} in $H$ if
\[
\left| \left\{
u \in N_{H}(v) \, : \, \Degree_{H}(u) \leq \Degree_{H}(v)
\right\} \right|
\, \geq \,
\Degree_{H}(v) / 3
\, .
\]
We use the well known graph theoretic fact (see, e.g.,
\cite[\Lem{}~4.4]{AlonBI1986fast}) that at least half of the edges in $E(H)$
are incident on a good node (at least one).
Let $U^{g}$ be the set of nodes that are good in $H$ and for a good node
$v \in U^{g}$,
let $F_{v}$ be the set of edges incident on $v$ in $H$.
Since
$c < \lambda(e) \leq 3 c$
for every edge
$e \in E(H)$,
it follows that
\[
\lambda \left( \bigcup_{v \in U^{g}} F_{v} \right)
\, > \,
\frac{1}{4} \lambda(E(G))
\, .
\]
To establish (\ref{equation:concrete-edges:incremental:progress}), we fix
a good node
$v \in U^{g}$
and prove that the following event occurs with probability
$\Omega (1)$:
\[
\text{event $S_{v}$:}
\qquad
\lambda'(F_{v})
\, \leq \,
\max \left\{
\frac{35}{36}, \frac{3 c - 1}{3 c}
\right\} \cdot \lambda(F_{v})
\, .
\]

Let
$F^{g}_{v} = \{
\{ u, v \} \in F_{v} \mid \Degree_{H}(u) \leq \Degree_{H}(v)
\}$,
recalling that
$|F^{g}_{v}| \geq \frac{1}{3} |F_{v}|$,
and let
\[
F^{g, c}_{v}
\, = \,
\left\{
e \in F^{g}_{v} \mid i \in C[e] \text{ for every } 1 \leq i \leq c - 1
\right\}
\, .
\]
\Obs{}~\ref{observation:concrete-edges:incremental:correctness}
implies that if
$e \in F^{g, c}_{v}$,
then
$C'(e) = c$
and
$\lambda'(e) = 0$.
Therefore, if
$|F^{g, c}_{v}| > \frac{1}{4} |F^{g}_{v}|$,
then
\[
\lambda'(F_{v})
\, < \,
\lambda(F_{v}) - c \cdot |F^{g, c}_{v}|
\, < \,
\lambda(F_{v}) - \frac{c}{4} |F^{g}_{v}|
\, \leq \,
\lambda(F_{v}) - \frac{c}{12} |F_{v}|
\, \leq \,
\lambda(F_{v}) - \frac{1}{36} \lambda(F_{v})
\, = \,
\frac{35}{36} \lambda(F_{v})
\, ,
\]
where
the first transition holds as
$\lambda(e) > c$
for every edge
$e \in F^{g, c}_{v}$
and the fourth transition holds as
$\lambda(e) \leq 3 c$
for every edge
$e \in F_{v}$.
We conclude that if
$|F^{g, c}_{v}| > \frac{1}{4} |F^{g}_{v}|$,
then event $S_{v}$ occurs with probability $1$.

So, assume hereafter that
$|\overline{F}^{g, c}_{v}| \geq \frac{3}{4} |F^{g}_{v}|$,
where
$\overline{F}^{g, c}_{v} = F^{g}_{v} - F^{g, c}_{v}$.
For an edge
$e \in \overline{F}^{g, c}_{v}$,
let $S'_{v}(e)$ be the event that $v$ accepts a proposal received over $e$.
Notice that $\PhaseProcedure_{\Alg}$ is designed so that event $S'_{v}(e)$
implies that $e$ becomes decided in the decision step with
$C'(e) = i$
for some
$i \in \{ 1, \dots, c - 1 \}$.
We argue that this causes every edge
$f \in F_{v}$
to lose at least one $\lambda$-unit, i.e,
$\lambda'(f) \leq \lambda(f) - 1$.
Indeed, if
$C'(e) = i$,
then
$C(\{ v, w \}) \neq i$
for every edge
$\{ v, w \} \in \DecidedSet(C)$
incident on $v$, hence the multiplicity $C[f](i)$ of $i$ in $C[f]$ is at most
$1$ for every
$f \in F_{v}$.
The argument follows by the definition of $\pi$'s potential coefficients as
$\min \{ C[f](i), 2 \} < 2$.
Since
$\lambda(f) \leq 3 c$
for every
$f \in F_{v}$,
we conclude that if event $S'_{v}(e)$ occurs, then
\[
\lambda'(F_{v})
\, \leq \,
\lambda(F_{v}) - |F_{v}|
\, \leq \,
\lambda(F_{v}) - \frac{1}{3 c} \lambda(F_{v})
\, = \,
\frac{3 c - 1}{3 c} \lambda(F_{v})
\, ,
\]
thus yielding event $S_{v}$.

It remains to prove that event
$S'_{v}
=
\bigvee_{e \in \overline{F}^{g, c}_{v}} S'_{v}(e)$
occurs with probability
$\Omega (1)$.
This is done by plugging
$A = \overline{F}^{g, c}_{v}$
in the following lemma.

\begin{lemma}
\label{lemma:concrete-edges:incremental:accepting-proposal}
Let
$A \subseteq F^{g}_{v}$
be an edge subset of size
$|A| \geq \frac{3}{4} |F^{g}_{v}|$.
Then, the event that $v$ accepts a proposal received over an edge in $A$
occurs
with probability
$\Omega (1)$.
\end{lemma}

The proof of
\Lem{}~\ref{lemma:concrete-edges:incremental:accepting-proposal}
relies on the following simple observation.

\begin{observation}
\label{observation:concrete-edges:incremental:no-proposal}
The event that $v$ does not receive any proposal over the edges in
$F_{v} - F^{g}_{v}$
occurs with probability
$\Omega (1)$. 
\end{observation}
\begin{proof}
Fix an edge
$\{ u, v \} \in F_{v} - F^{g}_{v}$.
Since
$\Degree_{H}(u) > \Degree_{H}(v)$,
it follows that the probability that $u$ sends a proposal to $v$ is smaller
than
$1 / \Degree_{H}(v)$.
The assertion follows by a standard calculation as
$|F_{v} - F^{g}_{v}| < |F_{v}| = \Degree_{H}(v)$.
\end{proof}

\begin{proof}[Proof of
\Lem{}~\ref{lemma:concrete-edges:incremental:accepting-proposal}]
Recall that $v$ becomes passive with probability
$1 / 2$;
condition hereafter on this event.
By \Obs{}~\ref{observation:concrete-edges:incremental:no-proposal}, we know
that $v$ does not receive any proposal over the edges in
$F_{v} - F^{g}_{v}$
with probability
$\Omega (1)$;
condition hereafter on this event.
Since
$|F^{g}_{v}| \geq \Degree_{H}(v)$
and since
$\Degree_{H}(u) \leq \Degree_{H}(v)$
for every edge
$\{ u, v \} \in F^{g}_{v}$,
we conclude by a standard calculation that $v$ receives at least one proposal
over an edge in $F^{g}_{v}$,
and thus, accepts a proposal received over an edge in $F^{g}_{v}$, with
probability
$\Omega (1)$;
condition hereafter on this event.

Let
$R \subseteq F^{g}_{v}$
be the subset of edges over which $v$ receives a proposal and notice that $R$
is a random variable depending on the coin tosses of $\PhaseProcedure_{\Alg}$
in step $0$.
For an edge
$e = \{ u, v \} \in F^{g}_{v}$,
define the random variable
\[
X_{e}
\, = \,
\begin{cases}
2^{\LogDeg(u)} \, , & e \in R \\
0 \, , & \text{otherwise}
\end{cases}
\]
and let
\[
X_{A}
\, = \,
\sum_{e \in A} X_{e}
\quad \text{and} \quad
X_{\overline{A}}
\, = \,
\sum_{e \in F^{g}_{v} - A} X_{e}
\, .
\]
The key observation now is that depending on the coin tosses of
$\PhaseProcedure_{\Alg}$ in step $1$, the probability that $v$ accepts a
proposal received over an edge in $A$ is exactly
$\frac{X_{A}}{X_{A} + X_{\overline{A}}}$.

To complete the proof, we argue that depending on the coin tosses of
$\PhaseProcedure_{\Alg}$ in step $0$, the event
$X_{A} \geq X_{\overline{A}}$
occurs with probability
$\Omega (1)$.
To this end, fix an edge
$e = \{ u, v \} \in F^{g}_{v}$
and notice that since
$\Degree_{H}(u) / 2
<
2^{\LogDeg(u)}
\leq
\Degree_{H}(u)$,
it follows that
$1 / 2
<
\Ex(X_{e})
\leq
1$.
As $X_{A}$ and $X_{\overline{A}}$ are sums of independent random variables
and recalling that
$|A| \geq 3 |F^{g}_{v} - A|$,
the argument follows by standard concentration bounds.
\end{proof}

\subsection{Maximal Matching}
\label{section:concrete-problems-edges:mm}
The following theorem is derived from
\Thm{}~\ref{theorem:concrete-edges:incremental-direct} by plugging
$c = 2$.

\begin{theorem} \label{theorem:concrete-edges:mm:application-incremental}
The MM problem admits a randomized self-stabilizing
algorithm that uses messages of size
$O (\log\log \Delta)$
whose fully adaptive run-time is
$O (\log (k + \Delta))$.
\end{theorem}

\subsection{Maximal Edge $c$-Coloring}
\label{section:concrete-edges:maximal}
In this section, we develop the self-stabilizing algorithms for maximal edge 
$c$-coloring listed in
Table~\ref{table:transformer:concrete-problems} and establish their fully 
adaptive run-time and message size bounds.
We start with the following theorem which is derived from 
\Thm{}~\ref{theorem:transformer:simulation-line-graph} by simulating the
maximal node $c$-coloring algorithm promised in 
\Thm{}~\ref{theorem:concrete-nodes:maximal:clone-graph}
on the line graph.

\begin{theorem} \label{theorem:concrete-edges:maximal:line-graph}
The maximal edge $c$-coloring problem admits a randomized self-stabilizing
algorithm that uses messages of size
$O (c \Delta \log\log \Delta)$
whose fully adaptive run-time is
$O (\log (k + \Delta))$.
\end{theorem}

The remainder of this section is dedicated to showing that the maximal edge
$c$-coloring problem meets
the
$(O (c))$-detectability criterion
(Criterion~\ref{criterion:edges:detectability}),
the
$(3, O (c + log\log \Delta))$-phased-based criterion
(Criterion~\ref{criterion:edges:phase-based}),
the respectful decisions criterion
(Criterion~\ref{criterion:edges:respectful-decisions}),
and
the
$(\Omega (1 / c), 3 c)$-progress criterion
(Criterion~\ref{criterion:edges:progress}).
\Thm{}~\ref{theorem:concrete-edges:maximal:direct} follows from
\Thm{}~\ref{theorem:edges:main} as we
have already established in \Sect{}~\ref{section:LCL-eligibility-criteria}
that the problem meets the
gap freeness criterion (Criterion~\ref{criterion:common:gap-freeness})
and
the $1$-bounded propagation criterion
(Criterion~\ref{criterion:common:bounded-propagation}).

\begin{theorem} \label{theorem:concrete-edges:maximal:direct}
The maximal edge $c$-coloring problem admits a randomized self-stabilizing
algorithm that uses messages of size
$O (c + \log\log \Delta)$
whose fully adaptive run-time is
$O (c \log (k + \Delta))$.
\end{theorem}

\subsubsection{Implementing the Detection Procedure}
\label{section:concrete-edges:maximal:detection}
Consider a graph
$G \in \mathcal{U}$.
We implement the detection procedure $\Detect$ for maximal $c$-coloring with
messages of size
$\mu_{D}(\Delta) = O (c)$.
To this end, a node
$v \in V(G)$
shares the content of
$\OutReg_{v}(u)$
with each neighbor
$u \in N_{G}(v)$
along with the vector
$b_{v, u} \in \{ 0, 1 \}^{c - 1}$
defined by setting
$b_{v, u}(i) = 1$
if and only if there exists a neighbor
$w \in N_{G}(v) - \{ u \}$
(at least one) such that
$\OutReg_{v}(w) = i$.
Clearly, this information allows $u$ and $v$ to determine if either
(I)
edge
$\{ u, v \}$
is port-inconsistent;
or
(II)
edge
$\{ u, v \}$
is uncontent.
Therefore, the maximal edge $c$-coloring problem meets the
$(O (c))$-detectability
criterion (Criterion~\ref{criterion:edges:detectability}) as promised.

\subsubsection{Implementing the Fault Free Algorithm}
\label{section:concrete-edges:maximal:algorithm}
The fault free algorithm $\Alg$ used for establishing
\Thm{}~\ref{theorem:concrete-edges:maximal:direct} is exactly the algorithm
developed
in \Sect{}~\ref{section:concrete-edges:incremental:algorithm} for incremental
edge $c$-coloring.
Although \Obs{}~\ref{observation:concrete-edges:incremental:correctness} is
established in \Sect{}~\ref{section:concrete-edges:incremental:algorithm} in
the context of incremental edge $c$-coloring, examining it again reveals that
the
$(3, O (c + \log\log \Delta))$-phase
based criterion (Criterion~\ref{criterion:edges:phase-based})
and
the respectful decisions criterion
(Criterion~\ref{criterion:edges:respectful-decisions})
are met with respect to $\Alg$ also by maximal edge $c$ coloring.
Moreover, since the two problems are defined over the same set of output
values, the edge separable potential function $\pi$ presented in
\Sect{}~\ref{section:concrete-edges:incremental:algorithm} in the context of
incremental edge $c$-coloring, also serves as an edge separable potential
function for maximal edge $c$-coloring.
Employing (\ref{equation:concrete-edges:incremental:progress}) and the fact
that the top potential coefficient of $\pi$ is
$\sigma_{0} = 3 c$,
we conclude that maximal edge $c$-coloring meets the
$(\Omega (1 / c), 3 c)$-progress
criterion (Criterion~\ref{criterion:edges:progress}) with respect to $\Alg$ as
well.

\subsection{Edge $c$-Coloring}
\label{section:concrete-problems-edges:coloring}
In this section, we develop the self-stabilizing algorithms for edge 
$c$-coloring listed in
Table~\ref{table:transformer:concrete-problems} and establish their fully 
adaptive run-time and message size bounds.
We start with the following theorem derived from
\Thm{}~\ref{theorem:concrete-edges:maximal:direct} by plugging
$c = 2 \Delta$.

\begin{theorem}
\label{theorem:concrete-edges:coloring-tight:application-maximal-direct}
The edge $(2\Delta - 1)$-coloring problem admits a randomized self-stabilizing
algorithm that uses messages of size
$O (\Delta)$
whose fully adaptive run-time is
$O (\Delta \log (k + \Delta))$.
\end{theorem}

The following theorem is derived from
\Thm{}~\ref{theorem:concrete-edges:maximal:line-graph} by plugging
$c = 2 \Delta$.

\begin{theorem}
\label{theorem:concrete-edges:coloring-tight:application-maximal-line-graph}
The edge $(2\Delta - 1)$-coloring problem admits a randomized
self-stabilizing
algorithm that uses messages of size
$O (\Delta^{2} \log \log \Delta)$
whose fully adaptive run-time is
$O (\log (k + \Delta))$.
\end{theorem}

The reminder of this section is dedicated to developing a self-stabilizing
algorithm for node
$\lceil (2 + \epsilon) \Delta \rceil$-coloring.
Fix some
$\Delta \in \Integers_{> 0}$
and a constant
$\epsilon > 0$
and assume, without loss of generality, that
$\epsilon \Delta \in \Integers_{> 0}$.
Recall that in the edge
$(2 + \epsilon) \Delta$-coloring
the set of output values $\mathcal{O}$
is
$[(1 + \epsilon) \Delta]$.

We will show that the edge
$((2 + \epsilon) \Delta)$-coloring
problem meets
the
$(O (\log \Delta))$-detectability criterion
(Criterion~\ref{criterion:edges:detectability}),
the
$(3, O (\log \Delta))$-phased-based criterion
(Criterion~\ref{criterion:edges:phase-based}),
the respectful decisions criterion
(Criterion~\ref{criterion:edges:respectful-decisions}),
and the
$(\Omega (1), 1)$-progress criterion
(Criterion~\ref{criterion:edges:progress}).
\Thm{}~\ref{theorem:concrete-edges:coloring-loose:direct} follows from
\Thm{}~\ref{theorem:edges:main} as we
have already established in \Sect{}~\ref{section:LCL-eligibility-criteria}
that the problem meets the
gap freeness criterion (Criterion~\ref{criterion:common:gap-freeness})
and the
$0$-bounded propagation criterion 
(Criterion~\ref{criterion:common:bounded-propagation}).

\begin{theorem} \label{theorem:concrete-edges:coloring-loose:direct}
The edge
$\lceil (2 + \epsilon) \Delta \rceil$-coloring
problem admits a  randomized
self-stabilizing algorithm that uses messages of size
$O (\log \Delta)$
whose fully adaptive run-time is
$O (\log (k + \Delta))$.
\end{theorem}

\subsubsection{Implementing the Detection Procedure}
\label{section:concrete-edges:2+epsilon:detection}
Consider a graph
$G \in \mathcal{U}$.
We implement the detection procedure $\Detect$ for
$((2 + \epsilon) \Delta)$-coloring
with messages of size
$\mu_{D}(\Delta) = O (\Delta)$.
To this end, a node
$v \in V(G)$
shares the content of
$\OutReg_{v}(u)$
with each neighbor
$u \in N_{G}(v)$.
In addition, if
$\OutReg_{v}(u) = i \in \{ 1, \dots, c \}$,
then $v$ also shares with $u$ a bit indicating whether color $i$ exists is some
$\OutReg_{v}(w)$
for some
$w \in N_{G}(v) - \{u\}$.
Clearly, this information allows $u$ and $v$ to determine if either
(I)
edge
$\{ u, v \}$
is port-inconsistent;
or
(II)
edge
$\{ u, v \}$
is uncontent.
Therefore, the edge
$((2 + \epsilon) \Delta)$-coloring
problem meets the
$(O (\log \Delta))$-detectability
criterion (Criterion~\ref{criterion:edges:detectability}) as promised.

\subsubsection{Implementing the Fault Free Algorithm}
\label{section:concrete-edges:2+epsilon:algorithm}
The fault-free algorithm is denoted by $\Alg$ and it
uses messages of size
$\mu_{A}(\Delta) = O (\log \Delta)$.
Being a phase-based algorithm, $\Alg$ is recounted through its phase procedure
$\PhaseProcedure_{\Alg}$ working with $3$-step phase.

Let $G$ be the graph on which $\PhaseProcedure_{\Alg}$ is invoked and let
$C : E(G) \rightarrow \{ 1, \dots, c \} \cup \{ \bot \}$
be the configuration associated with $\PhaseProcedure_{\Alg}$ at the beginning
of the execution.
Let
$H = G(\UnDecidedSet(C))$
be the subgraph induced on $G$ by the set of edges that are undecided in $C$.
Let
$e =\{v, u\} \in E(H)$,
let
$U_{v} = \{ u \in N_{G}(v) \mid \{v, u\} \in E(H) \}$,
and let
$D_{v} = N_{G}(v) - U_{v}$.

In step $0$, node $v$ select a color $c_{v}(u)$ u.a.r.\ from the pallet
$\mathcal{O}$
for every edge
$\{v, u\} \in E(H)$
and sends it to neighbor
$u \in U_{v}$.
In step $1$, both $v$ and $u$ determine the same candidate color for edge
$e$ defined to be
$c_{e} = 1 + ((c_{v}(u) + c_{u}(v)) \mod |\mathcal{O}|)$.
Moreover, preparing for the decision step,
node $v$ sends an ``accept'' or ``decline'' to every $u \in U_{v}$
according to the candidate colors and the colors in $\OutReg_{v}(w)$
for every $w \in D_{v}$.

Step $3$ is the decision step. Node $v$ sets
$\OutReg_{v}(u) = c_{\{v,u\}}$ only if $v$ sent an accept message to $u$ in the
previous step and $v$ received an accept message from $u$ (in that case $u$ will
also set
$\OutReg_{u}(v) = c_{\{v, u\}}$).

By design, algorithm $\Alg$ is decision-oblivious
phase-based algorithm, thus edge
$((2 + \epsilon) \Delta)$-coloring 
meets the
$(3, O (\log \Delta))$-phase-based
criterion (Criterion~\ref{criterion:edges:phase-based}) with respect to $\Alg$.
Consider an edge
$e \in \UnDecidedSet(C)$
and suppose that $e$ become decided during the execution of
$\PhaseProcedure_{\Alg}$ so that
$e \in \DecidedSet(C')$.
The procedure is designed to guarantee that
$C'(e) = i \in \mathcal{O}$
implies that
$C(f) \neq i$
for every node
$f \in N_{G}(f)$.
Therefore, edge
$((2 + \epsilon) \Delta)$-coloring
also meets the respectful decisions criterion
(Criterion~\ref{criterion:edges:respectful-decisions}) with respect to $\Alg$.

\paragraph{Edge Separable Potential Function.}
It remains to show that edge
$((2 + \epsilon) \Delta)$-coloring
meets the
$(\Omega (1), 1)$-progress
criterion (Criterion~\ref{criterion:edges:progress}) with respect to $\Alg$.
To this end, we introduce an edge separable potential function $\pi$ for
edge
$((2 + \epsilon) \Delta)$-coloring
with top potential coefficient
$\sigma_{0} = 1$
and prove that
\begin{equation} \label{equation:concrete-edges:2+epsilon:progress}
\Ex \left( \pi(G, C') \right)
\, \leq \,
(1 - \beta) \cdot \pi(G, C)
\end{equation}
for
$\beta = \Omega (1)$,
where the expectation is over $C'$.
The family
$\mathcal{I}
=
\{ \sigma(M) \}_{M \in \Multisets([(2 + \epsilon) \Delta])}$
of potential coefficients for $\pi$ is defined by setting
\[
\sigma(M)
\, = \, 1
\, .
\]
By definition, this family is monotonically non-increasing.
Moreover, the construction of $\mathcal{I}$ ensures that
\[
\sigma(M) \geq 1
\]
for every
$M \in \Multisets([(2 + \epsilon) \Delta])$,
hence it is a legitimate family of potential coefficients with
$\sigma_{0} = 1$

Consider some 
$G \in \mathcal{U}_{\Delta}$.
Let
$C : E(G) \rightarrow \mathcal{O} \cup \{\bot\} $
be an edge configuration such that
$(G,C) \in \ContConfGraphs(\Problem)$
and let
$e =\{u, v\} \in \UnDecidedSet(C)$.
We denote by
$A_{e}$
the event that edge $e$ is decided under edge configuration
$\PhaseProcedure_{\Alg}(G, C)$.
For a neighboring edge
$e' \in \UnDecidedSet(C) \cap N_{G}(e)$,
we denote by
$A_{e,e'}$,
the event that the candidate color $c_{e}$ equals to the candidate color
$e'$
chosen during the execution of
$\PhaseProcedure_{\Alg} (G, C)$.
For a neighboring edge
$e' \in \DecidedSet(C) \cap N_{G}(e)$
we denote by
$A_{e,e'}$
the event that
$C(e') = c_{e}$,
i.e., the candidate color chosen for edge $e$ during the execution of 
$\PhaseProcedure_{\Alg}(G, C)$ conflict with an existing decided neighboring 
edge color.

It is easily verifiable that for every
$e' \in N_{G}(e)$,
it holds that
$\Pr(A_{e,e'}) \leq \frac{1}{(2+\epsilon) \Delta}$.
By the union bound we get that
$\Pr\left(\bigcup_{e' \in N_{G}(e)} A_{e,e'}\right) \leq 
\frac{2}{2+\epsilon}$.
We can conclude that
$\Pr(A_{e}) \geq 1-\frac{2}{2+\epsilon}$. Hence,
for every
$e = \{u,v\} \in \UnDecidedSet(C)$,
the probability that $e$ is decided under the edge configuration
$\PhaseProcedure_{\Alg}(G, C)$ is at least
$\left(1-\frac{2}{2+\epsilon}\right) = \Omega(1)$.
Once $e$ becomes decided, the potential coefficient of edge $e$ (which
is $1$) is removed from $\pi$.

\section{Concrete Problems --- Node-LCLs}
\label{section:concrete-problems-nodes}
In this section, we develop self-stabilizing algorithms for the four
concrete node-LCLs listed in
\Sect{}~\ref{section:introduction:concrete-problems} and establish the fully
adaptive run-time and message size bounds promised in 
Table~\ref{table:transformer:concrete-problems} regarding node-LCLs.
To do so, we present the fault free algorithms and detection procedures
provided to our transformer and prove that the corresponding problems meet the
necessary eligibility criteria.
For all problems discussed in this section, we take the straightforward
detection procedure \Detect{} in which each node $v$ shares its output
register
$\OutReg_{v}$
with all its neighbors in every round.

\subsection{Incremental Node $c$-Coloring}
\label{section:concrete-problems-nodes:incremental-coloring}
This section is dedicated to showing that the incremental node $c$-coloring 
problem meets the
$(3, O (\log c + \log \log \Delta))$-phased-based criterion
(Criterion~\ref{criterion:nodes:phase-based}),
respectful decisions criterion
(Criterion~\ref{criterion:nodes:respectful-decisions}),
and
$(\Omega (1/c), 1, 2 c)$-progress criterion
(Criterion~\ref{criterion:nodes:progress}).
The incremental node $c$-coloring problem meets the
$(O (\log c))$-detectability criterion
(Criterion~\ref{criterion:nodes:detectability}) since each node shares its
output value with all of its neighbors.
\Thm{}~\ref{theorem:concrete-nodes:incremental:direct} follows from
\Thm{}~\ref{theorem:nodes:main} as we
have already established in \Sect{}~\ref{section:LCL-eligibility-criteria}
that the problem meets the
gap freeness criterion (Criterion~\ref{criterion:common:gap-freeness})
and
the
$(c - 1)$-bounded
propagation criterion
(Criterion~\ref{criterion:common:bounded-propagation}).

\begin{theorem} \label{theorem:concrete-nodes:incremental:direct}
The incremental node $c$-coloring problem admits a randomized self-stabilizing
algorithm that uses messages of size
$O (\log c + \log\log \Delta)$
whose fully adaptive run-time is
$O (c \log (k) + c^{2} \log (\Delta))$.
\end{theorem}

\subsubsection{Implementing the Fault Free Algorithm}
\label{section:concrete-nodes:incremental:algorithm}
The (fault free) algorithm $\Alg$ developed in the current section uses
messages of size
$O (\log c + \log \log \Delta)$
and follows a mode of operation similar to the (fault free) MIS
algorithm developed in the seminal work of
Alon et al.~\cite{AlonBI1986fast}.
Being a phase-based algorithm, $\Alg$ is recounted through its phase procedure
$\PhaseProcedure_{\Alg}$ that works with $3$-step phases.

Let $G$ be the graph on which $\PhaseProcedure_{\Alg}$ is invoked and let
$C : V(G) \rightarrow [c] \cup \{ \bot \}$
be the configuration associated with $\PhaseProcedure_{\Alg}$ at the beginning
of the execution.
Let
$H = G(\UnDecidedSet(C))$
be the subgraph induced on $G$ by the set of undecided nodes under $C$.

Consider a node
$v \in V(H)$.
In step $1$, if
$\Degree_{H}(v) > 0$,
then node $v$ marks itself w.p.\
$1 / \Degree_{H}(v)$, otherwise
it is marked w.p.\ $1$.
Node $v$ sends a message to its neighbors in $H$, indicating whether it is
marked or not, together with the value of
$\LogDeg(v) = \lfloor \log \Degree_{H}(v) \rfloor$,
using
$O (\log\log \Delta)$
bits.

Step $2$ is the decision step:
Node $v$ first calculates the size of the set
$\{ w \in N_{G}(v) \cap \DecidedSet(C) \mid C(w) \in \{ 1, \dots, c - 1\} \}$;
if this size is at least
$c - 1$,
then $v$ sets
$\OutReg_{v} \gets c$.
Following that, assuming that $v$ was not decided yet (i.e., $\OutReg_{v}$
still holds $\bot$), if $v$ is marked and
$\LogDeg(v) > \LogDeg(u)$
for every marked neighbor
$u \in N_{H}(v)$
of $v$, then $v$ sets
$\OutReg_{v} \gets i_{v}$,
where $i_{v}$ is the smallest non-conflicting color, that is,
$i_{v} = \min \{ 1 \leq i \leq c - 1 \mid i \notin C[v] \}$.
Notice that $i_{v}$ is well defined since if
$i \in C[v]$
for all
$1 \leq i \leq c - 1$,
then $v$ would have not remained undecided after the first part of the step.

Let
$C' : V(G) \rightarrow \{ 1, \dots, c \} \cup \{ \bot \}$
be the configuration associated with $\PhaseProcedure_{\Alg}$ when the
execution terminates and notice that this is a random variable depending on
the coin tosses of the procedure in step $1$.
The fact that incremental node $c$-coloring meets
the
$(3, O (\log\log \Delta))$-phase
based criterion (Criterion~\ref{criterion:nodes:phase-based})
and
the respectful decisions criterion
(Criterion~\ref{criterion:nodes:respectful-decisions}) with respect to $\Alg$
follows from \Obs{}~\ref{observation:concrete-nodes:incremental:correctness}.

\begin{observation}
\label{observation:concrete-nodes:incremental:correctness}
Algorithm $\Alg$ is a decision-oblivious phase-based algorithm.
Moreover, for each node
$v \in V(G)$,
if
$v \in \UnDecidedSet(C) \cap \DecidedSet(C')$,
then \\
(1)
$C'(v) = i \in \{ 1, \dots, c - 1 \}$
only if
$i \notin C'[v]$;
and \\
(2)
$C'(v) = c$
if and only if
$\sum_{i = 1}^{c - 1} C[v](i) \geq c - 1$.
\end{observation}

\paragraph{Node Separable Potential Function.}
It remains to show that incremental node $c$-coloring meets the
$(\Omega (1/c), 1, 2 c)$-progress
criterion (Criterion~\ref{criterion:nodes:progress}).
To this end, we introduce a node separable potential function $\pi$ for
incremental node $c$-coloring with top potential $\sigma$-coefficient
$\sigma_{0} = 1$
and top potential $\rho$-coefficient
$\rho_{0} = c$
and prove that
\begin{equation} \label{equation:concrete-nodes:incremental:progress}
\Ex \left( \pi(G, C') \right)
\, \leq \,
(1 - \beta) \cdot \pi(G, C)
\end{equation}
for
$\beta = \Omega (1/c)$,
where the expectation is over $C'$.
The families
$\mathcal{I}^{\sigma}
=
\left\{ \sigma(M) \right\}_{M \in \Multisets([c])}$
and
$\mathcal{I}^{\rho}
=
\left\{ \rho(M) \right\}_{M \in \Multisets([c])}$
of potential coefficients for $\pi$ are defined by setting
\[
\sigma(M)
\, = \,
1
\quad \text{and} \quad
\rho(M)
\, = \,
2 c - \min \left\{ c, \sum_{i = 1}^{c - 1} M(i) \right\}
\, .
\]
These families are clearly monotonically non-increasing with
\[
\sigma(M) = 1
\quad \text{and} \quad
c
\, \leq \,
\rho(M)
\, \leq \,
2 c
\]
for every
$M \in \Multisets([c])$.
hence they are legitimate families of potential $\sigma$-coefficients and
potential $\rho$-coefficients with
$\sigma_{0} = 1$
and
$\rho_{0} = 2 c$.

En route to establishing (\ref{equation:concrete-nodes:incremental:progress}),
we define the notation
\[
\lambda(\{ u, v \})
\, = \,
\begin{cases}
\rho(C[u]) + \rho(C[v]) \, , &
u, v \in \UnDecidedSet(C)
\\
0 \, , &
\text{otherwise}
\end{cases}
\]
and
\[
\lambda'(\{ u, v \})
\, = \,
\begin{cases}
\rho(C'[u]) + \rho(C'[v]) \, , &
u, v \in \UnDecidedSet(C')
\\
0 \, , &
\text{otherwise}
\end{cases}
\]
for each edge
$\{ u, v \} \in E(G)$.
This notation is extended to edge subsets
$F \subseteq E(G)$
by defining
$\sigma(F) = \sum_{e \in F} \sigma(e)$
and
$\sigma'(F) = \sum_{e \in F} \sigma'(e)$,
observing that
\[
\pi(G, C)
\, = \,
|V(H)| + \lambda(E(H))
\quad \text{and} \quad
\pi(G, C')
\, = \,
|\UnDecidedSet(C')| + \lambda'(E(H))
\, .
\]
We further observe that
\[
2 c
\, \leq \,
\lambda(e)
\, \leq \,
4 c
\]
for every edge
$e \in E(H)$.

Let
$U^{0} = \{ v \in V(H) \mid \Degree_{v}(H) = 0 \}$
be the set of nodes with no incident edges in $H$.
Since
$|V(H) - U^{0}| \leq 2 |E(H)|$,
it follows that
\[
|U^{0}| + \lambda(E(H))
\, \geq \,
\frac{c}{c + 1} \pi(G, C)
\, .
\]
The phase procedure $\PhaseProcedure_{\Alg}$ is designed so that each node in
$U^{0}$ is marked, and following that becomes decided, with probability $1$.
Therefore, to establish (\ref{equation:concrete-nodes:incremental:progress}),
it suffices to show that
\begin{equation} \label{equation:concrete-nodes:incremental:progress-lambda}
\Ex \left( \lambda'(E(H)) \right)
\, \leq \,
(1 - \beta') \cdot \lambda(E(H))
\end{equation}
for
$\beta' = \Omega (1 / c)$.

Recall (from \Sect{}~\ref{section:concrete-edges:incremental:algorithm}) that
a node
$v \in V(H)$
is good in $H$ if
$\left| \left\{
u \in N_{H}(v) : \Degree_{H}(u) \leq \Degree_{H}(v)
\right\} \right|
\geq
\Degree_{H}(v) / 3$
and that at least half of the edges in $E(H)$ are incident on a good node (at
least one).
Let $U^{g}$ be the set of nodes that are good in $H$ and for a good node
$v \in U^{g}$,
let $F_{v}$ be the set of edges incident on $v$ in $H$.
Since
$2 c < \lambda(e) \leq 4 c$
for every edge
$e \in E(H)$,
it follows that
\[
\lambda \left( \bigcup_{v \in U^{g}} F_{v} \right)
\, \geq \,
\frac{1}{3} \lambda(E(G))
\, .
\]
To establish (\ref{equation:concrete-nodes:incremental:progress-lambda}), we
fix a good node
$v \in U^{g}$
and prove that the following event occurs with probability
$\Omega (1)$:
\[
\text{event $S_{v}$:}
\qquad
\lambda'(F_{v})
\, \leq \,
\max \left\{
\frac{11}{12}, \frac{4 c - 1}{4 c}
\right\} \cdot \lambda(F_{v})
\, .
\]

Consider first that case that
\[
\sum_{i = 1}^{c - 1} C[v](i)
\, \geq \,
c - 1
\, .
\]
By \Obs{}~\ref{observation:concrete-nodes:incremental:correctness}, we know
that $v$ becomes decided with
$C'(v) = c$,
hence
$\lambda'(F_{v}) = 0$
which means that event $S_{v}$ occurs with probability $1$.

So, assume hereafter that
\[
\sum_{i = 1}^{c - 1} C[v](i)
\, < \,
c - 1
\, .
\]
Let
$U^{g}_{v} = \{
u \in N_{H}(v) \mid \Degree_{H}(u) \leq \Degree_{H}(v)
\}$,
recalling that
$|U^{g}_{v}| \geq \frac{1}{3} |N_{H}(v)|$,
and let
\[
U^{g, c}_{v}
\, = \,
\left\{
u \in U^{g}_{v} \mid \sum_{i = 1}^{c - 1} C[u](i) \geq c - 1 
\right\}
\, .
\]
\Obs{}~\ref{observation:concrete-nodes:incremental:correctness}
implies that if
$u \in U^{g, c}_{v}$,
then
$C'(u) = c$
and
$\lambda'(u) = 0$.
Therefore, if
$|U^{g, c}_{v}| > \frac{1}{2} |U^{g}_{v}|$,
then
\[
\lambda'(F_{v})
\, \leq \,
\lambda(F_{v}) - 2 c |U^{g, c}_{v}|
\, < \,
\lambda(F_{v}) - c |U^{g}_{v}|
\, \leq \,
\lambda(F_{v}) - \frac{c}{3} |F_{v}|
\, \leq \,
\lambda(F_{v}) - \frac{1}{12} \lambda(F_{v})
\, = \,
\frac{11}{12} \lambda(F_{v})
\, ,
\]
where
the first transition holds as
$\lambda(\{ u, v \}) \geq 2 c$
for every node
$u \in U^{g, c}_{v}$
and the fourth transition holds as
$\lambda(e) \leq 4 c$
for every edge
$e \in F_{v}$.
We conclude that if
$|U^{g, c}_{v}| > \frac{1}{2} |U^{g}_{v}|$,
then event $S_{v}$ occurs with probability $1$.

So, assume hereafter that
$|\overline{U}^{g, c}_{v}| \geq \frac{1}{2} |U^{g}_{v}|$,
where
$\overline{U}^{g, c}_{v} = U^{g}_{v} - U^{g, c}_{v}$.
For a node
$u \in \overline{U}^{g, c}_{v}$,
let $W_{u}$ be the event that $u$ is marked and no node in
$\{ w \in N_{H}(u) \mid \LogDeg(w) \geq \LogDeg(u) \}$
is marked (in step $1$ of
$\PhaseProcedure_{\Alg}$).
Since
$\Degree_{H}(u) \leq \Degree_{H}(v)$
for each node
$u \in \overline{U}^{g, c}_{v}$
and since
$|\overline{U}^{g, c}_{v}|
\geq
\frac{1}{6} \Degree_{H}(v)$,
we can use standard arguments (see, e.g., \cite{AlonBI1986fast}) to
conclude that with probability
$\Omega (1)$,
event $W_{u}$ occurs for at least one node
$u \in \overline{U}^{g, c}_{v}$.

Condition hereafter on event $W_{u}$ occurring
for a node
$u \in \overline{U}^{g, c}_{v}$.
The design of $\PhaseProcedure_{\Alg}$ ensures that
$C'(u) = i_{u}$
for some
$i_{u} \in \{ 1, \dots, c - 1 \}$.
hence
\[
\sum_{i = 1}^{c - 1} C'[v](i)
>
\sum_{i = 1}^{c - 1} C[v](i)
\, .
\]
Recalling that
\[
\sum_{i = 1}^{c - 1} C[v](i)
\, < \,
c - 1
\, ,
\]
we conclude that
$\rho(C'[v]) < \rho(C[v])$,
thus
$\lambda'(e) < \lambda(e)$
for every edge
$e \in F_{v}$.
Since
$\lambda(e) \leq 4 c$
for every edge
$e \in F_{v}$,
it follows that
\[
\lambda'(F_{v})
\, \leq \,
\frac{4 c - 1}{4 c} \lambda(F_{v})
\, ,
\]
yielding event $S_{v}$.

\subsection{MIS}
\label{section:concrete-problems-nodes:mis}
The following theorem is derived from
\Thm{}~\ref{theorem:concrete-nodes:incremental:direct} by plugging
$c = 2$.

\begin{theorem} \label{theorem:concrete-nodes:mis:application-incremental}
The MIS problem admits a randomized self-stabilizing algorithm that uses
messages of size
$O (\log\log \Delta)$
whose fully adaptive run-time is
$O (\log (k + \Delta))$.
\end{theorem}

\subsection{Maximal Node $c$-Coloring}
\label{section:concrete-problems-nodes:maximal-coloring}
In this section, we develop the self-stabilizing algorithms for maximal node 
$c$-coloring listed in
Table~\ref{table:transformer:concrete-problems} and establish their fully 
adaptive run-time and message size bounds.
We start with the following theorem.

\begin{theorem} \label{theorem:concrete-nodes:maximal:clone-graph}
The maximal node $c$-coloring problem admits a randomized
self-stabilizing
algorithm that uses messages of size
$O (c \log\log \Delta)$
whose fully adaptive run-time is
$O (\log (k + \Delta))$.
\end{theorem}

\begin{proof}
Consider a graph
$G \in \mathcal{U}$
and a positive integer
$\alpha \in \Integers_{> 0}$.
The \emph{clone} graph of $G_{\alpha}$ is the graph in which
$V(G_{\alpha}) = \{(v,i) \mid v \in V(G) ,i \in [\alpha] \}$
and
$E(G_{\alpha}) = \{ \{(v,i),(v,i')\} \mid v \in V(G), 1 \leq i < i' \leq 
\alpha  \}
\cup
\{ \{(v,i), (u,i)\} \mid \{v,u\} \in E(G), i \in [\alpha] \}$.
The clone graph was introduced in the seminal paper of
Luby~\cite{Luby1986simple}.
It is well known that the nodes of a communication network $G$, can 
simulate 
an execution of a fault free distributed algorithm on $G_{\alpha}$ for every
$\alpha \in \Integers_{> 0}$.
Moreover, if $\alpha$ is ``hard coded'' into the nodes of $G$ (i.e., cannot 
be
manipulated by the adversary), then the nodes of $G$ can simulate an 
execution
of a self-stabilizing algorithm on $G_{\alpha}$.
	
Consider some
$G \in \mathcal{G}$.
The self-stabilizing algorithm for maximal node $c$-coloring, is a 
simulation 
of our self-stabilizing MIS algorithm on $G_{c-1}$.
The output value of node
$v \in V(G)$
is set after all of the nodes
$(v ,i) \in V(G_{c-1})$
for
$1 \leq i \leq c-1$
are decided.
The output value is defined such that
node $v$ takes the color
$i$
if there exists
$(v, i)$
that is decided $\InMIS$. Otherwise, node $v$ takes the color $c$.
\end{proof}

\subsection{Node $c$-Coloring}
\label{section:concrete-problems-nodes:coloring}
In this section, we develop the self-stabilizing algorithms for node 
$c$-coloring listed in
Table~\ref{table:transformer:concrete-problems} and establish their fully 
adaptive run-time and message size bounds.
We start with the following theorem derived from
\Thm{}~\ref{theorem:concrete-nodes:maximal:clone-graph} by plugging
$c =\Delta + 2$.

\begin{theorem}
\label{theorem:concrete-nodes:coloring-tight:application-maximal-clone-graph}
The node $(\Delta + 1)$-coloring problem admits a randomized self-stabilizing
algorithm that uses messages of size
$O (\Delta \log \log \Delta)$
whose fully adaptive run-time is
$O (\log (k + \Delta))$.
\end{theorem}

The reminder of this section is dedicated to developing a self-stabilizing
algorithm for node
$\lceil (1 + \epsilon) \Delta \rceil$-coloring
Fix some constant
$\epsilon \geq 1/\Delta$
and assume, without loss of generality, that
$\epsilon \Delta \in \Integers_{> 0}$.
Recall that in the
node
$((1 + \epsilon) \Delta)$-coloring
the set of output values $\mathcal{O}$ is
$[(1 + \epsilon) \Delta]$.
We will show that the node 
$((1 + \epsilon) \Delta)$-coloring problem meets the
$(3, O (\log \Delta))$-phased-based
criterion
(Criterion~\ref{criterion:nodes:phase-based}),
respectful decisions criterion
(Criterion~\ref{criterion:nodes:respectful-decisions}),
and
$(\Omega (1), 1, 0)$-progress criterion
(Criterion~\ref{criterion:nodes:progress}).
The node
$((1 + \epsilon) \Delta)$-coloring
problem meets the
$(O (\log \Delta))$-detectability
criterion
(Criterion~\ref{criterion:nodes:detectability}) since each node shares its
output value with all of its neighbors.
\Thm{}~\ref{theorem:concrete-nodes:coloring-loose:direct} follows from
\Thm{}~\ref{theorem:nodes:main} as we
have already established in \Sect{}~\ref{section:LCL-eligibility-criteria}
that the problem meets the
gap freeness criterion (Criterion~\ref{criterion:common:gap-freeness})
and
the $0$-bounded propagation criterion
(Criterion~\ref{criterion:common:bounded-propagation}).

\begin{theorem} \label{theorem:concrete-nodes:coloring-loose:direct}
The node
$\lceil (1 + \epsilon) \Delta \rceil$-coloring
problem admits a randomized self-stabilizing algorithm that uses messages of
size
$O (\log \Delta)$
whose fully adaptive run-time is
$O (\log (k + \Delta))$.
\end{theorem}

\subsubsection{Implementing the Fault Free Algorithm}
\label{section:concrete-nodes:1+epsilon-coloring:algorithm}
The fault-free algorithm is denoted by $\Alg$ and it
uses messages of size
$\mu_{A}(\Delta) = O (\log \Delta)$.
Being a phase-based algorithm, $\Alg$ is recounted through its phase procedure
$\PhaseProcedure_{\Alg}$ working with $3$-step phases.
Let $G$ be the graph on which $\PhaseProcedure_{\Alg}$ is invoked and let
$C : V(G) \rightarrow \mathcal{O} \cup \{ \bot \}$
be the configuration associated with $\PhaseProcedure_{\Alg}$ at the beginning
of the execution.
Let
$H = G(\UnDecidedSet(C))$
be the subgraph induced on $G$ by the set of edge undecided in $C$.

Consider a node
$v \in V(H)$.
In step $1$, node $v$ chooses a color $c_{v}$
u.a.r. (uniformly at random) from the pallet
$\mathcal{O}$
and sends it to all of its undecided neighbors.
In the $2$ step (which is the decision step), node $v$ sets
$\OutReg_{v} = c_{v}$
only if $c_{v}$ differ from the color of all its decided neighbors and from
$c_{u}$ for every undecided neighbor $u$.
If all of $v$'s neighbors are decided, then $v$ sets
$\OutReg_{v}$ to some arbitrarily chosen non-conflicting color from the set
$\mathcal{O}$.

By design, algorithm $\Alg$ is decision-oblivious
phase-based algorithm, thus node $\lceil(1 + \epsilon)\Delta\rceil$-coloring 
meets the $(3, O (\log \Delta))$-phase based criterion
(Criterion~\ref{criterion:nodes:phase-based}) with respect to $\Alg$.
Consider a node
$v \in \UnDecidedSet(C)$
and suppose that $v$ become decided during the execution of
$\PhaseProcedure_{\Alg}$ so that
$v \in \DecidedSet(C')$.
The procedure is designed to guarantee that
$C'(v) = i \in \mathcal{O}$
implies that
$C(u) \neq i$
for every node
$u \in N_{G}(v)$.
Therefore, node $\lceil(1 + \epsilon)\Delta\rceil$-coloring also meets the
respectful decisions criterion
(Criterion~\ref{criterion:nodes:respectful-decisions}) with respect to $\Alg$.

\paragraph{Node Separable Potential Function.}
It remains to show that
node $\lceil(1 + \epsilon)\Delta\rceil$-coloring
meets the
$(\Omega (1), 1, 0)$-progress
criterion (Criterion~\ref{criterion:nodes:progress}).
To this end, we introduce a node separable potential function $\pi$ for
node $\lceil(1 + \epsilon)\Delta\rceil$-coloring
with top potential coefficients
$\sigma_{0} = 1$
and
$\rho_{0} = 0$
and prove that
\begin{equation} \label{equation:concrete-nodes:coloring:progress}
\Ex \left( \pi(G, C') \right)
\, \leq \,
(1 - \beta) \cdot \pi(G, C)
\end{equation}
for
$\beta = \Omega (1)$,
where the expectation is over $C'$.
The families
$\mathcal{I}^{\sigma}
=
\left\{ \sigma(M) \right\}_{M \in \Multisets(\{ \InMIS, \OutMIS \})}$
and
$\mathcal{I}^{\rho}
=
\left\{ \rho(M) \right\}_{M \in \Multisets(\{ \InMIS, \OutMIS \})}$
of potential coefficients for $\pi$ are defined by setting
\[
\sigma(M)
\, = \, 1
\quad \text{and} \quad
\rho(M)
\, = \, 0
\, .
\]
This family is clearly monotonically non-increasing with
$\sigma(\emptyset) = 1$
and
$\rho(\emptyset) = 0$
and
$\sigma(M) \geq 1$
for every
$M \in \Multisets(\mathcal{O})$,
hence it is a legitimate family of potential coefficients with
$\sigma_{0} = 1$
and
$\rho_{0} = 0$.

Consider a contently configured graph
$(G, C) \in \ContConfGraphs(\Problem)$
and let
$H = G(\UnDecidedSet(C))$
be the subgraph induced by the undecided nodes.
Let
$C' = \PhaseProcedure_{\Alg}(G, C)$
be the configuration obtained from applying the phase procedure
$\PhaseProcedure{}_{\Alg}$ to $G$ under node configuration $C$ and let
$v \in V(H)$.
We denote by
$A_{v}$
the event that node $v$ is decided under configuration $C'$.

For a neighbor
$u \in \DecidedSet(C) \cap N_{G}(v)$,
we denote by
$A_{v,u}$,
the event that node $v$ choose the same color during the execution of
$\PhaseProcedure_{\Alg}(G, C)$ as the color of neighbor $u$.
For a neighbor
$u \in \UnDecidedSet_{v}(C)$
we denote by
$A_{v,u}$
the event that nodes $v$ and $u$ choose the same color during the execution of
$\PhaseProcedure_{\Alg}(G, C)$.

It is easily verifiable that for every
$u \in N_{G}(v)$,
it holds that
$\Pr(A_{v,u}) \leq \frac{1}{(1+\epsilon) \Delta}$.
By the union bound we get that
$\Pr\left(\bigcup_{u \in N_{G}(v)} A_{v,u}\right) \leq \frac{1}{1+\epsilon}$.
We can conclude that
$\Pr(A_{v}) \geq 1-\frac{1}{1+\epsilon}$, hence
for every
$v \in V(H)$,
the probability that $v$ is decided under $C'$ is at least
$1-\frac{1}{1+\epsilon} = \Omega(1)$.

\section{An Improved MM Self-Stabilizing Algorithm}
\label{section:improved-MM}
The goal of this section is to present our $O(\log k)$-fully-adaptive
self-stabilizing MM algorithm referred to as \SSMM{}.
We define the LCL
$\Predicate$
of MM over the output values
$\mathcal{O} = \{ \Matched, \UnMatched \}$
such that for every
$M \in \Multisets(\mathcal{O})$
it holds that
(1)
$\Predicate(\Matched, M) = \true$
if and only if
$M(\Matched) = 0$; and
(2)
$\Predicate(\UnMatched , M) = \true$
if and only if
$M(\Matched) \geq 1$.

\subsection{Algorithm \SSMM{}}
\label{section:algorithm-ssmm}
We design \SSMM{} based on a slightly modified version of the classic maximal
matching (fault-free) algorithm of Israeli and Itai~\cite{IsraeliI1986fast} 
which is phase-based with phase length $\phi=3$.
Algorithm \SSMM{} uses the \PPS{} module presented in
\Sect{}~\ref{section:probabilistic-phase-synchronization} to simulate the steps
of the fault-free algorithm.

The key ``self-stabilizing'' feature of algorithm \SSMM{} lies in procedure
\ForceConsistency{} (see Algorithm~\ref{pseudocode:force-consistency})
which is invoked at the beginning of the local computation stage of every round.
The goal of this procedure is to set values to the output registers of the 
nodes according to messages they receive.
Algorithm \SSMM{} runs on an input graph
$G \in \mathcal{U}$
and is presented from the perspective of a node
$v \in V(G)$,
recalling that the messages sent to (resp., from) $v$
from (resp., to) a neighbor
$u \in N_{G}(v)$
is denoted by $\InMsg_{v}(u)$ (resp., $\OutMsg_{v}(u)$).

\begin{algorithm}[!t]
\caption{\label{pseudocode:force-consistency} \ForceConsistency{}(),
code for node $v$}
\begin{algorithmic}[1]
\State{$M \leftarrow \{u \in N(v) \mid \OutReg_{v}(u) = \Matched \}$}
\State{$\m_{v} \leftarrow$ arbitrary chosen neighbor from $M$}
\If{$|M| = 0$}
\ForAll{$u \in N(v)$}
\If{$\InMsg_{v}(u).\fieldConsistency = \MatchedToOther$}
\State{$\OutReg_{v}(u) \leftarrow \UnMatched$}
\Else
\State{$\OutReg_{v}(u) \leftarrow \bot$}
\EndIf
\EndFor
\If{$\forall u \in N(v), \OutReg_{v}(u) = \UnMatched$}
\State{$\m_{v} \leftarrow \UnMatched$}
\Else
\State{$\m_{v} \leftarrow \bot$}
\EndIf
\Else
\If{$\InMsg_{v}(\m_{v}).\fieldConsistency \neq \Matched$}
\State{$\m_{v} \leftarrow \bot$}
\ForAll{$u \in N(v)$}
\State{$\OutReg_{v}(u) \leftarrow \bot$}
\EndFor
\Else
\ForAll{$u \in N(v) - \{\m_{v}\}$}
\State{$\OutReg_{v}(u) \leftarrow \UnMatched$}
\EndFor
\EndIf
\EndIf
\end{algorithmic}
\end{algorithm}

\begin{algorithm}[!t]
\caption{\label{pseudocode:finalize}
\Finalize{}(), code for node $v$}
\begin{algorithmic}[1]
\If{$\m_{v} = \bot$ or $\m_{v} = \UnMatched$}
\ForAll {$u \in N(v)$}
{$\OutMsg_{v}(u). \fieldConsistency \leftarrow \OutReg_{v}(u)$}
\EndFor
\Else
\State{$\OutReg_{v}(\m_{v}) \leftarrow \Matched$}
\State{$\OutMsg_{v}(\m_{v}). \fieldConsistency \leftarrow 
\Matched$}
\ForAll {$u \in N(v)-\{\m_{v}\}$}
\State{$\OutReg_{v}(u) \leftarrow \UnMatched$}
\State{$\OutMsg_{v}(u). \fieldConsistency \leftarrow \MatchedToOther$}
\EndFor
\EndIf
\State{call $\AdvancePPS$}%
\Comment{updates the register $\Step_{v}$}
\ForAll{$u \in N(v)$} {$\OutMsg_{v}(u).\fieldPPS \leftarrow \Step_{v}$ }
\EndFor
\end{algorithmic}
\end{algorithm}

\begin{algorithm}
\caption{\label{pseudocode:ssmm} \SSMM{}(), code for node $v$}
\begin{algorithmic}[1]
\State{$\ForceConsistency()$}
\If {$\m_{v} = \bot$}
\If {$\Step_{v} = 0$}
\State {$\ChosenNeighbor_{v} \leftarrow \NIL$}
\State {$\Peers_{v} \leftarrow \{u \in N(v)
\mid
\OutReg_{v}(u) = \bot
\land
\InMsg_{v}(u). \fieldPPS = 0
\land
\InMsg_{v}(u). \fieldConsistency = \bot\}$}
\State {$\Active_{v}  \leftarrow \true \,\, w.p.\ 1/2; \, \false \,\, 
w.p.\
1/2$}
\If {$\Active_{v}$ and $|\Peers_{v}| > 0$}

\State {$\ChosenNeighbor_{v} \leftarrow$
node picked from $\Peers_{v}$ uniformly at random}
\State {$\OutMsg_{v}(\ChosenNeighbor_{v}). \fieldMM \leftarrow 
\MatchingRequest$}
\EndIf
\EndIf
\If {$\Step_{v} = 1$}
\If {$\Active_{v} = false$}
\State {$\Peers_{v} \leftarrow
\{ u \in \Peers_{v} \mid 
\InMsg_{v}(u). \fieldMM
= 
\MatchingRequest \}$}
\If{$|\Peers_{v}| > 0$}
\State {$\ChosenNeighbor_{v} \leftarrow$ node picked arbitrarily from 
$\Peers_{v}$}
\State {$\OutMsg_{v}(\ChosenNeighbor_{v}). \fieldMM \leftarrow \Accept$}
\EndIf
\EndIf
\EndIf
\If {$\Step_{v} = 2$}
\If {$\Active_{v}$}
\If{$\ChosenNeighbor_{v} \neq \NIL$ and
$\InMsg_{v}(\ChosenNeighbor_{v}). \fieldMM = \Accept$}
{$\m_{v} \leftarrow \ChosenNeighbor_{v}$}
\EndIf
\ElsIf {$\ChosenNeighbor_{v} \neq \NIL$}
{$\m_{v} \leftarrow \ChosenNeighbor_{v}$}
\EndIf
\EndIf
\EndIf
\State {$\Finalize{}()$}\label{line:ssmm:finalize}
\end{algorithmic}
\end{algorithm}

\subsection{Analysis}
\label{section:logk-mm:analysis}
Our goal in this section is to prove the following theorem.

\begin{theorem} \label{theorem:logk-mm}
Algorithm \SSMM{} is a self-stabilizing MM algorithm that uses
messages of size
$O (1)$
whose fully adaptive run-time is
$O (\log k)$.
\end{theorem}

Recall that the adversarial manipulations may include the addition/removal of
nodes/edges, hence the graph may change during the round interval
$[t^{\circ}, t^{*} - 1]$.
For
$t \geq t^{\circ}$,
let $G_{t}$ be the graph at time $t$.
In what follows, we reserve $G$
for the graph that exists at time $t^{*}$, recalling that this is also the
graph at any time
$t > t^{*}$.

We redefine the register values at time $t$ to be the register values after
the execution of procedure \ForceConsistency{} in round $t$.
As a direct result, the configuration of \SSMM{} at time
$t \geq t^{\circ}$
is the edge configuration
$C_{t} : E(G_{t}) \rightarrow \mathcal{O} \cup \{ \bot \}$,
defined by setting
$C_{t}(\{ u, v \}) = \OutReg_{v, t}(u) = \OutReg_{u, t}(v)$
if edge
$\{ u, v \} \in E(G_{t})$
is port-consistent at time $t$ and
$C_{t}(\{ u, v \}) = \bot$
otherwise.

Assume that \SSMM{} is in a legal configuration at time $t^{\circ}$ and that
the adversary manipulates
$k > 0$
nodes during the round interval
$[t^{\circ}, t^{*} - 1]$
and does not manipulate any node from round $t^{*}$ onward.
Our goal in this section is to establish \Thm{}~\ref{theorem:logk-mm} by 
proving
that \SSMM{} stabilizes to a legal configuration by time
$t^{*} + O (\log k)$
in expectation.
Recall that \SSMM{} uses the \PPS{} module with $\phi=3$.

\subsubsection{Roadmap}
\label{section:logk-mm:analysis-roadmap}
This section presents the general structure of \SSMM{}'s analysis.
It hinges on \Prop{}
\ref{proposition:logk-mm:un/matched-remains-un/matched}--
\ref{proposition:logk-mm:degrees-become-small}
whose combination yields \Thm{}~\ref{theorem:logk-mm}.
We say that a node
$v \in V(G_{t})$
is \emph{matched} at time
$t \geq t^{\circ}$
if there exists a neighbor
$u \in N_{G_{t}}(v)$
such that
$C_{t}(\{v,u\}) = \Matched$
and
$C_{t}(\{v, w\}) \neq \Matched$
for all 
$w \in N_{G_{t}}(v) - \{u\}$.
We say that a node
$v \in V(G_{t})$
is \emph{unmatched} at time
$t \geq t^{\circ}$
if for all
$u \in N_{G_{t}}(v)$
it holds that,
$C_{t}(\{v, u\}) = \UnMatched$
and
$u$ is matched.
Notice that edge configuration $C_{t}$ is legal if and only if all the nodes
are either matched or unmatched.
\Prop{}~\ref{proposition:logk-mm:un/matched-remains-un/matched}
ensuring that once the algorithm reaches a legal configuration, it stays in a
legal configuration.

\begin{proposition}
\label{proposition:logk-mm:un/matched-remains-un/matched}
For every
$t \geq t^{*} + 1$
and
$v \in V(G)$,
if $v$ is matched (resp., unmatched) at time $t$, then $v$ is
matched (resp., unmatched) at time
$t + 1$.
\end{proposition}

The rest of the analysis is devoted to showing that it takes
$O (\log k)$
rounds in expectation for the set
$\UnDecidedSet(C_{t^{*} + 1})$
of undecided edges at time $t^{*} + 1$
to become decided.
Let
$\tau = O (\phi^{3})$
be the parameter promised in
\Lem{}~\ref{lemma:pps:start-phase-probability}.
The next proposition implies that \SSMM{} stabilizes in
$O \left(
\log |E| + \log k
\right)$
rounds in expectation. This means in particular that it takes \SSMM{}
$O (\log k)$
rounds to stabilize in expectation once it reaches a configuration
with $\operatorname{poly}(k)$ undecided edges.

\begin{proposition}
\label{proposition:logk-mm:edge-removal}
Fix some time $t \geq t^{*} + 1$ and a global state of \SSMM{} at time $t$.
There exists a universal positive constant $\alpha$ such that
$\Ex\left(|\UnDecidedSet(C_{t+\tau+\phi})|\right) \leq 
(1-\frac{\alpha}{\phi^{2}}) 
\cdot 
|\UnDecidedSet(C_{t})|$.
\end{proposition}

Unfortunately, by manipulating $k$ nodes, the adversary can lead the algorithm
to a graph $G$ whose undecided edge set is arbitrarily large (and not 
polynomially bounded) with respect to $k$.
To resolve this obstacle, we prove that it takes the algorithm
$O (\log k)$
rounds in expectation to reduce the number of undecided edges in $G$ down to
$O (k^{2})$.
This relies on the following proposition derived from a simple combinatorial
property of MM.

\begin{proposition}
\label{proposition:logk-mm:three-sets}
Let
$S = \{ v \in V(G) \mid v \text{ is matched at time } t^{*} + 1 
\}$.
There exists a node set
$K \subseteq V - S$
of size
$|K| \leq 2 k$
such that
$I = V(G) - (S \cup K)$
is an independent set in $G$.
\end{proposition}

Consider the sets $K$ and $I$ promised in 
\Prop{}~\ref{proposition:logk-mm:three-sets}
and some positive constant
$\xi = \xi(\phi)$
whose value is determined later on.
For
$t \geq t^{*} + 1$,
let
$H_{t} = G(\UnDecidedSet(C_{t}))$
be the subgraph induced by the undecided edges at time $t$ and let
$P_{t} = \{ v \in K \cap V(H_{t})\} \mid \Degree_{H_{t}}(v) 
\geq \xi k \}$
and let $T$ be the random variable that takes on the earliest time $t$ such
that
$P_{t} = \emptyset$.
Since $I$ is an independent set, every edge in $E(H_{T})$ is incident on at 
least one vertex in $K$, thus
$|E(H_{T})| < |K| \cdot \xi k \leq 2 \xi k^{2}$.
The proof of
\Thm{}~\ref{theorem:logk-mm} is completed due to the following proposition
implying, by standard probabilistic arguments that
$T \leq O (\log k)$
in expectation.

\begin{proposition}
\label{proposition:logk-mm:degrees-become-small}
Fix some time
$t \geq t^{*} + 1$
and global state of \SSMM{} at time $t$.
There exists a universal positive constant $\delta$ such that
$\Pr \left( v \notin P_{t + 2 \phi} \right) \geq \delta 2^{-\phi}$
for every node
$v \in P_{t}$.
\end{proposition}

\subsubsection{Fault recovery}
\label{section:logk-mm:fault-recovery}
This section is dedicated to proving 
\Prop{}~\ref{proposition:logk-mm:un/matched-remains-un/matched}.
The proof directly follows from the following two structural observations 
regarding the operation of \SSMM{}.

\begin{observation}\label{observation:logk-mm:matched-remains-matched}
For every
$t \geq t^{\circ}$
and
$\{v, u\} \in V(G_{t})$,
if
(1) $v$ and $u$ were not manipulated by the adversary at time $t$;
(2) $v$ and $u$ are matched; and
(3) $C_{t}(\{v, u\}) = \Matched$,
then $v$ and $u$ are matched at time $t+1$ and
$C_{t+1}(\{v, u\}) = \Matched$.
\end{observation}
\sloppy
\begin{proof}
Since
$C_{t}(\{v, u\}) = \Matched$,
it holds that
$\OutReg_{v, t}(u) = \OutReg_{u, t}(v) = \Matched$,
$\m_{v, t} = u$,
and
$\m_{u, t} =v$.
The adversary did not manipulate $v$ nor $u$ at round $t$, hence
by procedure \Finalize{}
$\InMsg_{v, t+1}. \fieldConsistency = \Matched$,
$\InMsg_{u, t+1}(v). \fieldConsistency = \Matched$,
$\OutReg_{v, t+1}(w) = \UnMatched$
for every
$w \in N_{G_{t+1}}(v) - \{u\}$,
and 
$\OutReg_{u, t+1}(w) = \UnMatched$
for every
$w \in N_{G_{t+1}}(u) - \{v\}$
which implies that $v$ and $u$ are matched at time $t + 1$
and
$C_{t+1}(\{v, u\}) = \Matched$.
\end{proof}
\par\fussy

\begin{observation}
\label{observation:logk-mm:unmatched-remains-unmatched}
For every $t \geq t^{*} + 1$ and a node $v \in V(G)$, if $v$ is unmatched at
time $t$, then $v$ is unmatched at time $t + 1$.
\end{observation}

\begin{proof}
Node $v$ is unmatched at time $t$ which means that any neighbor
$u \in N_{G}(v)$
is matched and
$C_{t}(\{v, u\}) = \UnMatched$.
By \Obs{}~\ref{observation:logk-mm:matched-remains-matched} any neighbor
$u \in N_{G}(v)$
will remain matched (to a neighbor $w \in N_{G}(u)$ such that $w \neq v$)
and by procedure \Finalize{} will send $\MatchedToOther$ to $v$ which 
implies that $v$ remains unmatched.
\end{proof}

\subsubsection{Progress}
\label{section:log-k-mm-progress}
Our goal in this section is to establish \Prop{}
\ref{proposition:logk-mm:edge-removal}--
\ref{proposition:logk-mm:degrees-become-small},
starting with some additional definitions.
For every time
$t \geq t^{*} + 1$,
let
$S^{j}_{t} = \{v \in V(H_{t}) \mid \Step_{v, t} = j \}$
be the subset of nodes in $H_{t}$ with
$\Step_{v, t} = j$
for
$j \in \HoldSymbol \cup \{0, \cdots, \phi - 1\}$
recalling that
$\Step_{v, t}$
is the state of $v$'s \PPS{} module at time $t$.
Let
$\widetilde{H}_{t} = H_{t}(S^{0}_{t})$
be the graph induced on $H_{t}$ by $S^{0}_{t}$, i.e., the subgraph of $H_{t}$
induced by the nodes that start a phase in round $t$.

Consider some graph
$H$.
Node
$v \in V(H)$
is said to be \emph{good} in $H$ if at least
$1 / 3$
of its neighbors $u$ in $H$ have degree
$\Degree_{H}(u) \geq \Degree_{H}(v)$.
The following lemma is established by Alon et
al.~\cite[\Lem{}~4.4]{AlonBI1986fast}.

\begin{lemma}\label{lemma:logk-mm:half-of-the-edges-are-good}
Let $H$ be an undirected graph.
At least
$1 / 2$
of the edges in $H$ are incident on a good node.
\end{lemma}

\Lem{}~\ref{lemma:logk-mm:half-of-the-edges-are-good} is exploited in the
following two lemmas to show that sufficiently many edges becomes decided with a
sufficiently high probability;
\Prop{}~\ref{proposition:logk-mm:edge-removal} follows by a standard Markov 
inequality argument.

\begin{lemma} \label{lemma:logk-mm:good-matched-probability}
Fix some time $t \geq t^{*} + 1$ and the global state of \SSMM{} at time 
$t$.
There exists a universal positive constant $p_{g}$ such that
$\Pr\left(v \notin V(H_{t + \phi}) \right) \geq p_{g}$
for every good node $v \in V(\widetilde{H}_{t})$.
\end{lemma}
\begin{proof}
Denote by $d$ the degree of node $v$ in $\widetilde{H}_{t}$, i.e.,
$d=\Degree_{\widetilde{H}_{t}}(v) > 0$.
Node $v$ is good, thus there exist
$u_{1}, \dots, u_{\lceil d/3 \rceil} \in N_{\widetilde{H}_{t}}(v)$ such that
$d_{i} = \Degree_{\widetilde{H}_{t}}(u_{i}) \leq d$
for every
$1 \leq  i \leq \lceil d/3 \rceil$.
Recall that by the definition of $\widetilde{H}_{t}$, nodes $v$ and
$u_{1}, \dots, u_{\lceil d/3 \rceil}$
start a phase (in synchrony) at time $t$.
Node $v$ marks itself as passive in step $0$ of the phase with
probability
$1 / 2$;
condition hereafter on this event.
For
$1 \leq  i \leq \lceil d/3 \rceil$,
let $B_{i}$ be the event that $u_{i}$ marks itself as active and sends a
$\MatchingRequest$ message to $v$ in step $1$ of the phase, noticing
that
$\Pr(B_{i}) = \frac{1}{2 d_{i}} \geq \frac{1}{2d}$.
Since the events $B_{i}$ are independent, it follows that the probability 
that
none of them occurs is
up-bounded by
$(1 - 1 / (2 d))^{\lceil d / 3 \rceil} < e^{-1 / 6}$.
The assertion follows since the occurrence of any of the events $B_{i}$ 
implies
that $v$ becomes matched by the end of the phase that lasts for
$\phi$ rounds.
\end{proof}

\begin{proof}[Proof of \Prop{}~\ref{proposition:logk-mm:edge-removal}]
Denote
$m=|\UnDecidedSet(C_{t})|$
and recall that
$H_{t} = G(\UnDecidedSet({C_{t}}))$.
Let
$E_{0} = \{\{u,v\} \in E(G) \mid v,u \in S^{0}_{t + \tau} \}$
be the set of edges that both endpoints are is state $0$ of the \PPS{} 
module at time $t +\tau$..
For every edge
$e = \{v, u\} \in \UnDecidedSet(C_{t})$,
let $A_{e}$ be the event that
$e \in E_{0}$.
By \Lem{}~\ref{lemma:pps:start-phase-probability},
we have
\begin{equation}\label{eq:edge-start-phase-prob}
\Pr\left(A_{e}\right)
\geq
\frac{1}{4\phi^{2}}.
\end{equation}
Hence, 
\begin{equation}\label{eq:expected-start-phase-edges}
\Ex\left(|E_{0}|\right) \geq \frac{1}{4\phi^{2}}m.
\end{equation}

After $E_{0}$ was determined, we revel the set $E_{0}$ to an adversary. 
Conditioned on the event that $E_{0}$ is already determined, the adversary 
is allowed to determine the coin tosses of all the nodes
during the round interval
$[t, t + \tau - 1]$.

We partition the set $E_{0}$ into two (possibly empty) disjoint sets,
$E_{0} \cap E(H_{t+\tau})$
and
$E^{c}_{0} = E_{0} \cap \left(E(H_{t}) - E(H_{t + \tau}) \right)$.
Notice that by definition
$E(\widetilde{H}_{t + \tau}) = E_{0} \cap E(H_{t+ \tau})$.	
By \Lem{}~\ref{lemma:logk-mm:half-of-the-edges-are-good},
in the (random) graph $\widetilde{H}_{t+\tau}$ at least half of edges are
incident on at least one good node. Every good node in
$\widetilde{H}_{t+ \tau}$
is removed with probability at least $p_{g}$, where $p_{g}$ is the constant
promised by \Lem{}~\ref{lemma:logk-mm:good-matched-probability}. Hence, the 
expected number of removed edges in the round interval
$[t+ \tau,t+ \tau + \phi]$
can be low-bounded as follows
\begin{align*}
\Ex\left(|E(H_{t+\tau}) - E(H_{t+\tau+\phi})|\right)
\, = \, &
\Ex\left(\Ex\left(|E(H_{t+\tau}) - 
E(H_{t+\tau+\phi})| \bigm|
|E(\widetilde{H}_{t+\tau})|\right)\right)
\\
\geq \, &
\frac{1}{2}\cdot p_{g} \cdot 
\Ex\left(|E(\widetilde{H}_{t+\tau})|\right).
\end{align*}
By definition, 
$|E_{0}|=|E^{c}_{0}| + |\widetilde{H}_{t+\tau}|$, hence
$\Ex\left(|E_{0}|\right)=\Ex\left(|E^{c}_{0}|\right) +
\Ex\left(|\widetilde{H}_{t+\tau}|\right)$ and by
\Eq{}~\ref{eq:expected-start-phase-edges},
$\Ex\left(|E^{c}_{0}|\right) + \Ex\left(|\widetilde{H}_{t+\tau}|\right)
\geq \frac{1}{4\phi^{2}}m$.
We complete the proof by low-bounding the expected number of edges removed 
in the round interval
$[t, t + \tau + \phi]$
by
\begin{equation*}
\begin{split}
\Ex\left(|E(H_{t}) - E(H_{t + \tau + \phi})| \right)
&\geq
\Ex\left(|E^{c}_{0}|\right) + \Ex\left(|E(H_{t+\tau}) - 
E(H_{t + \tau + \phi})|\right) \\
&\geq
\Ex\left(|E^{c}_{0}|\right) + \frac{1}{2} \cdot p_{g} \cdot 
\Ex\left(|\widetilde{H}_{t+\tau}|\right)\\
& =
\frac{1}{2} \cdot p_{g}\left(\Ex\left(|E^{c}_{0}|\right) + 
\Ex\left(|\widetilde{H}_{t+\tau}|\right)\right) \\
& =
\frac{1}{8\phi^{2}}\cdot p_{g} \cdot m.
\end{split}
\end{equation*}
\end{proof}

We now turn to establishing \Prop{}~\ref{proposition:logk-mm:three-sets}.

\begin{proof}[Proof of \Prop{}~\ref{proposition:logk-mm:three-sets}]
Let $A$ be the set of nodes in
$V(G) - S$
that are manipulated (by the adversary) during the round interval
$[t^{\circ}, t^{*} - 1]$.
Let $R$ be the set of nodes in
$V(G) - (S \cup A)$
that are matched at time $t^{\circ}$, referred to hereafter as
\emph{orphans}.
We define
$K = A \cup R$
and establish the assertion by proving that
(1)
$|K| \leq 2 k$;
and
(2)
$I = V(G) - (S \cup K)$
is an independent set in $G$.

For every orphan
$v \in R$,
let $w(v)$ be the node with which $v$ is matched at time 
$t^{\circ}$
, i.e., $C_{t^{\circ}} (\{v, w(v)\}) = \Matched$.
Note that $w(v)$ does not necessarily exist in $G$ as it may have been 
removed during the round interval
$[t^{\circ}, t^{*} - 1]$.
Since $v$ and $w(v)$ are no longer matched at time $t^{*} + 1$ and
since $v$ is not manipulated during the round interval
$[t^{\circ}, t^{*} - 1]$,
\Obs{}~\ref{observation:logk-mm:matched-remains-matched} implies
that $w(v)$ must be manipulated during that time interval.
We conclude that
$|R| \leq k$
as the mapping defined by $w$ is injective.
The bound
$|K| \leq 2 k$
follows since
$|A| \leq k$.

It remains to show that $I$ is an independent set in $G$.
This is done by arguing that every node
$v \in I$
is unmatched at time $t^{\circ}$.
This establishes the assertion recalling that by definition, the nodes in 
$I$
are not manipulated during the round interval
$[t^{\circ}, t^{*} - 1]$,
hence the adversary does not introduce new edges in
$I \times I$.
To that end, assume by contradiction that there exists some node
$v \in I$
that is not unmatched at time $t^{\circ}$.
Since \SSMM{} is in a legal configuration at time $t^{\circ}$, it follows 
that $v$ must be strongly matched at that time.
But by definition, the nodes in
$V(G) - S$
that are matched at time $t^{\circ}$ belong to either $R$ or $A$, 
in contradiction to
$v \in I = V(G) - (S \cup A \cup R)$.
\end{proof}

\begin{corollary}\label{corollary:deg-v-3k-good}
Fix some time $t \geq t^{*} + 1$ and the global state of \SSMM{} at time 
$t$.
If
$v \in V(\widetilde{H}_{t}) \cap K$
and
$\Degree_{\widetilde{H}_{t}}(v) \geq 3k$,
then $v$ is good in $\widetilde{H}_{t}$.
\end{corollary}

\begin{proof}
Let $d =\Degree_{\widetilde{H}_{t}}(v)$. Node $v$ has at most
$|K| - 1 < 2 k$ neighbors from the set $K$,
thus at least
$d - 2 k \geq (1 / 3) d$
of $v$'s neighbors in
$\widetilde{H}_{t}$
belong to
$V(\widetilde{H}_{t}) - K = I \cap V(\widetilde{H}_{t})$.
By \Prop{}~\ref{proposition:logk-mm:three-sets}, every node $u \in I \cap
V(\widetilde{H}_{t})$ has degree at most $2k$ which completes the proof.
\end{proof}

The analysis is completed by establishing
\Prop{}~\ref{proposition:logk-mm:degrees-become-small}.

\begin{proof}[Proof of \Prop{}~\ref{proposition:logk-mm:degrees-become-small}]
For every node $v \in V(H_{t})$, let $t \leq t(v)$ be the first time after
time $t$ such that
$\Step_{v,t(v)} = \HoldSymbol$.
According to the definition of the \PPS{} module it must hold that
$t \leq t(v) \leq t + \phi$
and notice that $t(v)$ is fully determined by $\Step_{v,t}$. Denote by
$A_{v}$
the event that
$\Step_{v, t'} = \HoldSymbol$
for every
$t(v) \leq t' \leq t + \phi$.
By \Obs{}~\ref{observation:transformer:step-counter}, it holds that the 
event $A_{v}$ is independent of any coin toss of $v$ prior to time
$t(v)$ and of any coin toss of all other nodes and that
\begin{equation}\label{eq:logk-mm:remain-in-Hold-prob}
\Pr\left(A_{v}\right) \geq 2^{-\phi}.
\end{equation}

For every node $v \in V(H_{t})$, we augment the power of the adversary by 
allowing it choose the outcome of any coin toss in the round interval
$[t,t(v)-1]$.
Notice that this adversary can only choose the outcome of coin tosses that 
are within a phase and cannot choose the outcome of coin tosses of the 
\PPS{} module.
Let
$S = \{v \in V(H_{t}) \mid v \in V(H_{t(v)}) \}$
be the set of nodes
$v \in V(H_{t})$
that are not matched or unmatched at time $t(v)$ and let
$H^{S} = H_{t}(S)$ be the 
graph induced on $H_{t}$ by $S$.

Fix some node $v \in P_{t}$. If node $v$ is matched or unmatched at time 
$t(v)$ or
$\Degree_{H^{S}}(v) < \xi k$,
then
$v \notin P_{t + 2\phi}$
with probability $1$. Otherwise ($v$ is not matched or unmatched at time 
$t(v)$ 
and more than $\xi k$ of its neighbors are in $S$), we will show that with
probability
$\Omega(2^{-\phi})$
node $v$ is matched at time
$t+2\phi$
which implies that $v \notin P_{t + 2\phi}$.

Let $B$ be the event that
$\Degree_{\widetilde{H}_{t+\phi}}(v) \geq 3k$.
We start by showing that there exists
$\alpha = \alpha(\phi)$
such that
$\Pr\left(B\right) \geq \alpha$.
For every
$u \in S$,
let $\widetilde{A}_{u}$ be the event that $A_{u}$ occurred and
$\Step_{v,t+\phi} = 0$.
By \Eq{}~\ref{eq:logk-mm:remain-in-Hold-prob}
and
\Obs{}~\ref{observation:transformer:step-counter}
we conclude that
$\Pr\left(\widetilde{A}_{u}\right) = \Pr\left(A_{u}\right) \cdot (1/2) \geq 
2^{-(\phi+1)}$.
Moreover, the events in $\{\widetilde{A}_{u} \bigm| u \in S\}$ are 
independent.

Denote $d = \Degree_{H^{S}}(v)$ and let $u_{1}, \dots, u_{d}$ be the 
neighbors
of $v$ is $H^{S}$. Let $Y$ be the random variable that counts the number of 
occurrences of events
$\widetilde{A}_{u_{i}}$.
Notice that if $Y \geq 3k$ and $\widetilde{A}_{v}$ occurs, then event $B$ 
occur;
moreover, events
$Y \geq 3k$
and
$\widetilde{A}_{v}$ are independent and
$\Ex\left(Y\right) \geq \xi k 2^{-(\phi+1)}$
since
$d \geq \xi k$.

By choosing
$\xi = 4/2^{-(\phi + 1)} = 2^{\phi+3}$
and applying Chernoff's (lower tail) bound we conclude that
\begin{equation*}
\begin{split}
\Pr\left(Y<3k\right)
=
\Pr \left( Y<(1-1/4) \cdot \xi k 2^{-(\phi+1)} \right)
\leq
\Pr\left(Y<(1-1/4) \cdot \Ex\left(Y\right)\right)
<
e^{-k/8}
\leq e^{-1/8} \, .
\end{split}
\end{equation*}
Thus
$\Pr\left(B\right) \geq (1-e^{-1/8})\cdot 2^{-(\phi+1)}$.
By \Cor{}~\ref{corollary:deg-v-3k-good}
and
\Lem{}~\ref{lemma:logk-mm:good-matched-probability},
occurrence of $B$ implies that $v$ is good in $\widetilde{H}_{t + \phi}$
and a good node is removed with at least a constant probability by the end
of the phase, i.e., at time
$t + 2\phi$.
We conclude that
\begin{equation*}
\begin{split}
\Pr\left(v \notin P_{t + 2\phi}\right)
& = \Pr\left(B \land v \notin V(H_{t + 2\phi}) \right) \\
& \geq \Pr\left(v \notin V(H_{t + 2 \phi}) \bigm| B\right) \cdot
\Pr\left(B\right) = p_{g} \cdot (1-e^{-1/8})\cdot 2^{-(\phi+1)} \, ,
\end{split}
\end{equation*}
thus establishing the assertion.
\end{proof}

\section{Infinitely Many Faults}
\label{section:infinite-faults}
Consider a node/edge-LCL
$\Problem = \langle
\mathcal{O}, \mathcal{G}, \Predicate
\rangle$
and let $\SelfStabAlg$ be a self-stabilizing algorithm for $\Problem$
generated by our transformer.
Recall that the graph
$G \in \mathcal{G}$
on which $\SelfStabAlg$ runs may be (countably) infinite and the fully
adaptive run-time guarantees promised in \Thm{} \ref{theorem:edges:main} and
\ref{theorem:nodes:main} hold if the adversary manipulates a finite node
subset.
Given the deterministic
$\psi + \phi + 2$
bound on the ``radius of influence'', we can actually allow the adversary to
manipulate an infinite node subset
$S \subset V(G)$
as long as there exist node clusters
$U_{1}, U_{2}, \dots \subseteq V(G)$
such that \\
(1)
$S \subset \bigcup_{i \geq i} U_{i}$;
\\
(2)
$U_{i} \cap S = k_{i}$,
where
$k_{i} \in \Integers_{> 0}$,
for every
$i \geq 1$;
and \\
(3)
if
$v \in U_{i} \cap S$,
then
$\Distance_{G}(v, V(G) - U_{i}) \geq \Omega (\psi + \phi)$.
\\
\Thm{} \ref{theorem:edges:main} and \ref{theorem:nodes:main} guarantee that
the expected stabilization time of
$G(U_{i})$ is at most
\[
O \left(
\left( \phi^{5} / \beta \right)
\cdot
\left(
\log (k_{i}) + (\psi + \phi) \log (\Delta) + \log (\sigma_{0})
\right)
\right)
\]
for each
$i \geq 1$.

\section{Non-Eligible LCL Problems}
\label{section:non-eligibile-problems}
As an additional insight into the power and limitations of the transformer
developed in the current paper, it may be helpful to state an interesting
necessary condition for a node/edge-LCL
$\Problem = \langle
\mathcal{O}, \mathcal{G}, \Predicate
\rangle$
to be eligible (with any combination of eligibility parameters).
It turns out that $\Problem$ cannot be eligible if there exist a graph
$G \in \mathcal{G}$
and an incomplete node/edge configuration
$C : X(G) \rightarrow \mathcal{O} \cup \{ \bot \}$
such that
\\
(1)
$C$ is content;
and
\\
(2)
any configuration
$C' : X(G) \rightarrow \mathcal{O} \cup \{ \bot \}$,
with
$\DecidedSet(C') \supset \DecidedSet(C)$,
that agrees with $C$ on every
$x \in \DecidedSet(C)$
is not content.

To see why our transformer cannot be applied to $\Problem$, notice that on the
one hand, as $C$ is content, the detection procedure does not identify
anything wrong and should therefore return $\true$ for every output register.
On the other hand, as $C$ cannot be augmented in a content manner, we know
that an application of the phase procedure cannot assign any new output values
without violating the respectful decisions criterion (Criteria
\ref{criterion:edges:respectful-decisions} or
\ref{criterion:nodes:respectful-decisions}).
More concretely, the
$(\beta, \sigma_{0})$-criterion
(Criterion~\ref{criterion:edges:progress})
or
$(\beta, \sigma_{0}, \rho_{0})$-criterion
(Criterion~\ref{criterion:nodes:progress})
cannot be met for any
$\beta > 0$.
Examples for LCL problems that belong to this non-eligible class include
$2$-hop MIS,
node/edge $2$-coloring of a path,
and node $4$-coloring of a rectangular grid.

\section{Additional Related Work and Discussion}
\label{section:related-work}

\paragraph{The Most Related Issues.}
We start with a summary of the most important points which are
(1)
previous adaptive algorithms
\cite{DBLP:journals/cjtcs/DolevH97,
	kutten1997time,afek1997self,
	herman2000phase,
	ghosh2002scalable,
	azar2003distributed,
	demirbas2004hierarchy,
	burman2005asynchronous,
	arora2006lsrp,
	DBLP:conf/wdag/KuttenM07,
	dolev2009empire,ghosh1996fault,
	DBLP:journals/dc/GhoshGHP07,
	dasgupta2007probabilistic,
	kohler2012fault,
	beauquier2006necessary,
	herman2000error,
	beauquier20061,
	turau2018computing,
	beauquier1999optimal,
	genolini2002lower,
	beauquier20061,
	balliu2021local}
were not fully adaptive;
(2)
previous transformers does not produce algorithms that are adaptive, especially 
outside the $\mathcal{LOCAL}$ model, since many has large messages, e.g., 
\cite{KatzPerry,
	awerbuch1991distributed,
	afek2002local,
	azar2003distributed,
	lenzen2009local,
	balliu2021local};
(3)
previous work found it hard for a transformer to handle randomized algorithms 
\cite{lenzen2009local,
	awerbuch1991distributed,
	aastrand2010fast,
	barenboim2018locally,
	Turau2019making,
	beauquier20061};
(4)
the algorithms produced by the transformer given here, even when 
taking $n=k$ (ignoring adaptivity), compare well to previous work on local self 
stabilizing algorithms
\cite{barenboim2018locally,
	Turau2019making,
	guellati2010survey,
	hsu1992self,
	ikeda2002space,
	kosowski2006self,
	sur1993self,
	tel1994maximal,
	cohen2016self,
	cohen2018self,
	hedetniemi2001maximal,
	goddard2003robust,
	goddard2003self,
	manne2009new,
	shi2004anonymous,
	asada2016efficient,
	kimoto2010time,
	arapoglu2019asynchronous,
	Turau2007linear,Hedetniemi2003self,
	Hedetniem2003coloring}
in terms of time complexity, message size, being size uniform, and more; and
(5)
the assumptions in many of the previous papers are stronger when speaking of 
self stabilizing algorithms (e.g., a global clock whose values are synchronized 
everywhere \cite{balliu2021local,dolev2009empire}, or that the adversary cannot 
change a unique ID, or the value of $n$  
\cite{barenboim2018locally,turau2018computing}).
Related to the definition of the propagation radius, we also bring a discussion 
of definitions of related radii in the literature.

Note that the more detailed survey of previous transformers in this section 
speaks of converting algorithms for related goals (e.g. self stabilization) but 
not for full adaptivity, randomized algorithms, size uniformity, or algorithms 
that are local in bandwidth constrained models, since no transformer satisfying 
these property has been presented before
(See Table~\ref{table:related-work-comparison}). Because of a similar reason, 
we survey adaptivity (rather than full adaptivity) including adaptivity in 
related 
models, and not just in self stabilization.
\begin{table}[t]
\centering
\begin{tabular}{|p{1.8cm}|p{1.0cm}|p{1.2cm}|p{1.1cm}|p{1.3cm}
|p{1.0cm}|p{1.2cm}|p{1.0cm}|p{1.3cm}|}
\hline
& (1)  \mbox{Local} under {\tiny \bf LOCAL} & (2) Fully-adapt.
& (3)
Adapt.
&
(4)  
\mbox{\tiny \bf CONGEST}  & 
(5)  \mbox{Const.} msg size 
& (6) \;\;\; Size \mbox{uniform} & (7)  Async. 
&
(8) \mbox{\small  General} 
\\
\hline
This \;\; paper& \CH & \CH & \CH &\CH  & \CH & \CH &  &from a class
\\
\hline
\cite{KatzPerry} +~e.g. \cite{afek1990memory} 	&  &  &  &  &  &  &  
\CH & \CH \\
\hline
\cite{awerbuch1991distributed} Rollback		& \CH &   &  &  &  & \CH 
&   & det. \\
\hline
\cite{awerbuch1991distributed} Resynch		&  &  &  &  \CH &  &  &  & 
det.\\
\hline
\cite{awerbuch1991self}		&  &  &  & \CH &  &  & \CH & from a class \\
\hline
\cite{awerbuch1994self}		&  &  &  & \CH &  &  & \CH & \CH \\
\hline
\cite{DBLP:journals/cjtcs/DolevH97}		&   & for topo. changes &  &   
&  &  &  & \CH \\
\hline
\cite{kutten1997time}		&  &  & $k<\frac{n}{2}$  &  &  &  & \CH & 
\CH \\
\hline
\cite{afek2002local}
&  &  &  &  &  &  &  & \CH \\
\hline
\cite{ghosh2002scalable}
&  & \CH &  &  &  &  &  &  on trees \\

\hline
\cite{azar2003distributed}
&  &  & \CH &  &  &  &  & \CH \\
\hline
\cite{burman2005asynchronous}		&  &  & \CH &  &  &  & \CH & \CH  \\

\hline
\cite{beauquier20061}		& \CH  &  & \CH &  &  &  & \CH & from a 
class  \\
\hline
\cite{lenzen2009local}	
& \CH &  &  &  &  & \CH &  & det. \\
\hline
\cite{kohler2012fault}
& \CH &  & for a single fault  & \CH &  \CH & \CH  & \CH & \CH \\
\hline
\cite{balliu2021local}
& \CH &  &  &  &  & \CH &  & \CH \\
\hline
\end{tabular}
\caption{\label{table:related-work-comparison}%
Comparing transformers for self stabilization:
(1)
Produces local algorithms under the $\mathcal{LOCAL}$ model;
(2)
Produces fully-adaptive algorithms;
(3)
Produces adaptive algorithms;
(4)
Produces $O(\log n)$ message size algorithms;
(5)
Produces constant message size algorithms;
(6)
Produces size uniform algorithms;
(7)
Converts asynchronous algorithms;
(8)
Transforms general algorithms.
}
\end{table}

\paragraph{Locally Checkable Labelings and Local algorithms.}
Distributed functions whose legal state can be expressed as a conjunction of 
local node states, one per node \cite{afek1990memory} or one per edge  
\cite{awerbuch1991self} where first introduced in the context of self 
stabilization.
Their appeal was that the detection of faults could be done in one time unit.
In \cite{naor1995can,mayer1995local}, who coined the name LCL for such 
functions, it was noted that the computation of such a function (to bring the 
configuration to be legal) could still take time that depends on the diameter 
of the network.\footnote{In \cite{naor1995can,mayer1995local}, as well as in 
some later studies, the network graphs were assumed to be of some constant 
degree; we consider and arbitrary degree that is bounded by some $\Delta$ that 
could be $n-1$ if the graph is finite.} Thus, they initiated the very rich 
study of the time it could take to compute an LCL in a fault free case, see,
e.g., a survey in \cite{DBLP:conf/swat/Suomela20}. Many such algorithms are 
``local'' in the sense that they terminate in time that is less than the 
diameter \cite{linial1987distributive}

The transformer presented here makes use of the fact
that phase based algorithms are rather common for LCLs, so the applicability of 
the transformer
presented here is rather wide. Some examples for additional phase based 
algorithms (beyond those converted in the current paper or mentioned elsewhere 
in this section) are \cite{awerbuch1994efficient,wattenhofer2004distributed}.

\paragraph{Non-Adaptive Self Stabilization for Local Algorithms.}
The research on self stabilization in general is far too wide to mention here. 
Refer to 
\cite{dolev2000self,altisen2019introduction,dubois2011taxonomy,schneider1993self,tixeuil2009self}.
For local algorithms, \cite{barenboim2018locally} presented self-stabilizing 
distributed algorithms for $\Delta+1$
vertex-coloring, $2\Delta-1$ edge-coloring, maximal independent
set and maximal matching that stabilize in $O(\Delta+\log^{\ast} n)$ time.
Let us note that the stabilization time of the algorithms presented here - even 
if we take $k$ to be $n$, is not comparable to that of  
\cite{barenboim2018locally} (since it would be $O(\log(\Delta +k))$).
In \cite{Turau2019making}, a self stabilizing randomized algorithm for MIS is 
presented, with stabilization time that is $O(\log n)$
where each node is required to know an upper bound on $n$ (no such requirement 
is needed in the algorithm presented here). That paper also present an $O(\log 
n)$  stabilization time randomized algorithm for maximal matching.
(It is worth mentioning that the author of \cite{Turau2019making} stated that a
transformer for a wide class of phase-based algorithms is the ultimate goal, a 
goal addressed in the current paper.)

Previously-known self stabilizing algorithms for such problems only promised
$O(n)$ or larger stabilization time 
\cite{guellati2010survey,hsu1992self,ikeda2002space,kosowski2006self,sur1993self,tel1994maximal,cohen2016self,
cohen2018self,
hedetniemi2001maximal, goddard2003robust, goddard2003self,
manne2009new,shi2004anonymous, asada2016efficient,
kimoto2010time,
arapoglu2019asynchronous,
Turau2007linear,Hedetniemi2003self,
Hedetniem2003coloring}.  See also
Table $1$ in \cite{ileri2019self}.
The algorithms of \cite{barenboim2018locally} are deterministic and make use of 
assumptions we do not used in the current paper, namely, that each node has a 
unique identity and knowledge of the maximum number of nodes $n$. Moreover, it 
is assumed that these values could not be changed by the adversary. Thus, these 
algorithms are not adaptive, not size uniform, not anonymous, not randomized, 
and cannot be used in infinite graphs even if $\Delta$ is bounded.

\paragraph{Adaptive Run-Time and Local (sublinear) Adaptive Run-Time (not 
necessarily self stabilizing).}
Research on adaptive algorithms outside of distributed computing is too wide to 
survey here, see, e.g., \cite{eppstein1999dynamic} or, specifically,
\cite{baswana2015fully, baswana2015fully, neiman2016simple,barenboim2017fully} 
for maximal matching (or approximated maximum matching), 
\cite{assadi2019fully,assadi2019fully,chechik2019fully} for MIS, 
\cite{barba2019dynamic,bhattacharya2018dynamic,solomon2020improved} for node 
coloring, and \cite{barenboim2017fully} for edge coloring.

Outside the scope of self stabilization but already in the area of distributed 
computing (including fault tolerance), various (often incomparable) notions of 
adaptivity were addressed. See for example
\cite{mayer1995local,kutten1995fault,DBLP:conf/focs/KuttenP95,
demirbas2004hierarchy,lotker2009distributed,konig2013local,
censor2016optimal,parter2016local,bamberger2018local,censor2019fast,
balliu2021local}.
The combination of adaptivity and locality was also addressed:
Very informally, given a local algorithm, e.g., one with complexity $(\log n)$,
a tight fault local algorithm strives for recovery times that was, e.g., 
$O(\log k)$ (rather than e.g. $O(k)$
\cite{DBLP:conf/focs/KuttenP95}.


In the context of self stabilization, multiple algorithms were presented that 
were adaptive
but not fully adaptive. In particular,
to obtain adaptivity, all the faults were assumed to occur in a single
batch before the mending starts. No additional faults were assumed to occur 
until the mending is done. The time between batches of faults
was called the \emph{fault gap} \cite{ghosh1996fault} (where a proof is also 
given that obtaining an $O(1)$ fault gap necessarily increases the 
stabilization time by at least a constant factor, see also 
\cite{kohler2012fault}).
In some of those papers, $k$ was assumed to be some known constant, or, at 
least, known in advance, and the size of the messages grew fast with $k$.
Such
adaptive self stabilizing algorithms whose stabilization times were at least 
linear in the number of faults (even in the $\mathcal{LOCAL}$ model)
were presented, e.g., in 
\cite{DBLP:journals/cjtcs/DolevH97,
	kutten1997time,
	afek1997self,
	herman2000phase,
	ghosh2002scalable,
	azar2003distributed,
	demirbas2004hierarchy,
	burman2005asynchronous,
	arora2006lsrp,
	DBLP:conf/wdag/KuttenM07}.
An adaptive MIS algorithm with $O(\log k)$ stabilization time
(when started from a legal configuration) is presented in 
\cite{dolev2009empire} for graphs with constant degree nodes.
To stabilize in $O(\log k)$ time, it is also assumed in \cite{dolev2009empire}
that nodes are synchronized on the same value of a clock. Recall that the 
\PPS{} method is presented here to avoid such an assumption.
A self-stabilizing paradigm to handle a single fault in constant time was 
suggested in
\cite{ghosh1996fault,DBLP:journals/dc/GhoshGHP07,dasgupta2007probabilistic,kohler2012fault,beauquier2006necessary,herman2000error,beauquier20061,turau2018computing}.
The more general case that $k$ may be larger than $1$ but still known in 
advance, was addressed in
\cite{beauquier1999optimal,genolini2002lower,beauquier20061}.

\paragraph{General Parameters for Adaptivity.}
Before mentioning parameters used in the literature, let us describe an 
intuition behind the influence number, defined and used extensively in the 
current paper. Intuitively, it reflects the following process: Let $v$ be a 
node for which the LCL predicate does not hold. Reset $v$'s output to $\bot$. 
As a result, the LCL predicate at some of $v$'s neighbors may stop holding; 
reset their inputs too and so forth. Let $u$ be the furthest node from $v$ 
whose value was reset in this process. The distance between $v$ and $u$ is the 
propagation radius of this predicate. Note that the propagation radius is a 
function of the (predicate of the) LCL problem. Let us note that resetting the 
variables of a node who noticed a violation of the LCL predicate is a common 
practice in self stabilization; to do while not resetting the whole network, a 
second, more relaxed LCL predicate is often defined to be transiently legal (to 
say that the algorithm is working on making the LCL predicate correct too) 
\cite{afek1997local,DBLP:journals/cjtcs/DolevH97}.

Let us note that one of the motivations stated in \cite{naor1995can} was that 
given an algorithm that can compute an LCL problem in a constant time would 
enjoy the following property:
a fault
at a processor $p$ could only affect processors in some bounded region around
$p$. This motivation seems similar to ours, though they have not given an exact 
definition of that region, nor shown how to use that for adaptivity.

A related parameter is the ``radius of the labeling'' for an LCL function 
\cite{mayer1995local}, defined in the context of dynamic graphs (without self 
stabilization).
First, they generalize the predicate of an LCL for a node $v$ to be evaluated 
for the output values
of all the nodes in the ball of some radius $r$ around $v$ (including $v$) 
where $r$ is a part of the revised definition of the LCL problem. Note that any 
radius $r$ can be defined for an LCL that way. However, for the radius to be 
meaningful for the local mending of in incorrect output, there should also 
exist an algorithm to correct the output such that this algorithm is ``robust'' 
for that radius.
Very intuitively, an algorithm is \emph{stable} in a node $v$ at time $t$ for 
radius $r$ if at time $t$ for which no changes occur in nodes in the $r$-ball 
around $v$, the algorithm does not change the output of $v$ at time $t$. An 
algorithm is $r$ robust if it is stable for any node with legal output (with 
respect to $v$'s $r$ ball). The exact relation between this radius and 
propagation radius is not clear.

Yet another parameter that sounds related is the ``contamination number'' which 
is the worst case number of processes that change their output value during 
recovery 
from a configuration with one fault into a legal configuration 
\cite{ghosh1996fault}. 
This parameter seems more related to measures such as messages complexity or 
the aggravated loss of work; its relation to time complexity seems more 
indirect. This may be the motivation of the variation of ``contamination 
radius'' used in \cite{turau2018computing}.

For another parameter, consider a partial configuration where some nodes have 
decided on their outputs while others have not. Consider a node $v$ whose 
output is not determined in the partial configuration. To assign an output to 
$v$, one may need to change also the outputs of some nodes in some radius $r$ 
around $v$. The maximum such radius for a given LCL and a given graph family is 
called the ``mending radius'' in \cite{balliu2021local}. Let us note that there 
exist LCLs with constant propagation radii and  large
mending radii and vice versa. 
Intuitively, process defined above to be captured by the propagation radius 
removes the inconsistencies but leaves the network with a configuration that is 
partial (some nodes are undecided). The mending radius captures a (not 
necessarily distributed) process that takes a partial configuration and 
completes it to a legal one, possibly needing to change some of the already 
assigned values.

\paragraph{Transformers.}

From the more theoretical point of view, a transformer is used to show the
reduction (or possibly the equivalence) of some source class of algorithms
(and/or problems) to some target class intended for problems that initially
seem ``harder''. Some known such transformers in the context of distributed
computing are those that allow the execution of synchronous algorithms on
asynchronous networks and those that allow the use of algorithms for static
networks to be performed on one that may undergo (detectable) topological
changes
(e.g., \cite{awerbuch1985complexity,afek1987applying}). Transformers
can also ease the design of algorithms in the target class
\cite{awerbuch1985complexity,KatzPerry}.
Some of the transformers mentioned below are more general than the one
presented in the current paper in either transforming algorithms where nodes
may also start with some input other than the graph, or algorithms for non-LCL
problems, or
even algorithms that never terminate and interact with users outside the
network.
Outside the realm of self stabilization, a transformer to make algorithms 
adaptive on the average (but not self stabilizing) is given in 
\cite{parter2016local}.
The first transformer to convert algorithms to be self stabilizing was 
presented already in 1990. All the full states of all the nodes were collected 
in \cite{KatzPerry} to a leader, and such a leader was elected in a self 
stabilized manner in \cite{afek1990memory} together with a spanning tree upon 
which to perform a reset if necessary. More efficient leader election 
algorithms and spanning tree construction algorithms were developed since then 
(e.g., \cite{afek1997self,datta2011self,awerbuch1993time}), but the 
stabilization time remained at least the diameter of the network graph. This 
``global'' stabilization time is also inherent in transformers that detect 
faults locally, using local checking, but globally ``reset'' all the nodes to 
restarts the converted algorithm from a legal configuration 
\cite{arora1990distributed,afek1990memory,awerbuch1993time,awerbuch1994self}.
Similarly, the ``local checking with local correction'' 
\cite{awerbuch1991self}, and the general ``superstabilizer''
\cite{DBLP:journals/cjtcs/DolevH97}
are inherently global in their stabilization time 
(depending on the distribution of faults) for a similar reason - the correction 
may touch all the nodes even if only a small number of them is faulty.

The ``rollback compiler'' in \cite{awerbuch1991distributed}
logs in each node, all the local history of the algorithm it transforms.
Moreover, each node exchanges its log with its neighbors in every round. That
way, the nodes can simulate the history repeatedly and verify that all the
actions were taken correctly, or correct them if they were not.
If the time complexity of a fault-free algorithm to be transformed is
$T$, then the stabilization time of the transformed algorithm is also $T$.
Indeed they mention as applications local algorithms such as maximal
independent set and $\Delta^2$ coloring.
A more detailed description of how this transformer applies to local algorithms
appears in \cite{lenzen2009local} who observed in  
\cite{awerbuch1991distributed,mayer1995local}
the strong connection between self stabilization and the preservation of the 
locality of the algorithms.
(In fact, it seems that the transformer described in \cite{lenzen2009local} is 
somewhat more economical than that of \cite{awerbuch1991distributed}.)
However, the transformer of \cite{awerbuch1991distributed,lenzen2009local} is 
only local under the $\mathcal{LOCAL}$ model, since the messages are very
large. Note that the resulting algorithm is not necessarily adaptive.

In addition, it is noted in \cite{lenzen2009local} that this transformer does 
not work for randomized algorithms. In fact, they state that to the best of 
their knowledge, little is known about which
randomized local algorithms can be made self-stabilizing efficiently. Several 
other papers mention this fact; for example, they mention that obtaining 
deterministic algorithms was useful since those algorithms could be made self 
stabilizing, see, e.g., \cite{aastrand2010fast,barenboim2018locally}.
This general difficulty motivated \cite{Turau2019making} to try to demonstrate 
how to convert non-stabilizing randomizing algorithms to stabilizing ones. The 
conversions was done manually for two specific algorithms. Interestingly, it is 
noted in \cite{Turau2019making} that it is difficult to convert phase based 
algorithms, for similar reasons to those explained in the current paper, see 
\Sect{}~\ref{section:transformer-overview}. That is why the algorithms in 
\cite{Turau2019making} are converted manually, and also why the method uses 
tweaks with phases in the original algorithms. Recall that we address this 
issue of phases in the current paper in a black box fashion. In 
\cite{boczkowski2017minimizing}, a self stabilizing (but not adaptive) 
algorithm is presented to synchronize nodes' phases synchronization in the PULL 
model in $O(\log n)$ expected stabilization time (for phases of constant 
length). This means that the diameter of the network graph is $1$, but each 
round, each node $v$ can communicate only with a constant number of other nodes 
(but $v$ is free to select which other nodes). 

Returning to the above mentioned idea of exchanging complete logs; this was 
pushed further in
\cite{afek2002local} to not only exchange logs with neighbors, but also to
flood the network with the log of every node in every round of computation
This idea can be applied to algorithms that never terminate and interact 
repeatedly with users outside the system. Such interactive algorithms are 
adaptive in a different sense than the one addressed in the current paper. That 
is, the stabilization time is stated to be proportional to the diameter in some 
cases, even if the number of faults is small.
However, assuming that faults occur in a node according to some probability, 
the probability of bad cases is small.
In addition, the duration of the time each node participates in the correction 
process is at most linear with the number of faults (though some nodes may be 
corrected long after that).

It may or may not be possible to develop also a time adaptive transformer based 
on the ideas behind the transformer of
\cite{afek2002local}.
However,
even if it does turned out possible, this transformer would generate
algorithms that are fully adaptive only in the $\mathcal{LOCAL}$ model (because 
of the huge messages) and with fully adaptive run time only for deterministic 
algorithms.

Another transformer that makes use of the locality of the converted fault-free 
algorithm to is sketched in \cite{balliu2021local} as an application of the
``mending radius'' they present (see discussion in the current section). The 
transformer maintains at each node $v$, a snapshot of the ball of nodes around 
$v$ whose radius is the mending radius. This approach is very heavy in 
communication requirements too and thus, yields algorithms that are not local 
in bandwidth constrained model.
Even in the $\mathcal{LOCAL}$ model, the transformer requires every node to 
know what is the current step (time slot) number in a phase whose length is not 
known in advance, but may be up to $(\Delta^{mending \; \; radius})$. This kind 
of synchronization requires $\Omega(diameter)$ time to stabilize 
\cite{boulinier2004graph} or, alternatively, requires an extra assumption that 
the phases are synchronized.
Even under such an assumption, the stabilization time is $\Omega(\min\{ 
\Delta^{mending \;\; radius},n\})$ (which may no longer be local).
Recall that the current paper introduces the \PPS{} method to avoid such 
assumptions.

Other 'brute force'' transformers for deterministic algorithms
by replicating states and using huge messages are those of 
\cite{kutten1997time,burman2005asynchronous}.
Those do obtain time adaptive self stabilizing algorithms but again,
only under the
$\mathcal{LOCAL}$ model. Moreover, they were not \emph{fully} adaptive. In 
addition, they were not tight fault local - the dependence of the stabilization 
time on the number of faults is linear (rather than logarithmic, for our 
algorithms).

When restricting the attention to networks whose graphs are trees, a 
transformation method was presented in \cite{ghosh2002scalable} that uses 
possibly less space (and smaller communication bandwidth, or messages)
- $O(\Delta(\Delta k + \log^2 n))$  but possibly longer time ($O(k^2)$ for some 
problems, $O(k)$ for others).

A transformer of any silent \cite{dolev1999memory} self stabilizing algorithm 
into a 1-fault containing one is presented in  \cite{kohler2012fault}. Recall 
that an error confined algorithm stabilizes faster when suffering a single 
fault, in a sense, such an algorithm is partially adaptive. A methodology for 
designing such a transformer is given in \cite{beauquier2006necessary}. An 
actual transformer that also provides the extra property of strong confinement  
is given in \cite{beauquier20061}. (Strong confinement means that a non-faulty 
node has the same behavior with or without the present of faults elsewhere in 
the network.) They also discuss a direction for expanding the result to a 
larger number $k$ known in advance. The stabilization time with $k$ faults 
would be linear in $k$ in the $\mathcal{LOCAL}$ model but would be heavy in 
communication since every node needs to maintain a snapshot to a distance $k$ 
continuously.

\clearpage
\bibliographystyle{alpha}
\bibliography{references}

\clearpage
\appendix
\begin{figure}[!t]
{\centering
\Large{APPENDIX}
\par}
\end{figure}

\section{Proving \Lem{}~\ref{lemma:pps:start-phase-probability}}
\label{appendix:proof:lemma:pps:start-phase-probability}
\Lem{}~\ref{lemma:pps:start-phase-probability} is established by plugging
$\epsilon = 1 - \frac{\phi + 2}{2 \phi}$
into the following more general lemma.

\begin{lemma}\label{lemma:transformer:start-phase-probability}
Fix some time
$t_{0} \geq t^{*}$,
node $v$,
and
$j_{0} \in \mathcal{S}_{\phi}$.
For every
$0 < \epsilon < 1$,
there exists a time
$\hat{t} = t_{0} + O (\log (1 / \epsilon) \cdot \phi^{3})$
such that for every
$t \geq \hat{t}$,
it holds that
\[
\Pr \left( \Step_{v, t} = j \mid \Step_{v, t_{0}} = j_{0} \right)
\, \geq \,
\begin{cases}
\frac{2}{\phi + 2} \cdot (1 - \epsilon) \, ,
&
\text{if }
j = \HoldSymbol \\
\frac{1}{\phi + 2} \cdot (1 - \epsilon) \, ,
&
\text{otherwise}
\end{cases} \, .
\]
This holds independently of any coin toss of $v$ prior to time $t_{0}$ and
of any coin toss of all other nodes.
\end{lemma}
\begin{proof}
To avoid cumbersome notation, we assume throughout this proof that
$t_{0} = 0$.
Let
$\{X_{t}\}_{t=0}^{\infty}$ be a stochastic process such that $X_{t} =
\Step_{v,t}$. The stochastic process $\{X_{t}\}_{t=0}^{\infty}$ can be
defined
by the discrete time Markov chain $M_{\phi}$ with state space $\mathcal{S}_{\phi}$ as
in
\Fig{}~\ref{figure:transformer:markov-chain}.

Let
$P_{\phi} \in \Reals^{\mathcal{S}_{\phi} \times \mathcal{S}_{\phi}}_{\geq 0}$
be the time-homogeneous transition matrix of the Markov chain $M_{\phi}$ and
$P_{\phi}^{t}$ the matrix $P_{\phi}$ to the power of $t$. Denote entry
$(i,j)
\in \mathcal{S}_{\phi} \times \mathcal{S}_{\phi}$ of $P_{\phi}$ by $P_{\phi}(i,j)$ and recall
that
for every $t,s \in \mathbb{Z}_{\geq 0}$ such that $t \geq 1$, it holds that
$P_{\phi}^{t}(i,j)=\Pr\left(X_{t+s} = j \mid X_{s} = i\right)$.
For convenience sake, let $n=|\mathcal{S}_{\phi}|=\phi + 1$.

The chain $M_{\phi}$ is ergodic, thus there exists a unique stationary
distribution on the state space $\mathcal{S}_{\phi}$, which we denote by the size $n$
vector
$\boldsymbol{\pi} \in \Reals_{>0}^{\mathcal{S}_{\phi}}$. It is easily verifiable that
$\boldsymbol{\pi}(\HoldSymbol)=\frac{2}{n+1}$
and
$\boldsymbol{\pi}(j)=\frac{1}{n+1}$
for every
$j \in \mathcal{S}_{\phi}-\{\HoldSymbol\}$.
We will prove that there exists
$\hat{t} = O(\log(1/\epsilon)\cdot n^{3})$
such that for every $t \geq \hat{t}$,
it holds that
\begin{equation}\label{eq:transformer:markov-chain-in-state-hbar}
P_{\phi}^{t}(j_{0},\HoldSymbol) \geq \frac{2}{n+1}(1-\epsilon)
\end{equation}
and
\begin{equation}\label{eq:transformer:markov-chain-in-state}
P_{\phi}^{t}(j_{0},j) \geq \frac{1}{n+1}(1-\epsilon), \forall j \in
\mathcal{S}_{\phi}-\{\HoldSymbol\}.
\end{equation}

The chain $M_{\phi}$ is a special case of the \emph{coin-flip} chain as 
defined
in \cite{wilmer1999exact} with $p=q=1/2$.
By Proposition $5.8$, Theorem $3.6$ and Lemma $2.4$ in 
\cite{wilmer1999exact},
for every positive $\epsilon < 1$ there exists a time
$t(n) = O(n^{3})$ such that for every
$t \geq O(\log(1/\epsilon) \cdot t(n))$, and $i \in \mathcal{S}_{\phi}$
\begin{equation*}
\max \limits_{j \in \mathcal{S}_{\phi}}
\left|
\frac{P_{\phi}^{t}(i,j)}{\boldsymbol{\pi}(j)} -1
\right|
\leq \epsilon.
\end{equation*}
Thus, for every $t \geq O(\log(1/\epsilon)\cdot n^{3})$ we can low-bound
\Eq{}~\ref{eq:transformer:markov-chain-in-state-hbar}--\ref{eq:transformer:markov-chain-in-state}
by,
\begin{equation*}\label{eq:transformer:markov-chain-in-state-low-bound}
P^{t}_{\phi}(j_{0},j)
\geq
\boldsymbol{\pi}(j) \cdot (1-\epsilon).
\end{equation*}
The proof is completed by the value of $\boldsymbol{\pi}$.
\end{proof}

\clearpage
\begin{figure}[!t]
{\centering
\Large{FIGURES}
\par}
\end{figure}

\begin{figure}
\begin{center}
\begin{tikzpicture} [->,>=stealth,shorten >=1pt,auto,node distance=3cm,
thick,state/.style={circle,draw,font=\Large\bfseries, minimum size 
=0.2cm}]
\def\radius {2.3cm}
\node [state] (v0) at (18:\radius)  {$ 0 $};
\node [state] (v1) at (306:\radius)  {$ 1 $};
\node [circle] (vd) at (234:\radius)  {\LARGE{$\vdots$}};
\node [state] (vp) at (162:\radius)  {\tiny{$\boldsymbol{\phi-1}$}};
\node [state] (vh) at (90:\radius)  {$ \HoldSymbol $};
\draw (vh) to [out=120,in=60,looseness=12] node[above]{\small{$1/2$}}
(vh);
\draw [bend left=25] (vh) to node[above]{\small{$1/2$}} (v0);
\draw [bend left=25] (v0) to node[right]{\small{$1$}} (v1);
\draw [bend left=25] (v1) to node[below]{\small{$1$}} (vd);
\draw [bend left=25] (vd) to node[left]{\small{$1$}} (vp);
\draw [bend left=25] (vp) to node[above]{\small{$1$}} (vh);
\end{tikzpicture}
\end{center}
\caption{\label{figure:transformer:markov-chain}%
The underlying Markov chain of \PPS{} with state space
$\mathcal{S}_{\phi} = \{ \HoldSymbol, 0, \dots, \phi - 1 \}$.}
\end{figure}

\begin{figure}
\centering
\begin{subfigure}[b]{0.4\textwidth}
\centering
\scalebox{.4}{\includegraphics[width=\textwidth]{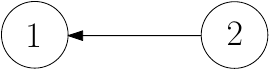}}
\caption{MIS/MM}
\label{figure:supportive-mis-mm}
\end{subfigure}
\hfill
\begin{subfigure}[b]{0.4\textwidth}
\centering
\includegraphics[width=\textwidth]{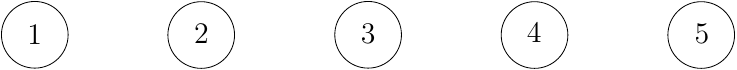}
\caption{node/edge $5$-coloring}
\label{figure:supportive-coloring}
\end{subfigure}
\hfill
\begin{subfigure}[b]{0.4\textwidth}
\centering
\includegraphics[width=\textwidth]{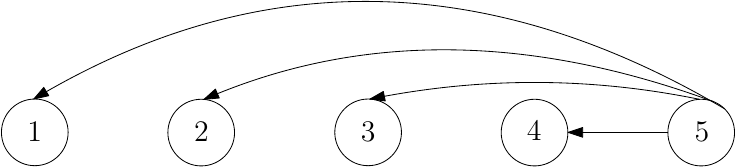}
\caption{maximal node/edge $5$-coloring}
\label{figure:supportive-maximal}
\end{subfigure}
\hfill
\begin{subfigure}[b]{0.4\textwidth}
\centering
\includegraphics[width=\textwidth]{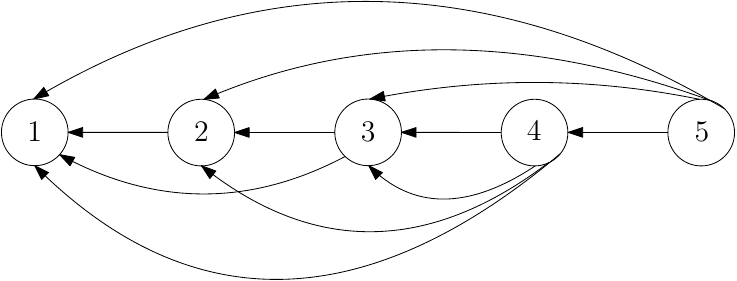}
\caption{incremental node/edge $5$-coloring}
\label{figure:supportive-incremental}
\end{subfigure}
\caption{\label{figure:supportive-digraphs}%
The supportive digraphs corresponding to the four concrete LCL predicates
listed in \Sect{}~\ref{section:introduction}.}
\end{figure}

\end{document}